\documentclass[12pt]{article}
\usepackage[utf8]{inputenc}
\usepackage{graphicx}
\usepackage{bbm}
\usepackage{threeparttable}
\usepackage{amssymb, amsmath}
\usepackage{mathtools}
\usepackage{bbm}
\usepackage{tablefootnote}
\usepackage{caption}
\usepackage{etoolbox}
\appto\TPTnoteSettings{\footnotesize}

\def\cirnum#1{\raisebox{.5pt}{\textcircled{\raisebox{-.9pt} {#1}}}}

\newcommand{\blind}{0}

\DeclareMathAlphabet{\mathpzc}{OT1}{pzc}{m}{it}
\usepackage[colorlinks = true,
            linkcolor = blue,
            urlcolor  = blue,
            citecolor = blue,
            anchorcolor = blue]{hyperref}
\usepackage{subfig}
\usepackage{float}
\usepackage[toc,page,titletoc]{appendix}
\usepackage{pdfpages}

\newtheorem{definition}{Definition}
\newtheorem{remark}{Remark}

\newcommand{\obtrade}{\texttt{(t)}}
\newcommand{\obinsert}{\texttt{(i)}}
\newcommand{\obcancel}{\texttt{(c)}}
\newcommand{\floor}[1]{\lfloor #1 \rfloor}

\DeclareMathOperator*{\argmin}

\begin{document}
\def\spacingset#1{\renewcommand{\baselinestretch}%
{#1}\small\normalsize} \spacingset{1}

\if0\blind
{
    \title{\bf Order Book Queue Hawkes-Markovian Modeling}
    \author{
    Philip Protter \thanks{
    \textit{The research of Dr. Philip Protter is partially supported by NSF Grant DMS-2106433.}}\hspace{.2cm} \\
    Department of Statistics, Columbia University\\
    and\\
    Qianfan Wu\\
    Kelley School of Business, Indiana University Bloomington\\
    and\\
    Shihao Yang\thanks{
    \textit{Corresponding Author. Email: shihao.yang@isye.gatech.edu. The authors thank Dr. Dan Christina Wang for providing access to the LOBSTER data used in this study.}}\hspace{.2cm}\\
    H. Milton Stewart School of Industrial and Systems Engineering\\
    Georgia Institute of Technology}
    \maketitle

} 

\if0\blind
{
  \bigskip
  \bigskip
  \bigskip
  \begin{center}
    {\LARGE\bf Order Book Queue Hawkes-Markovian Modeling}
\end{center}
  \medskip
} \fi

\bigskip

\newpage
\begin{abstract}
This article presents a Hawkes process model with Markovian baseline intensities for high-frequency order book data modeling. We classify intraday order book trading events into a range of categories based on their order types and the price changes after their arrivals. To capture the stimulating effects between multiple types of order book events, we use the multivariate Hawkes process to model the self- and mutually-exciting event arrivals. We also integrate a Markovian baseline intensity into the event arrival dynamic, by including the impacts of order book liquidity state and time factor to the baseline intensity. A regression-based non-parametric estimation procedure is adopted to estimate the model parameters in our Hawkes+Markovian model. To eliminate redundant model parameters, LASSO regularization is incorporated in the estimation procedure. Besides, model selection method based on Akaike Information Criteria is applied to evaluate the effect of each part of the proposed model. An implementation example based on real LOB data is provided. Through the example, we study the empirical shapes of Hawkes excitement functions, the effects of liquidity state as well as time factors, the LASSO variable selection, and the explanatory power of Hawkes and Markovian elements to the dynamics of the order book.
    
\end{abstract}

\noindent%
\vfill

\newpage
\spacingset{1.5}
\section{Introduction}
\addtolength{\textheight}{.5in}%

An Electronic Limit Order Book (LOB) is a list of electronic orders that a trading venue uses to record the interest of buyers and sellers in a particular financial instrument. The modern financial market has witnessed unprecedented increases in trading volume and frequency during the recent decades, with the global total value of stock traded escalating and with the average stock holding period plunged significantly from 1990 to 2018 \cite{WB:2019, Maloney2019lengthening}. Therefore, understanding the dynamics of the LOB has become increasingly significant in the analysis of liquidity, transaction costs, and the regulation of the modern global financial market. 

A Hawkes process~\cite{hawkes1971spectra,hawkes1974cluster}, also called ``self-exciting and mutually(cross)-exciting point process'', is a type of stochastic point process whose essential property is that the arrival/\\occurrence of any event will increase the arrival probability of further events. As electronic LOB data is typically complex, large-scale, and high-frequency, Hawkes process has become increasingly popular in fitting LOB dynamics due to its ability to capture the complex stimulating effects between event flows. Bacry \& Muzy \cite{bacry2015hawkes} and Hawkes \cite{ hawkes2018hawkes} have provided comprehensive reviews on Hawkes process applications in finance, especially in modeling LOB data, including related works in estimation procedure \cite{kirchner2017estimation, bacry2014hawkes}, Hakwes process generalizations \cite{blanc2017quadratic}, and model modifications of minor details \cite{clements2015modelling,ferriani2020dynamics}. In addition, Hawkes process with exponential kernels is among the most popular model specifications for LOB modeling in recent works \cite{muni2011modelling,abergel2015long,morariu2018state,kirchner2017estimation}.

Besides the self and cross-exciting property, LOB data can also be viewed as a chain of transitions from one state to another based on different price and order size levels as different types of orders arrive. Based on this perspective, Markov models can be applied to LOB data, assuming the transition from one LOB state to another as an event arrives depends only on the state attained in the previous event. For example, Huang et al. \cite{huang2015simulating} and Huang \& Rosenbaum \cite{huang2017ergodicity} propose a simulation and analytical framework for Markov models on LOB data and demonstrates empirical findings on various LOB event types; Kelly \& Yudovina \cite{kelly2018markov} discuss a tractable Markov LOB model on short time scales and provides applications in high-frequency trading; Morariu-Patrichi \& Pakkanen \cite{morariu2018state} estimates a state-dependent Hawkes model for LOB modeling in which the exponential excitement kernel depends on the state process that switches state whenever events arrive. 
\subsection*{Our contribution}

Most of the previous works mentioned above apply Hawkes process and Markov models independently to LOB data. The effectiveness of combining the two models in LOB modeling is not well studied. Moreover, assuming a parametric form of Hawkes excitement kernels, such as an exponential distribution, gives less flexibility for estimating kernels that exhibit more complex shapes other than the assumed parametric distribution. In this paper, we introduce a Hawkes+Markovian model to capture the dynamics of electronic LOB data. We apply a multi-dimensional Hawkes process on a comprehensive range of event types derived from LOB movements. In addition, by integrating the Markovian model on the LOB data, our model captures the intuition that the stimulating effects among events also depend on the liquidity state and time factor of the LOB right before event arrives. Instead of assuming the Hawkes kernels to follow parametric distributions from the exponential family, we implement a regression-based non-parametric method \cite{kirchner2017estimation} for model parameter estimation. The main idea of the estimation procedure is to allocate the event arrival sequences into a series of fixed-size bins of discretized short-period, and then obtain the Hawkes kernel estimators as a step function. In contrast to previous works \cite{kirchner2017estimation, morariu2018state} that do not consider order sizes, we do take order sizes into account in the arrival sequence construction to better capture the stimulating effects of large and small orders. As the number of estimated parameters is large, we also incorporate the use of LASSO regularization \cite{tibshirani1996regression} in our model estimation. Furthermore, we propose a model selection method based on Akaike Information Criterion (AIC) to analyze the contribution of the Hawkes stimulation part, the Markovian part, and the LASSO part to model explanatory power. Finally, we demonstrate an implementation example using real order book data.

The rest of this paper is organized as follows: Section \ref{bigsec:order book representation} demonstrates a brief mathematical introduction of order book data representations;  Section \ref{bigsec:model specification} introduces our proposed model, with detailed illustrations on event classification and model structures; Section \ref{bigset:estimation procedure} provides detailed descriptions of the non-parametric model estimation as well as model selection; Section \ref{bigsec:example} showcases some aggregated empirical results from an implementation example based on real LOB data.

\section{Order book representation}\label{bigsec:order book representation}

\begin{figure}[H]
\centering
\includegraphics[scale=0.30]{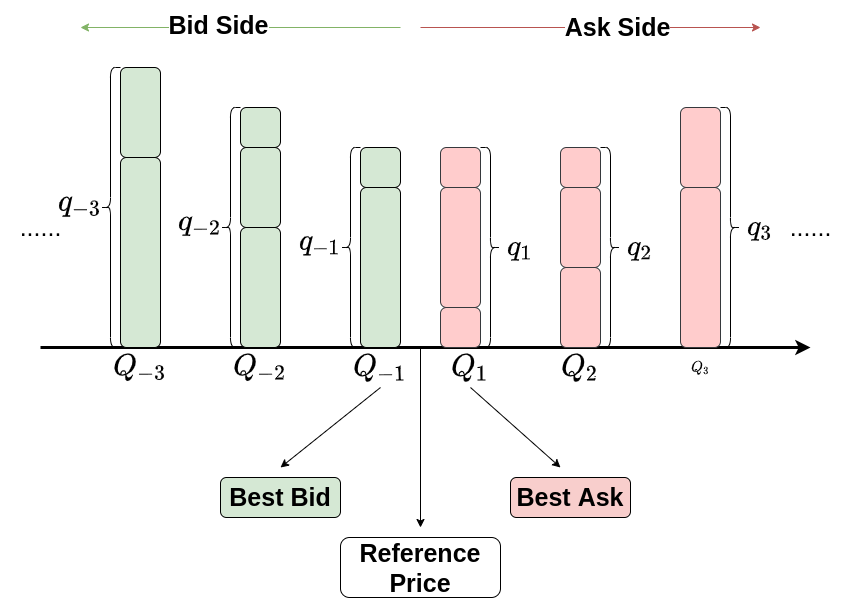}
\caption{A simple order book representation. ($Q_{-1}$, $Q_{-2}$, $Q_{-3}$) / ($Q_{1}$, $Q_{2}$, $Q_{3}$) represent the first, second, and third price level on the bid/ask side, respectively; ($q_{-1}$, $q_{-2}$, $q_{-3}$) / ($q_{1}$, $q_{2}$, $q_{3}$) represent the number of orders on the first, second, and third price level on the bid/ask side, respectively. }
\label{fig: order book representation}
\end{figure}

The limit order book is mainly constructed by two elements. The first element is the shape of the order book, consisting of all the orders at which prices the market wants to buy (bid price) and the market wants to sell (ask price). The bid/ask prices form the bid/ask queues and must be multiples of the tick size, which is the measure of the minimum upward or downward movement of security prices (currently the tick size for all U.S exchanges is \$0.01). The second element is the center position of the order book between the best bid price (the highest price the market wants to buy) and the best ask price (the lowest price the market wants to sell). The center position of LOB is often referred to as the ``reference price'' and the distance between the best bid and best ask is referred to as ``bid-ask spread''. The easiest way to approximate the reference price is to define it as the midpoint of the best bid and best ask, also known as the ``mid price''. The order book also includes information on the size of each order, which is the quantity of shares an order attempts to execute. See Figure \ref{fig: order book representation} for an example.

\paragraph{Mathematical Representation} 
Huang et al. \cite{huang2015simulating} and Huang \& Rosenbaum \cite{huang2017ergodicity} demonstrate a comprehensive overview of the mathematical representations of LOB. Recall the reference price $p_{\text{ref}}(t)$ must lie strictly between the best bid and best ask. Let $q_i(t)$ be the size of the ask orders at price level $Q_i$ that is the $i$-th tick strictly above $p_{\text{ref}}(t)$. Symmetrically, $q_{-i}(t)$ is the negation of size of the bid orders at price level $Q_{-i}$ that is the $i$-th tick strictly below $p_{\text{ref}}(t)$.

Formally, let $\alpha$ be the single tick value, then $Q_1 := \min\{n\alpha: n\alpha > p_{\text{ref}}, n \in \mathbb{Z}\}$, $Q_i := Q_1 + (i-1)\alpha$, $\forall i > 1$. Similarly, $Q_{-1} := \max\{n\alpha: n\alpha < p_{\text{ref}}, n \in \mathbb{Z}\}$, $Q_{-i} := Q_{-1} - (i-1)\alpha$, $\forall i > 1$. See Figure \ref{fig: order book representation} and its caption for an example.

The complete order book’s shape at time $t$ is an infinite vector for the current size at all prices $q(t) = [\ldots,q_{-k}(t),\ldots,q_{-1}(t),q_1(t) ,\ldots,q_k(t),\ldots]$ with $q_i \in \mathbb{Z}$ denoting the size at each price. Note $q_i < 0$ if these orders are bid orders and $q_i > 0$ if they are ask orders. If $q_i = 0$ there is no orders at price level $Q_i$. Since $q_{i}$ can be $0$, the best bid and best ask prices are defined as the nearest price levels to the reference price with non-empty orders sizes:
\[
Q_\text{best-bid} = Q_{max\{i: i<0 \text{ and } |q_i|\neq0\}} \text{ , } Q_\text{best-ask} = Q_{min\{i: i>0 \text{ and } |q_i|\neq0\}}
\]

The reference price, $p_{\text{ref}}(t)$, is often the mid-price, with some technicality detailed in Supplementary Material Section \ref{sec_support: reference price}. The LOB information at time $t$ is therefore fully represented by $[q(t),p_{\text{ref}}(t)]$, $t \ge 0$. To restrict the dimensions of $[q(t),p_{\text{ref}}(t)]$, we consider only $K$ limits on each side, and thus have now $q(t) = [q_{-K}(t),\ldots,q_{-1}(t),q_1(t),\ldots,q_K(t)]$, which we shall call ``level-$K$ order book''.

\section{Model specifications}\label{bigsec:model specification}
Following the framework we outlined above, we propose a specific event arrival dynamic for empirical modeling of the level-$K$ order book.

We consider the LOB event arrival process as $6\times(K + 1)$ dimensional, indicating $6\times(K + 1)$ types of events are studied in the level-$K$ order book. They are generally grouped into two categories: (1) order book events that do not change reference price, and (2) order book events that change the reference price. The way we classify LOB events is mainly extended from previous work on the ``Queue-Reactive'' LOB model \cite{huang2015simulating,huang2017ergodicity}. Compared to the simpler LOB event classifications used in \cite{kirchner2017estimation,morariu2018state}, our classification not only separates events belonging to each of the $K$ order book levels, but also classifies events in more detailed groups when the reference price changes, enabling our proposed model to capture more complex dynamics of event stimulation.

\subsection{Order book events that do not change reference price} \label{sec: price not change events}
When studying the level-$K$ order book, each order book queue can have the following events:
\begin{itemize}
	\item a trade, or market order, which is denoted as \obtrade
	\item an insertion of new limit order, which is denoted as \obinsert
	\item a cancellation of existing limit order, which is denoted as \obcancel
\end{itemize}
For level-$K$ order book, we have in total $2K$ number of queues:
\begin{itemize}
	\item $K$ ask queues, denoted as $+1, +2, \ldots, +K$, which are the 1st tick above reference price, 2nd tick above reference price, ..., $K$-th tick above reference price, respectively.
	\item $K$ bid queues, denoted as $-1, -2, \ldots, -K$, which are the 1st tick below reference price, 2nd tick below reference price, ..., $K$-th tick below reference price, respectively.
\end{itemize}
Therefore, we have in total $3\times 2K$ number of events that do not change reference price. For example, ``\texttt{+1\obtrade}'' denotes the event that a trade order arrives at the first ask queue, ``\texttt{-2\obinsert}'' denotes the event that a new order arrives at the second bid queue and is inserted, and ``\texttt{+3\obcancel}'' denotes the event that an existing order is canceled at the third ask queue. In our framework, higher level trade orders such as ``\texttt{+2\obtrade}'', ``\texttt{+3\obtrade}'', ``\texttt{-2\obtrade}'', and ``\texttt{-3\obtrade}'' are possible when there is no order above/below that order book trade level. For instance, ``\texttt{+2\obtrade}'' is possible when the queue at the first ask queue (1 tick above the reference price) is empty, meaning the price at the 2nd-level (2 ticks above the reference price) is the current best execution price.

\subsection{Order book events with reference price change} \label{sec: price change events}
This section focuses on the modeling for another 6 types of events that shift the reference price. The reference price can increase due to the following event:
\begin{itemize}
\item Trade at the best ask price that depletes the queue of best ask. The event of reference price increase due to trade is denoted as \texttt{p+(t)}
\item Cancellation of all orders at the best ask price. The event of reference price increase due to cancellation is denoted as \texttt{p+(c)}
\item Insertion of bid order at a higher price than the current best bid offer, which is only possible when the bid-ask spread is strictly larger than one tick. The event of reference price increase due to insertion is denoted as \texttt{p+(i)}
\end{itemize}
On the flip side, the reference price can decrease due to the following event:
\begin{itemize}
\item Trade at the best bid price that depletes the queue of best bid. The event of reference price decrease due to trade is denoted as \texttt{p-(t)}
\item Cancellation of all orders at the best bid price. The event of reference price decrease due to cancellation is denoted as \texttt{p-(c)}
\item Insertion of ask order at a lower price than the current best ask offer, which is only possible when the bid-ask spread is strictly larger than one tick. The event of reference price decrease due to insertion is denoted as \texttt{p-(i)}
\end{itemize}

\subsection{Hawkes+Markovian model of the order book} \label{sec:model of the order book}

The order book event processes of dimension $6K+6$ at time $t$ are represented as $X_{i}(t),i = 1,2,\ldots,6K+6$, with $X_{1},X_{2}, \ldots,X_{6K+6}$ representing the arrival process for the following events:
\[
\begin{pmatrix}
X_{1} \\ X_{2} \\ \vdots \\ X_{6K+6}
\end{pmatrix}
 = 
\begin{pmatrix}
\underbrace{(\texttt{-K(i)}, \texttt{-K(c)}, \texttt{-K(t)}, \ldots, \texttt{-1(i)}, \texttt{-1(c)}, \texttt{-1(t)})^\intercal}_{3K} \\
\underbrace{(\texttt{+1(i)}, \texttt{+1(c)}, \texttt{+1(t)}, \ldots, \texttt{+K(i)}, \texttt{+K(c)}, \texttt{+K(t)})^\intercal}_{3K} \\
\underbrace{(\texttt{p-(t)}, \texttt{p-(c)}, \texttt{p-(i)}, \texttt{p+(t)}, \texttt{p+(c)}, \texttt{p+(i)})^\intercal}_{6}
\end{pmatrix}_{(6K+6) \times 1}
\]
Basically, $X_{i}(t)$ represents the cumulative size of the corresponding event. For example, $X_1(t)$ corresponds to event \texttt{-K(i)}, the insertion of limit order at the bid queue $K$-th tick below reference price; $X_{6K+6}(t)$ corresponds to event \texttt{p+(i)}, price increase due to insertion of bid above current best bid.

\subsubsection{Hawkes part}
Define the instantaneous rate of event $i$'s arrival to be
\[
\lambda_{i}(t) = \lim_{\Delta t \to 0+} \frac{\mathbb{E}(X_i(t+\Delta t) - X_i(t) | \mathcal{F}_t)}{\Delta t}
\]
$\forall i = 1, 2, \ldots, 6K+6$, where $\mathcal{F}_t$ is the filtration generated by $X$ at time $t$.
Then the intensity of event $i$ for a plain-vanilla multivariate Hawkes process model for LOB data can be represented as:
\begin{equation}\label{eq:hawkes-part}
\lambda_{i}(t) = \eta_{i} + \sum_{j=1}^{6K+6} \int_{0}^{t} \phi_{ji} (t-s) dX_j(s), \quad \forall i = 1,2,\ldots,6K+6.
\end{equation}
in which $\eta_{i}$ represents the baseline intensity, $\phi_{ji}(\cdot)$ represents the Hawkes kernel for event $j$ stimulating event $i$.

\subsubsection{Markovian part}
\begin{figure}[H]
\centering
\includegraphics[scale=0.35]{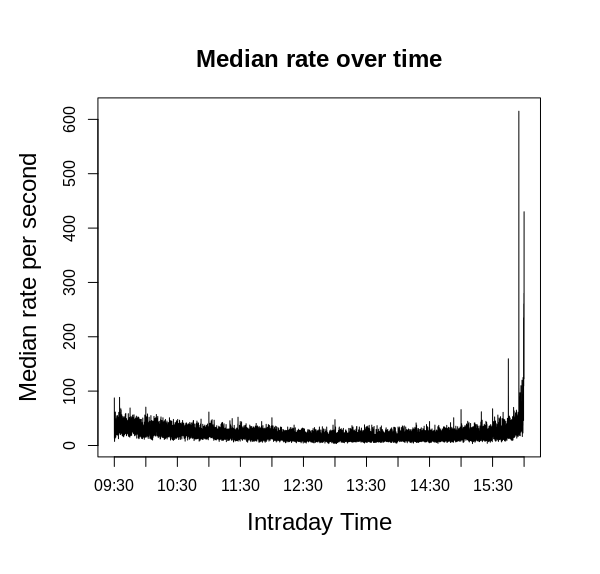}
\caption{An illustration of the order arrival rate for LOB data. The figure demonstrates median order arrival rates per second for 20 trading days from 1/2/2019 to 1/31/2019 based on LOB data for Apple.Inc.}
%We choose $K=3$ so that 24 types of events $\{\texttt{-K(+)}, \texttt{-K(-)}, \texttt{-K(t)}, \ldots,  \texttt{+K(+)}, \texttt{+K(-)}, \texttt{+K(t)}, \texttt{p-(t)}, \texttt{p-(-)}, \texttt{p-(+)}, \texttt{p+(t)}, \texttt{p+(-)}, \texttt{p+(+)}\}$ are considered in data processing.
\label{fig: rough arrival rate}
\end{figure}

To further extend on the plain-vanilla multivariate Hawkes process, we integrate two factors derived from LOB queues to the baseline intensity (intercept) $\eta_i$ of the Hawkes process: 
\begin{itemize}
    \item \textbf{The current LOB liquidity state:} 
    The liquidity state includes the number of existing orders on the associated price of the event, the number of orders on the best bid/ask price, and the bid-ask spread.
    \item \textbf{Time clustering:} Figure \ref{fig: rough arrival rate} indicates that the trading frequency for real LOB data differs between time groups throughout trading hours. The last 30 minutes have significantly more activities, and the first 30 minutes see slightly higher activities, while other periods during the day seem to be tranquil. 
\end{itemize}
We call the above two factors ``Markovian" factors because these factors carry the present information only, which is the current state of LOB, represented by liquidity and time. Under this setting, we assume that the dynamics of the LOB is affected by the LOB liquidity and time only through the current state. This type of modeling is consistent with the LOB Markovian models proposed by \cite{huang2015simulating, huang2017ergodicity}.
Therefore, we model the baseline event arrival intensities $\eta_i$ with a Markovian structure that depends on the current liquidity state and time clustering:
\begin{equation}\label{eq:markovian-part}
\eta_i(t) =  M_i(l_{i}(t)) + \Theta_i(t)
\end{equation}
where $M_{i}(\cdot)$ and $\Theta_i(\cdot)$ denote specified functions on liquidity state $l_{i}(t)$ and time $t$. 

Next we shall define the liquidity state $l_i(t)$. Note $l_{i}(t)$ only changes if an event arrives. 

For all events that do not change the reference price (i.e., \texttt{-K(i)}, \texttt{-K(c)}, \texttt{-K(t)}, \ldots, \texttt{+K(i)}, \texttt{+K(c)}, \texttt{+K(t)}), the liquidity state $l_i(t)$ is intuitively $q_{k_{i}}(t)$, the accumulated order size right before time $t$ on the $k_{i}$-th price level where event $i$ belongs. For example, events {\texttt{+1(i)},\texttt{+1(c)},\texttt{+1(t)}} have $k_i = 1$ and $l_i(t) = q_{1}(t)$.

Then we consider the events changing the reference price. For events \texttt{p-(t)} and \texttt{p-(c)},  $l_{i}(t)$ is defined as the queue size at the best bid price. For  events \texttt{p+(t)} and \texttt{p+(c)}, $l_{i}(t)$ is defined as the queue size at best ask price; this is because the queue at the best bid/ask must be either consumed or canceled for these event types. For events \texttt{p+(i)} and \texttt{p-(i)}, we consider order insertions within the bid-ask spread, and therefore the bid-ask spread is considered as the liquidity state $l_{i}(t)$ for these two types of events. \texttt{p+(i)} and \texttt{p-(i)} are the only two types of events whose liquidity state is based on price (bid-ask spread) while the liquidity state of the rest of the events is based on existing order size at the associated price.

Formally, we have the mathematical definition: 
\begin{align*}
l_i(t)  =\begin{cases}
q_{k_{i}}(t), ~ i \in \{\texttt{-K(i)}, \texttt{-K(c)}, \ldots,\texttt{+K(i)},\texttt{+K(c)},\texttt{+K(t)} \} \text{ (queue size)}\\
q_\text{best-bid}(t), ~ i \in \{\texttt{p-(t)}, \texttt{p-(c)}\} \text{ (size of the best-bid queue)}\\
q_\text{best-ask}(t), ~ i \in \{\texttt{p+(t)}, \texttt{p+(c)}\} \text{ (size of the best-ask queue)}\\
Q_\text{best-ask}(t) - Q_\text{best-bid}(t), ~ i \in \{\texttt{p+(i)}, \texttt{p-(i)}\} \text{ (bid-ask spread)}\\
\end{cases}
\end{align*}
where $k_{i}$ denote the queue to which event $i$ belongs to.

\subsubsection{Final form of Hawkes+Markovian combined model} \label{sec: final form model}
Our final model on the instantaneous rate $\lambda_{i}(t)$ is a combination of Hawkes part Eq.\eqref{eq:hawkes-part} and Markovian part Eq.\eqref{eq:markovian-part}:
\begin{equation}\label{equ: final form model}
\lambda_{i}(t) = M_i(l_{i}(t)) + \Theta_i(t) + \sum_{j = 1}^{6K+6} \int_{0}^{t} \phi_{j,i} (t-s) dX_j(s), \quad \forall  i = 1,2,\ldots,6K+6
\end{equation}

\section{Estimation procedure}\label{bigset:estimation procedure}
Given the above hybrid Hawkes+Markovian model, we employ a non-parametric regression-based approach to estimate the model parameters. Our approach is extended from \cite{kirchner2017estimation} as well as \cite{bacry2014hawkes}.

Inspired by \cite{kirchner2017estimation} and \cite{bacry2014hawkes}, we approximate the whole intensity function as a standard vector-valued linear autoregressive time series. The estimation procedure discretizes a continuous point process into multiple fixed-size bins on the time domain and thereby fits a vector autoregression model to the discretized samples.

To outline the intensity function estimation, we first define discretization bin-size $\Delta$ and maximum support $s$. The maximum support represents the maximum duration (in seconds) during which the arrival of one event can have stimulating effects on the arrival of other events. The bin-size $\Delta$ defines the short period of time during which the Hawkes self- and cross-exciting function stays unchanged. Take $s = 20 \text{ seconds}, \Delta=0.25 \text{ seconds}$ as an example: this setting indicates the maximum duration that an event $i$ can stimulate the intensities of future event arrivals is 20 seconds, while event $i$'s self- and cross-stimulating functions stay the same within each 0.25-second short period. That is to say, the 80 of 0.25-second windows combined define the 20-second stimulating horizon for event $i$ and its stimulating function shrinks to 0 beyond 20 seconds after event $i$'s arrival. 

\begin{figure}[H]
\centering
\includegraphics[scale=0.40]{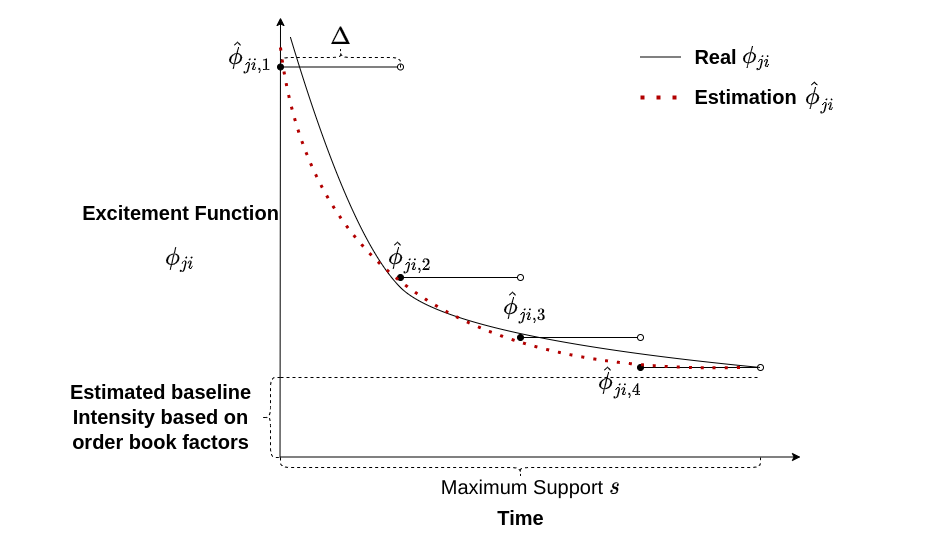}
\caption{An example illustration of Hawkes excitement function estimation. The example demonstrates the estimation for a continuous function $\phi_{ji}$. Suppose $s = 4\Delta$ and therefore $p=4$. The estimation idea is that $\phi_{ji}$ can be approximated by $\hat{\phi}_{ji} = (\hat{\phi}_{ji,1},\hat{\phi}_{ji,2},\hat{\phi}_{ji,3},\hat{\phi}_{ji,4})$. $\hat{\phi}_{ji,1}$ represents the fixed function value over $[0,\Delta]$, $\hat{\phi}_{ji,2}$ represents the fixed function value over $[\Delta,2\Delta]\ldots$.Though the main objective is to obtain the discrete estimators $\hat{\phi}_{ji}$, we can further fit a cubic smoothing spline to the discrete estimators to derive a continuous estimated function as shown by the red dotted line. }
\label{fig: discrete example}
\end{figure}

Given an appropriate choice of bin-size $\Delta$ and maximum support $s$, a continuous Hawkes excitement function between two events can be approximated by a piece-wise constant function of $p =\floor{s/\Delta}$ steps, with each step standing for the constant function value over the short period $\Delta$. To complete the estimation, we need to apply some smoothing methods over the estimated point-wise function. We adopt the cubic smoothing spline in our method. The cubic smoothing spline is a smoothing technique such that the curve spanning each data interval is represented by a cubic polynomial. The cubic smoothing spline is achieved by minimizing the curvature of the smoothed function and is not required to pass all data points. The overall illustration of the estimation method is demonstrated in Figure \ref{fig: discrete example}.

\subsection{Bin construction} \label{sec: bin construction}
To obtain our proposed discretized estimators, the data needs to be discretized under the $(s,\Delta)$-framework as the first step. The estimation requires allocating LOB event arrival and LOB state sequences into fixed-size bins with length $\Delta$ over the time horizon and counting the number of realizations in each bin. Afterward, these bin-count sequences can be used as sufficient statistics to obtain model estimates.

Recall in Section \ref{sec:model of the order book}, we denote the order book event process at time $t$ as $X_{i}(t)$, where $i$ is index for events. % in \texttt{-K(+)}, \texttt{-K(-)}, \texttt{-K(t)}, \ldots,  \texttt{+K(+)}, \texttt{+K(-)}, \texttt{+K(t)}, \texttt{p-(t)}, \texttt{p-(-)}, \texttt{p-(+)}, \texttt{p+(t)}, \texttt{p+(-)}, \texttt{p+(+)}. 
For LOB data, each type of event outlined in Section \ref{sec: price not change events} and \ref{sec: price change events} comes with an order size indicating the quantity of shares the LOB event attempts to execute. For example, a \texttt{+1\obinsert} event with order size 100 indicates the event at the first ask queue to insert 100 shares.

Since all regular order book events \texttt{-K(i)},\-\texttt{-K(c)},\-\texttt{-K(t)},\ldots, \-\texttt{+K(i)}, \-\texttt{+K(c)}, \-\texttt{+K(t)} are additive, big events can be considered as the sum of the same events with smaller sizes in a short period of time. As an intuitive example, one insertion order of size 100 shares is assumed to be equivalent to two insertion orders of size 50 shares happening at the same time and the same price. However, the events \texttt{p-(t)}, \-\texttt{p-(c)}, \-\texttt{p-(i)}, \-\texttt{p+(t)}, \-\texttt{p+(c)}, \-\texttt{p+(i)} are not addictive because the reference price change caused by these events shift the LOB queue distribution. Specifically, the queue size value $q_{k}$ can shift to its neighbours when the reference price (center of LOB distribution) goes up and down. Therefore, an event causing reference price changes cannot be thought simply as the sum of the same event with a smaller size. Overall, the size of an LOB event is modeled in a dichotomous approach for $X_i$: for all regular addictive events, $X_{i}$ increases by its order size at event arrival; for all non-addictive events, $X_{i}$ by 1 at event arrival, treating as pure point processes.

\begin{figure}[H]
\centering
\includegraphics[scale=0.55]{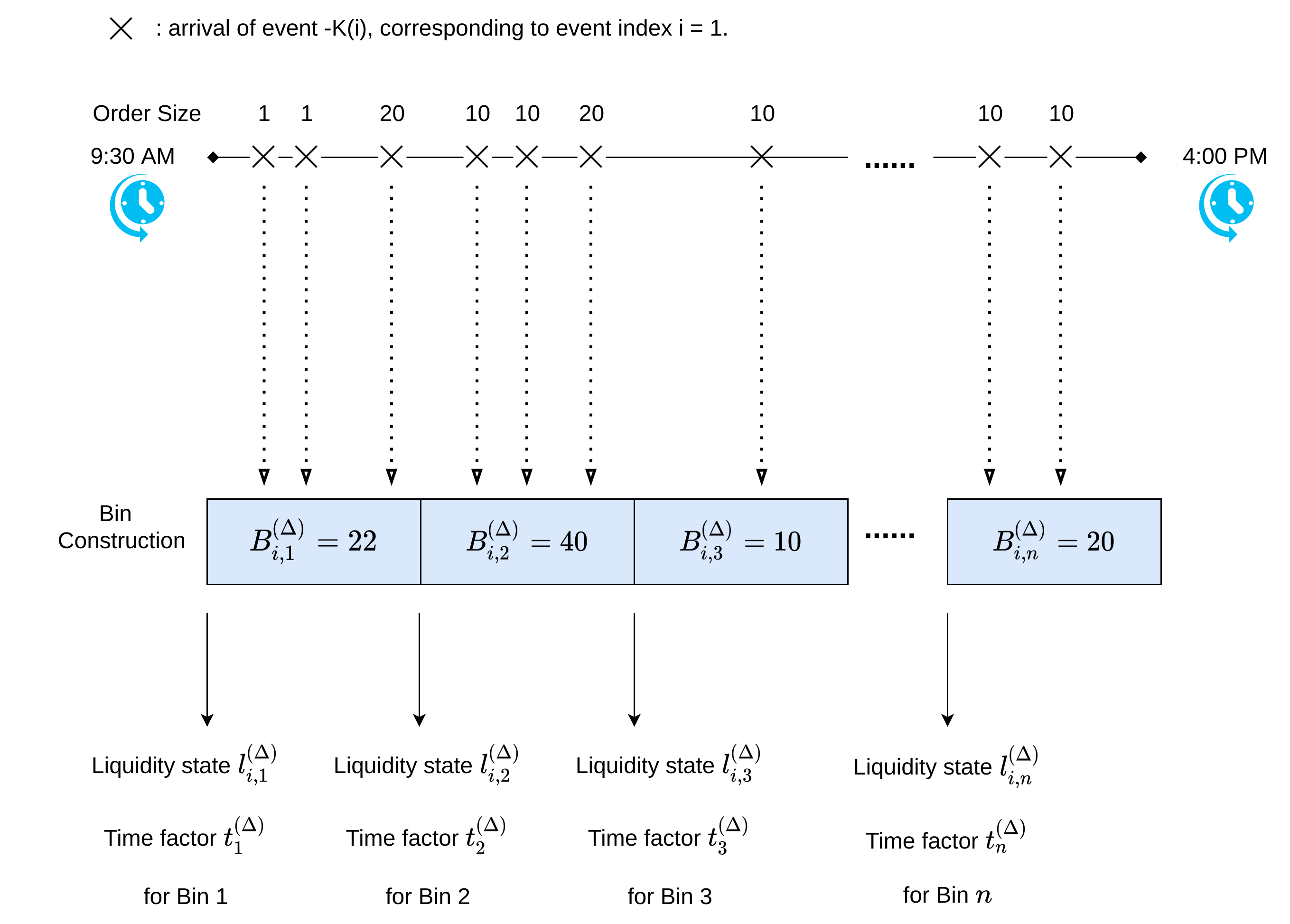}
\caption{An illustration example on bin-count sequence construction for event \texttt{-K(i)}. Order size at the cross mark are allocated into bins $1,2,\ldots,n$ based on event arrival time. }\label{fig:bin_construction_withsize} 
\end{figure}

We can then construct a series of fixed-size bins over the total time horizon $(0,T]$, and sum the realizations of $X_{i}$ during each bin to obtain the bin-count sequences, as illustrated in Figure \ref{fig:bin_construction_withsize}. 
Formally, for $t \in (0,T]$, for some bin-size $\Delta>0$, we construct the $(6K+6)$-dimensional bin-count, liquidity state, and time factor sequences as:
\[
B_{k}^{(\Delta)} = \left(B_{i,k}^{(\Delta)}\right)_{i=1}^{6K+6} , \quad B_{i,k}^{(\Delta)} :=    X_{(i)} \big(   \left((k-1)\Delta, k\Delta \right]        \big), 
\]
\[
l_{k}^{(\Delta)} = \left( l_{i,k}^{(\Delta)} \right)_{i=1}^{6K+6}, \quad l_{i,k}^{(\Delta)} := l_{i}\big(  (k-1)\Delta  \big),
\]
\[
t_{k}^{(\Delta)} = \left( t_{i,k}^{(\Delta)} \right)_{i=1}^{6K+6}, \quad t_{i,k}^{(\Delta)} := (k-1)\Delta,
\]
where $i = 1,2,\ldots,6K+6$,  $k = 1,2,...,n$, and $n := \floor{T/\Delta}$.

\subsection{Bucketize LOB state factors} \label{sec:estimation-baseline}
As mentioned in Section \ref{sec:model of the order book}, we assume that the baseline intensity function of the proposed model is controlled by functions based on LOB liquidity state $l_{i}(t)$ and intraday time $t$. In the previous section, we have discussed discretizing and transforming sequences $l_{i}(t)$ and $t$ into bin-based sequences $l_{i,k}^{(\Delta)}$ and $t_{i,k}^{(\Delta)}$ for a selected bin-size $\Delta$. Consequently, we have indeed transformed $M_{i}(l_{i}(t))$ and $\Theta_{i}(t)$ into $M_{i}(l_{i,k}^{(\Delta)})$ and $\Theta_{i}(t_{i,k}^{(\Delta)})$ for estimation purpose.

In order to approximate the non-parametric function $M_{i}(\cdot)$ and $\Theta_{i}(\cdot)$, we treat $l_{i,k}^{(\Delta)}$ and $t_{i,k}^{(\Delta)}$  as categorical variables to obtain the linear parameters. Therefore, $M_{i}(\cdot)$ and $\Theta_{i}(\cdot)$ become step functions that depend only on the specified categories of the discretized liquidity state and time factor. We use the following method for bucketization:
\begin{itemize}
    \item Liquidity state $l_{i,k}^{(\Delta)}$: we bucketize $l_{i,k}^{(\Delta)}$ into 10 categories $\{L_{1}, L_{2}, \ldots,L_{10}\}$ \footnote{These categories apply to all event types $\forall i = 1, \ldots, 6K+6$, and therefore the $i$ subscript is eliminated from the $\{L_{1}, L_{2}, \ldots,L_{10}\}$ notation.}with $L_{1} = [0,100), L_{2} = [100,200), \ldots, L_{10} = [900,+\infty)$, which represents the number of existing orders on the corresponding price queue (except for events \texttt{p+(i)}, \texttt{p-(i)}) right before event arrives. For example, $l_{i,k}^{(\Delta)} \in L_{1}$ means that the number of orders on the queue is between 0 and 100 shares. For the special case of events \texttt{p+(i)}, \texttt{p-(i)}, the $l_{i,k}^{(\Delta)}$ represents the bid-ask spread in market price (Section \ref{sec:model of the order book}). The group $l_{i,k}^{(\Delta)} \in L_{1}$ means that the bid-ask spread is between \$0 and \$0.01.
    
    \item Time factor $t_{i,k}^{(\Delta)}$: major U.S. electronic stock exchanges trade from 9:30 am ET to 4:00 pm ET. As illustrated in Figure \ref{fig: rough arrival rate}, the LOB event arrival frequency exhibits time clustering effects at the beginning and ending intraday 30-minutes window (9:30 - 10:00 am ET and 3:30 - 4:00 pm ET). Therefore, we construct 1-minute categories at the beginning and ending 30-minutes to better capture the event arrival volatility. For the rest of the period between 10:00 am ET to 3:30 pm ET, we bucketize time $t_{k}^{(\Delta)}$ into 5-minute categories since the event arrivals are more tranquil. Consequently, we construct 126 time categories consisting of 60 1-minute categories and 66 5-minute categories, denoted as $\{T_{1}, T_{2}, \ldots, T_{126}   \}$.
\end{itemize}

The above category creation is an implementation choice. Other choices on the number of categories for the liquidity state and time factor can also be applied as long as they do not impact implementation feasibility.

\subsection{Non-parametric estimation} \label{sec:nonparametric estimation}

After defining the bin-count sequence and the LOB state/time categories, we fit a vector-valued autoregression model to the bin-count sequences to obtain the Hawkes excitement functions, together with the Markovian liquidity state and time factor parameters. For each event-to-event pair, the auto-regression model is implemented with lag $p :=\floor{s/\Delta}$ on the bin-count sequence data $B_{k}^{(\Delta)}$ as well as the Markovian liquidity state and time factor sequence $l_{k}^{(\Delta)}$ and $t_k^{(\Delta)}$ across all event types. 

Kirchner \cite{kirchner2017estimation} has provided a detailed framework for autoregressive non-parametric estimation for Hawkes process with constant baseline intensity. Our estimation is an extension of Kirchner's method with modifications that the baseline intensity is varying and controlled by $l_{i}(t)$ and $t$. An illustration of the estimation procedure is given in Figure \ref{fig: Design Matrix}.

The mathematical representation detail on the estimation procedure is given in Supplementary Material Section \ref{sec_support:estimation math}. As shown Eq.(\ref{eq3}) in the Supplementary Material Section \ref{sec_support:estimation math}, $\hat{\Phi}^{(\Delta,s)}\in \mathbb{R}^{(6K+6)\times((6K+6)p+10+126)}$ represents all the parameters of our model for a choice of bin-size $\Delta$ and maximum support $s$. To summarize, the estimation procedure for our proposed model estimates: A total of $(6K+6)\times(6K+6)$ Hawkes excitement functions $\phi(\cdot)$, each as a step function with $p=\floor{s/\Delta}$ distinct levels (See Figure \ref{fig: discrete example}); A total of $6K+6$ liquidity state functions $M(\cdot)$, each as a step function with 10 distinct levels on the 10 liquidity state baskets; A total of $6K+6$ time factor functions $\Theta(\cdot)$, each as a step function with $126$ distinct levels on the 126 time factor baskets.

\begin{figure}[H]
\centering
\includegraphics[scale=0.50]{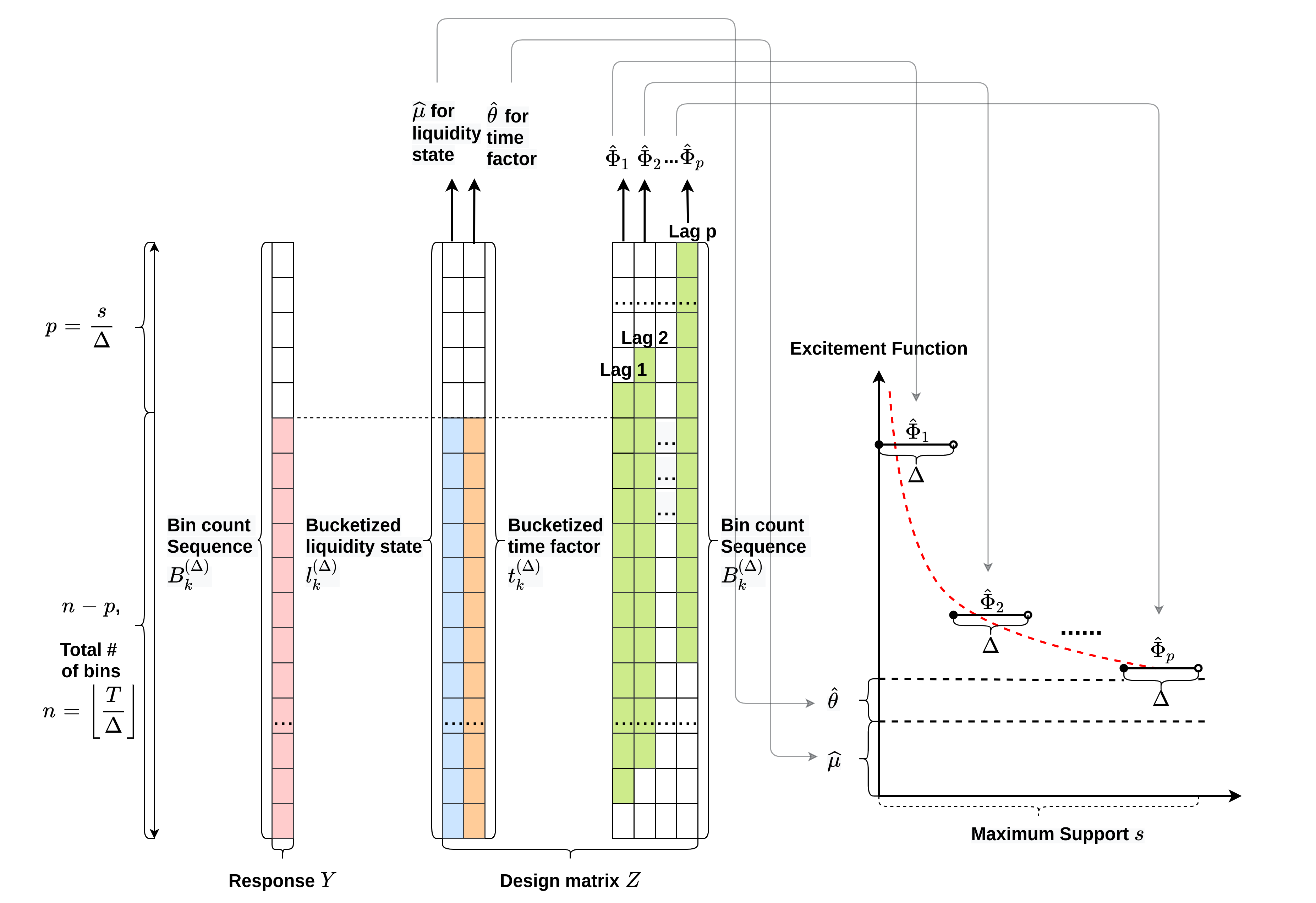}
\caption{An illustration of the non-parametric estimation procedure. Over a time horizon $(0,T]$, given a bin-size $\Delta$ and a maximum support $s$, the parameters of our proposed model can be estimated using an autoregressive framework with lag $p=\floor{s/\Delta}$. There are totally $n  = \floor{T/\Delta}$ bins. The autoregression response $Y$ ranges from the $(p+1)$-th bin to the $n$-th bin from the bin-count sequence. The design matrix $Z$ consists of two parts: the first part contains the bucketized liquidity state and time factor sequences controlling the regression intercept; the second part contains lag-$p$ sequences constructed from the bin-count sequence. The estimated parameters $\hat{\mu}$ and $\hat{\theta}$ for liquidity state and time factor control the baseline intensity of the estimated excitement functions. Parameters $\left( \hat{\Phi}_{1},\ldots,\hat{\Phi}_{p} \right)$ are the estimates for the constant excitement function value during each short period of bin-size $\Delta$.}
\label{fig: Design Matrix}
\end{figure}

\subsection{LASSO regularization}\label{sec: add LASSO}
Many previous works have discussed the exponential decaying shape of the LOB excitement functions \cite{muni2011modelling,abergel2015long,morariu2018state,kirchner2017estimation}. Therefore, We believe some of the estimators for the excitement function are likely to be very close to zero, especially at the tail part when the support $s$ is large and the bin-size $\Delta$ is small. As discussed in Remark \ref{remark: OLS-implement} from Supplementary Material Section \ref{sec_support:estimation math} and Section \ref{sec:model of the order book}, for a single event type the OLS estimation procedure outputs a large $(6K+6)\times p$ (recall $p=\floor{s/\Delta}$) number of estimators for the Hawkes stimulating functions $\phi(\cdot)$ for a given combination of $(s,\Delta)$. To fit linear models with such a large design matrix in a more robust way, we consider adding LASSO regularization to the non-parametric estimation of excitement function to shrink the estimations to zero. 

On the other hand, we don't regularize any LOB liquidity state and time factor parameters $(\hat{\mu}, \hat{\theta} )$ since we believe most of the redundant parameters tend to appear in the Hawkes excitement function part, while the liquidity state and time factor should have smooth and non-zero effects. 
The mathematical representation detail of the LASSO is given in Definition \ref{def: LASSO math} in Supplementary Material Section \ref{sec_support: LASSO math}.

\subsection{Model selection using AIC (Akaike Information Criterion)} \label{sec: AIC model selection}
AIC is among the common approaches for linear model selection and diagnosis. Kirchner \cite{kirchner2017estimation} uses multivariate AIC to determine the ideal model fit using both simulated and real LOB data. In our proposed methodology, AIC can be used to evaluate the effects of different parts of the proposed model in Section \ref{sec: final form model}.

Given a selected bin size $\Delta = \Delta_{0}$, the estimation lag equals $p = \floor{s/\Delta_{0}}$ for any maximum support $s>0$. Similarly, we denote the total number of the bins as $n_{0} := \floor{T/\Delta_{0}}$. AIC calculation involves the total number of effective parameters (model degree-of-freedom) in its formula. When the estimated parameters are not effective (exactly zero, or not estimable), they cannot be counted as degree-of-freedom. For example, a number of redundant estimators shrink to zero under LASSO regularization. Also, when there is no liquidity state observation in some of the bucketized categories among $(L_{1}, L_{2},\ldots,L_{10})$, the parameter $\hat{\mu}^{(\Delta,s)}$ will not be estimable. Consequently, these parameters are not part of the AIC calculation.

Denote the $(6K+6)\times1$ regression residual vector as $\boldsymbol{\hat{u}_{k}}^{(\Delta_{0},p)} = \left(  \hat{u}_{1,k},\ldots,\hat{u}_{6K+6,k}   \right)^\top$ for $\forall k = (p+1),\ldots,n_{0}$. The multivariate AIC is given as:
\[
AIC^{(\Delta_{0})}(p) = \text{log}\left(  \text{det} \textstyle \Hat{\Sigma}^{(\Delta_{0})}(p)  \right) + \frac{2 \times d_{e}}{(n_{0}-p)}, \quad \hat{\Sigma}^{(\Delta_{0})}(p) = \sum^{n_{0}}_{k = p+1}(\boldsymbol{\hat{u}_{k}}^{(\Delta_{0},p)})(\boldsymbol{\hat{u}_{k}}^{(\Delta_{0},p)})^\top\\/ (n_{0} - p)
\]
where $d_{e}$ is the number of effective parameters involved in the estimation.

\section{Implementation Example}\label{bigsec:example}

In this section, we give a detailed example of the implementation of the proposed model and its estimation illustrated in Section \ref{bigsec:model specification} and Section \ref{bigset:estimation procedure}. We apply our model on real order book data obtained from Lobster Data (\url{http://lobsterdata.com}). The data is reconstructed from the ITCH data \cite{NASDAQ:ITCH} provided by NASDAQ and accurately delivers order book flows and distributions in nanoseconds. Detailed processing, cleaning, and reconstruction methods for the Lobster Data is shown in \cite{huang2011lobster}. Order book data of Apple.Inc from 9:30 AM ET to 4:00 PM ET from 01/02/2019 to 01/31/2019 was used in our implementation. We choose to model the level-3 order book, i.e., setting $K=3$, so that there are $6K+6 = 24$ events considered. They are: 

\[
\begin{pmatrix}
\underbrace{(\texttt{-3(i)}, \texttt{-3(c)}, \texttt{-3(t)},\texttt{-2(i)}, \texttt{-2(c)}, \texttt{-2(t)}, \texttt{-1(i)}, \texttt{-1(c)}, \texttt{-1(t)})^\intercal}_{9} \\
\underbrace{(\texttt{+1(i)}, \texttt{+1(c)}, \texttt{+1(t)}, \texttt{+2(i)}, \texttt{+2(c)}, \texttt{+2(t)}, \texttt{+3(i)}, \texttt{+3(c)}, \texttt{+3(t)})^\intercal}_{9} \\
\underbrace{(\texttt{p-(t)}, \texttt{p-(c)}, \texttt{p-(i)}, \texttt{p+(t)}, \texttt{p+(c)}, \texttt{p+(i)})^\intercal}_{6}
\end{pmatrix}_{24 \times 1}
\]

The order book estimation for every single day can be considered as an independent realization of event arrivals. Since the support and bin size values are fixed throughout the days in consideration, we can obtain the unbiased estimator of the estimation parameters by taking the mean of each element from single-day parameters. The aggregated results across many days can smooth out single-day extreme values and therefore is a clear representation of estimation results over a certain period.

Suppose we have $N$ days of order book data for a company and the estimator $\bar\hat{\Phi}^{(\Delta,s)}_{n}$ for each day with $n = 1,2,\ldots,N$, the aggregated estimator can be represented as:
\[
\bar{\hat{\Phi}}^{(\Delta,s)}_{1:N} := \frac{\sum^{N}_{n=1} \hat{\Phi}^{(\Delta,s)}_{n}}{N}
\]

In addition, a small LASSO regularization parameter $\lambda_{i} = 0.0005$ has been implemented for our estimations, as a large parameter may drastically alter the estimation results. The main findings and all the observed features demonstrated in the rest of Section \ref{bigsec:example} are based on the 20-day aggregation method mentioned above.

\subsection{Main findings} \label{main findings}
Through our non-parametric estimation over 20-day LOB data of Apple.Inc, the main findings include:

\begin{itemize}
    \item Most of the estimated Hawkes excitement functions exhibit exponential shape with trailing zeros. However, there are some exceptions and most of them appear on market (trade) orders on higher levels (the 2nd and 3rd best bid/ask price) of the LOB queue.
    \item The estimated Hawkes excitement functions are similar with respect to insertion and deletion events on the 1st level of LOB queue (the 1st best bid/ask price). 
    \item For most of the LOB events that do not change the price (reference price, which is the center of the order book), the event arrival intensity increases with the current order size in the order book queue. % right before event arrives move in the same direction, with exceptions on order insertion events on the 2nd and 3rd level of the LOB queue.
    \item The arrival intensities for almost all events elevate during the beginning and ending 30-minutes window of daily trading hours. The intensity increases drastically before market close between 15:55 to 16:00 pm ET.
    \item Our qualitative result is not sensitive to the discretization size in time and the handling of the size of the order. 
    \item Through model selection analysis using AIC (Akaike Information Criterion), the inclusion of the Hawkes excitement functions, the LOB liquidity state, and LOB time factor to the event arrival intensity all contribute to the improved model fitting. Besides, reducing the bin-size and adding LASSO regularization to our estimation both contribute to better model fitting.
    
\end{itemize}

The details of the observations listed above will be discussed in the rest of this section. 

\subsection{Estimated excitement functions} \label{sec:estimated excitement functions: ask}
\begin{figure}[H]
    \centering
    \subfloat[\texttt{+1(i)} stimulate \texttt{+1(i)}]{{\includegraphics[scale = 0.16]{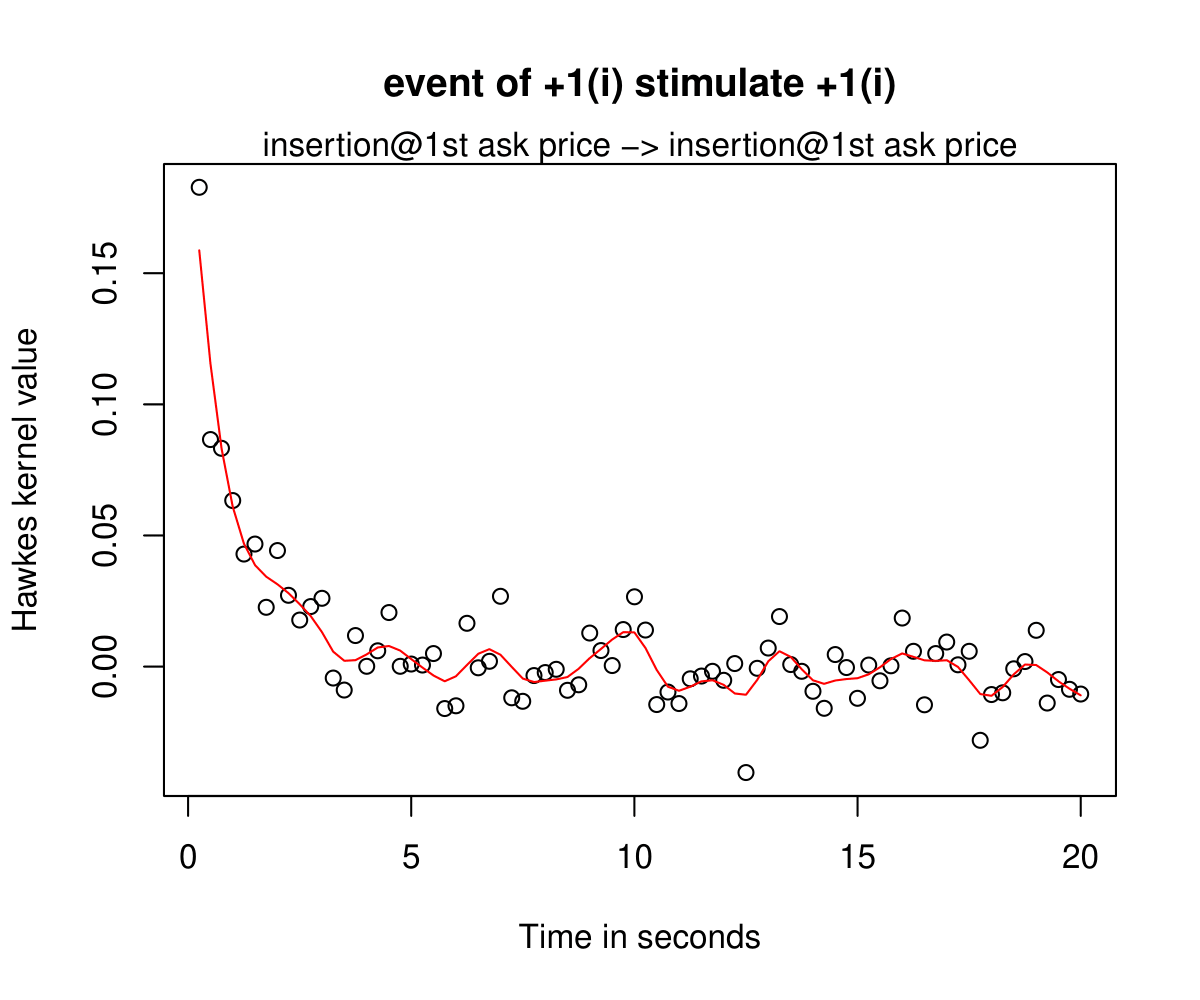} }}%
    \qquad
    \subfloat[\texttt{+1(i)} stimulate \texttt{+1(c)}]{{\includegraphics[scale =0.16]{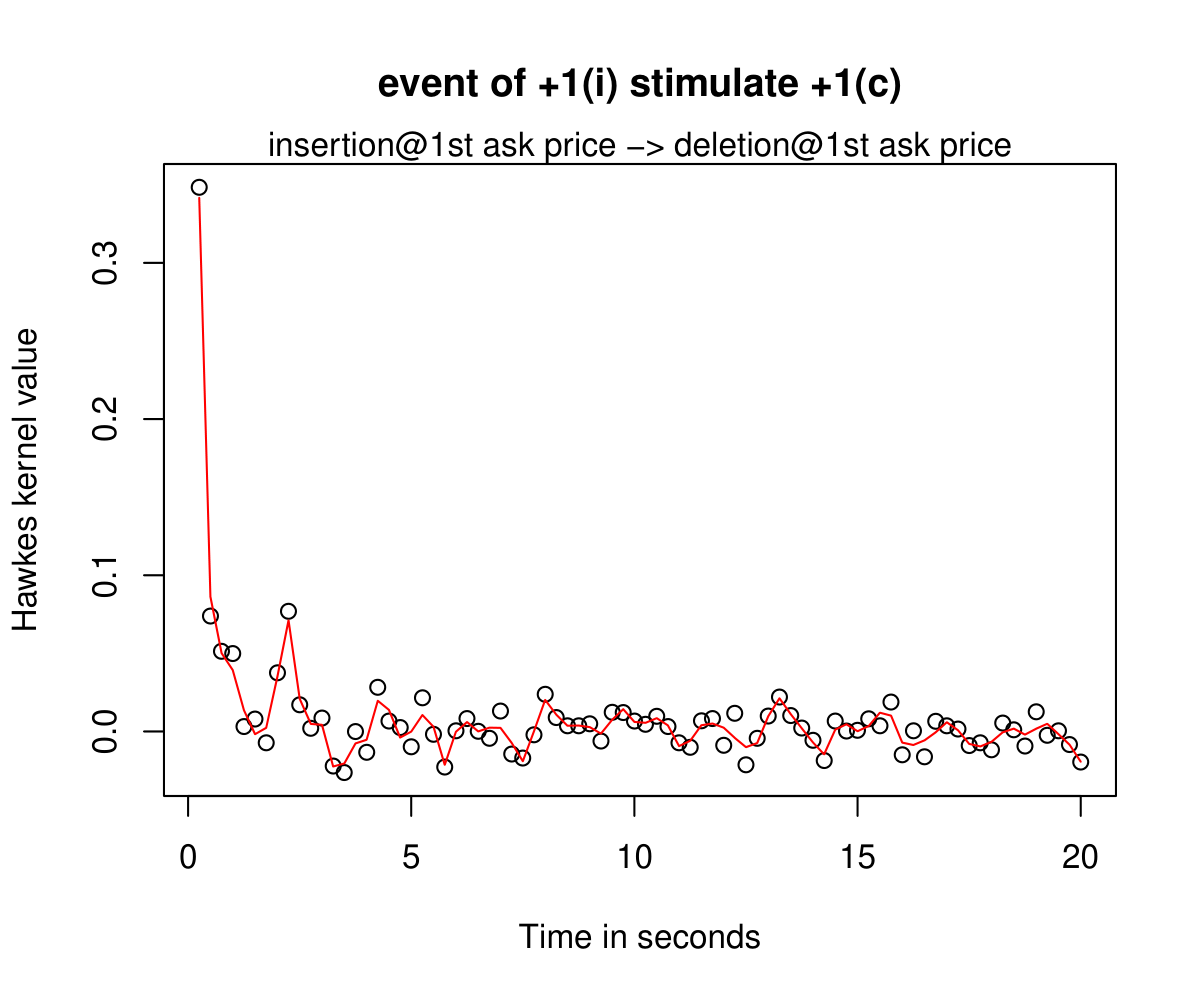} }}%
    \caption{Aggregated Hawkes excitement function estimations under $(s=20 \text{ seconds}, \Delta=0.25 \text{ seconds})$ with LASSO regularization. The points illustrate the discrete function estimators. The red line illustrates the cubic smoothing spline for the points.}
    \label{fig: first-level similar}
\end{figure}

Figure \ref{fig: first-level similar} illustrates the estimated excitement functions of insertion event at 1st ask (event \texttt{+1(i)}) stimulating insertion and cancellation at the 1st ask (event \texttt{+1(i)} and \texttt{+1(c)}). 

We observe the excitement function to have a time-decaying shape in general: the stimulation is highest in a very short time following the event arrival, and then gradually decaying to zero. 

We also observe similarities between Fig.\ref{fig: first-level similar}(a) and Fig.\ref{fig: first-level similar}(b): the figures both exhibit exponential decaying shapes with slight spikes around 5 and 13 seconds after event \texttt{+1(i)} arrives. This observation suggests that the estimated excitement functions are very similar for the effect towards the insertion and cancellation at the 1st ask (i.e., effect towards \texttt{+1(i)} and \texttt{+1(c)}). We have similar observations for 1st bid as well demonstrated in Supplementary Material Section \ref{sec_support: estimated excitement functions bid}, suggesting such stimulation behavior exists at the 1st level of the LOB queue.

Further examination shows that this feature (similarity at the 1st ask/bid) becomes less obvious for the 2nd and 3rd levels of the LOB queue. One possible explanation of this feature is that market participants may use a sequence of insertion/deletion orders in a short period of time to drive up/down the price for specific purposes. Insertion/deletion on the 1st level is typically less risky than insertion/deletion on any higher levels especially when the short-term price movement is unpredictable.

\subsection{Exponential and non-exponential shape of excitement functions} \label{sec: exponential and Non-exponential shape of excitement function}
From the aggregated estimation results we observe that a large proportion of Hawkes excitement functions exhibit exponential features with their function values converging to zero as time elapses on the horizontal axis. For example, we can see the excitement functions demonstrated in Figure \ref{fig: first-level similar} exhibit this feature. This observation agrees with the existing works that assume the stimulation effects are exponential-decaying and use Hawkes process with exponential kernels \cite{muni2011modelling,abergel2015long,morariu2018state,kirchner2017estimation}.

\begin{figure}[H]
    \centering
    \subfloat[\texttt{-1(t)} stimulate \texttt{+3(t)}]{{\includegraphics[scale = 0.16]{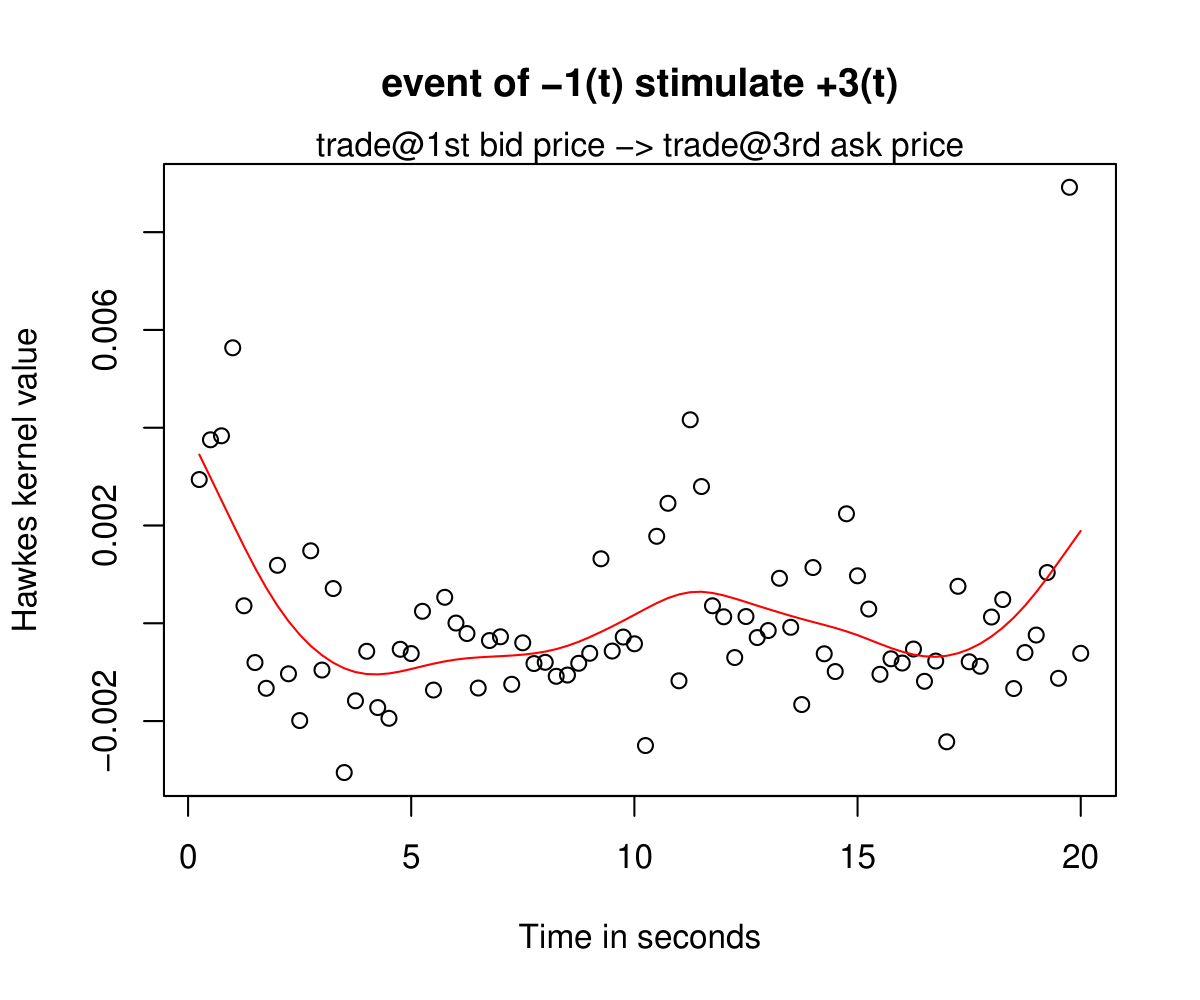} }}%
    \qquad
    \subfloat[\texttt{p+(t)} stimulate \texttt{-3(t)}]{{\includegraphics[scale =0.16]{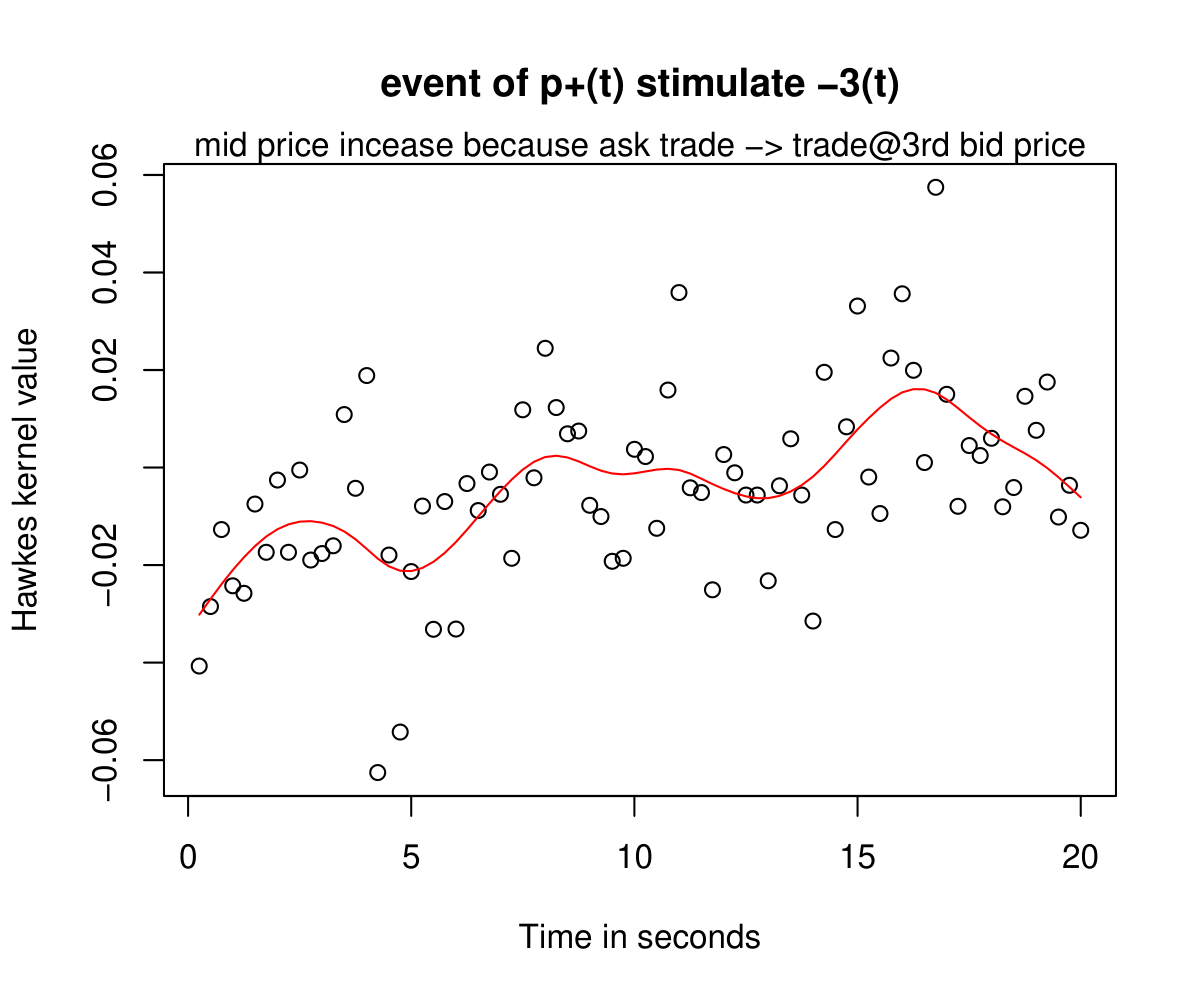} }}%
    \caption{Aggregated Hawkes excitement function estimation under $(s=20 \text{ seconds}, \Delta =0.25 \text{ seconds})$ with LASSO regularization. The points illustrate the discrete function estimators. The red line illustrates the cubic smoothing spline for the points. } 
    \label{fig: non-exponential kernels}
\end{figure}

However, we have also observed that some Hawkes excitement functions exhibit other shapes. Figure \ref{fig: non-exponential kernels} shows that the excitement functions for specific events can deviate from the exponential shape. These non-exponential shaped functions appear more frequently for market(trade) order events on higher levels of the LOB queue when they are the ``being stimulated" part of the function, such as events \texttt{-3(t)}, \texttt{-2(t)}, and \texttt{+3(t)}. These ``high-level" trade orders typically arrive when the bid-ask spread is large. One possible reason for this non-exponential feature is that, for a heavily traded stock like Apple.Inc, some high-frequency algorithmic market participants may use complex execution and risk management strategies to deal with the increased risk brought by the enlarged bid-ask spread. Therefore, our estimated Hawkes function can generate complicated shapes possibly because of this layer of complexity. These non-exponential shaped functions indicate the advantage of our non-parametric method since it might be problematic to adopt exponential shape assumption for stimulating functions throughout all events. More examples on the non-exponential shapes are demonstrated in Supplementary Material Section \ref{sec_support: more non_exponential}.

\subsection{Liquidity state} \label{sec: liquidity state}
\begin{figure}[H]
    \centering
    \subfloat[Liquidity state for \texttt{+1(c)} ]{{\includegraphics[scale = 0.43]{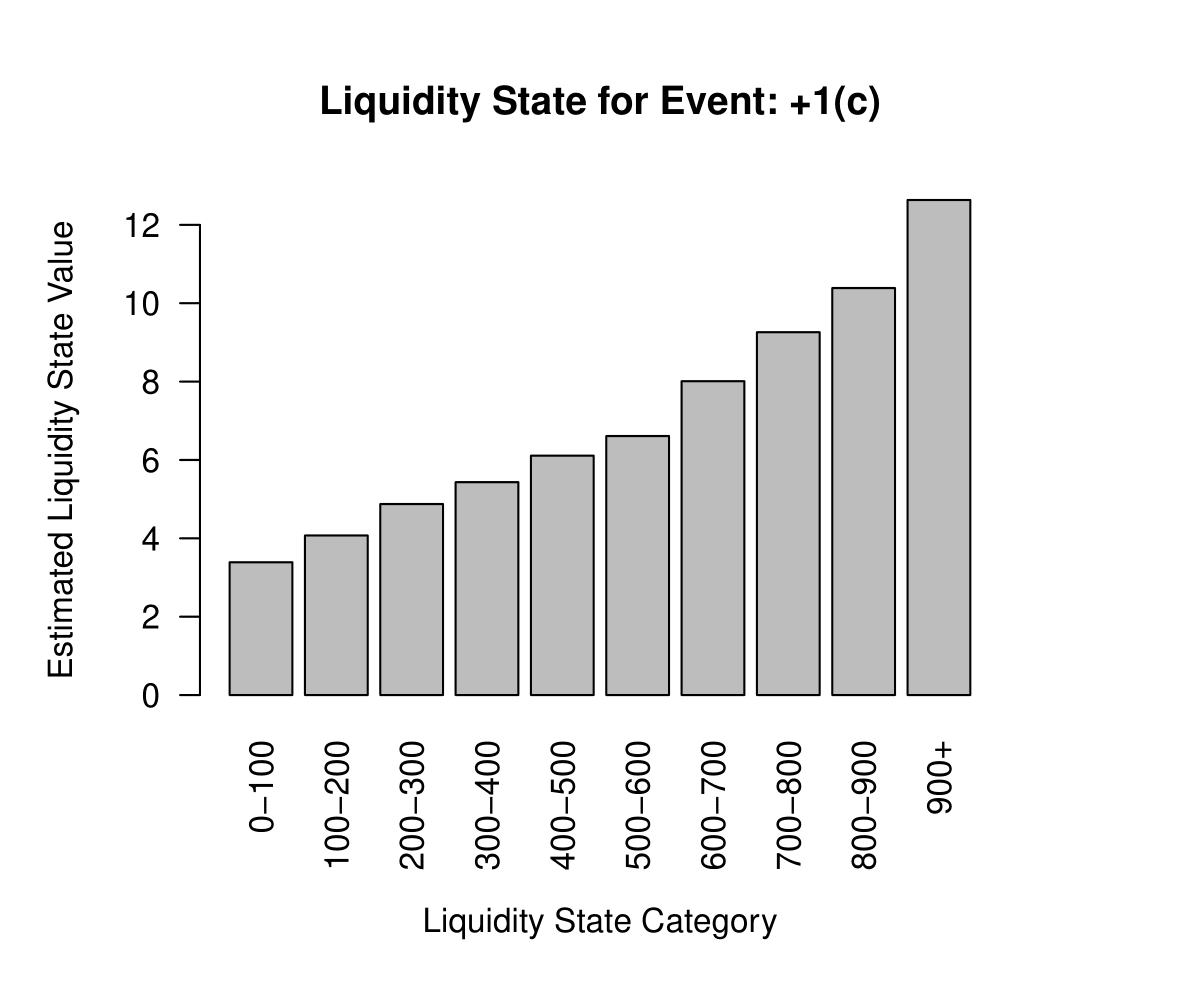} }}%
    \qquad
    \subfloat[Liquidity state for \texttt{-3(i)} ]{{\includegraphics[scale = 0.43]{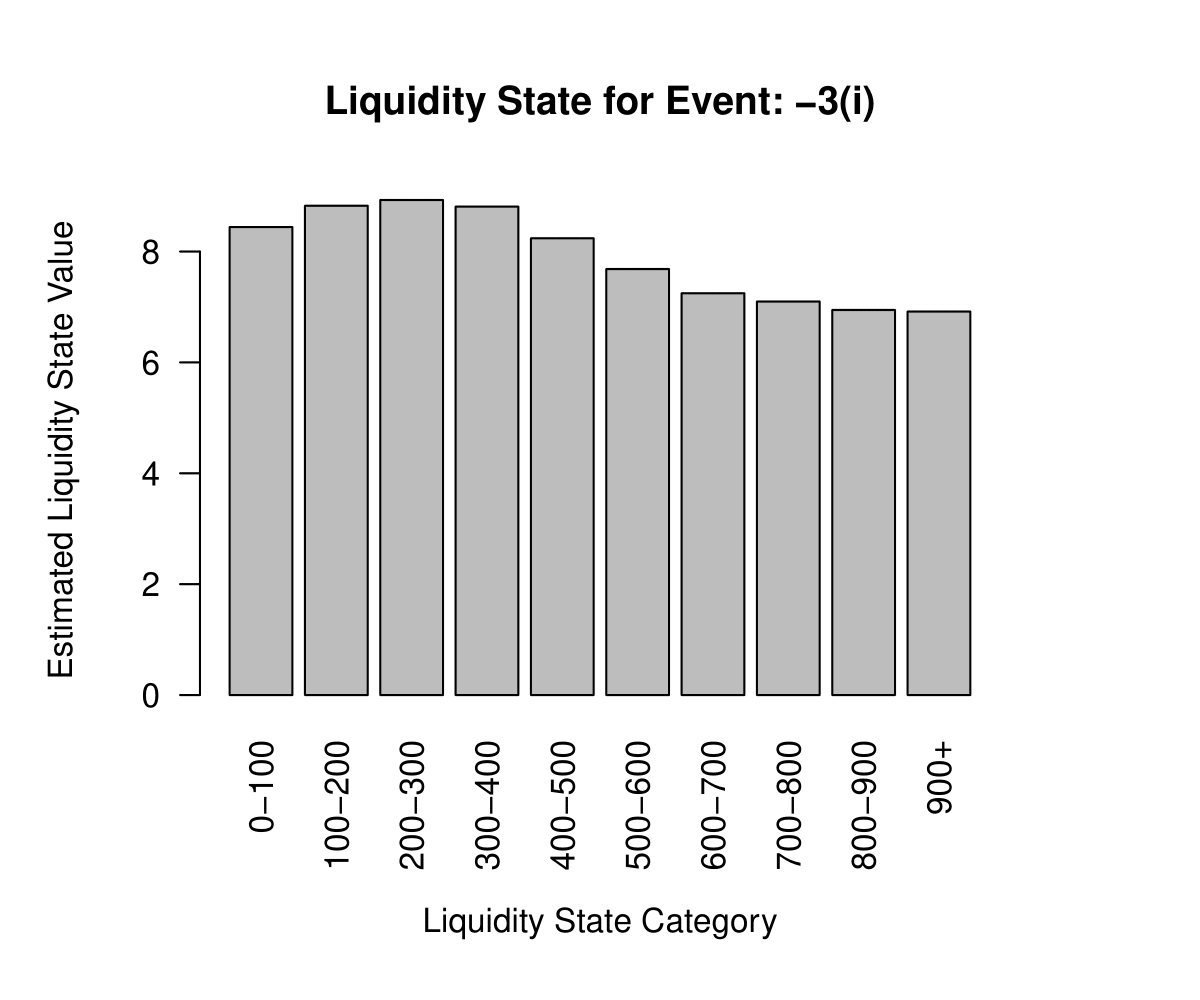} }}%
    \caption{Aggregated estimation result for liquidity state for selected events under $(s=20 \text{ seconds}, \Delta=0.25 \text{ seconds})$. In general, the arrival intensity increases/decreases as the liquidity state increases for event \texttt{+1(c)} and \texttt{-3(i)}, respectively.}
    \label{fig: state_example}
\end{figure}

For liquidity state results, we have observed that for event types that do not change the reference price, the arrival intensity increases as the liquidity state increases for all trade/cancellation events, as well as insertion events on the 1st level (i.e., \texttt{-3(c)}, \texttt{-3(t)}, \texttt{-2(c)}, \texttt{-2(t)},\texttt{-1(i)}, \texttt{-1(c)}, \texttt{-1(t)},\texttt{+1(i)}, \texttt{+1(c)}, \texttt{+1(t)},\texttt{+2(c)}, \texttt{+2(t)}, \texttt{+3(c)}, \texttt{+3(t)}). An example is shown in Figure \ref{fig: state_example}(a). This is consistent with the intuition that more trade/cancellation events are likely to happen when the number of existing orders on the corresponding queue is large, because the trade/cancellation are actions on existing orders. It also suggests that more insertion events at 1st level LOB are likely to happen when the number of existing orders on the 1st level is large, which could be potentially explained by the popularity of the stock -- when the stock is popular, naturally the number of existing orders at 1st level LOB is high, yet market participants are willing to insert more orders at 1st level LOB, and vice versa.

On the other hand, the arrival intensity generally decreases as the liquidity state increases for the insertion events at 2nd and 3rd level (i.e., \texttt{+3(i)}, \-\texttt{-3(i)}, \-\texttt{+2(i)}, \-\texttt{-2(i)}). An example is demonstrated in Figure \ref{fig: state_example}(b). The observation implies that market participants are less likely to insert orders on the 2nd and 3rd level of LOB when the number of existing orders is large on these levels. This behavior is possibly due to the increased risk that the orders on the 2nd and 3rd level may fail to be filled as orders accumulate in these queues, and maybe the market participants focus more on the 1st level when such a situation occurs.

More examples demonstrating the above patterns are presented in Supplementary Material Section \ref{sec_support: more liquidity state}.

\subsection{Time factor} \label{sec: time factor}
\begin{figure}[H]
    \centering
    \subfloat[Time factor for event \texttt{+1(i)}]{{\includegraphics[scale = 0.43]{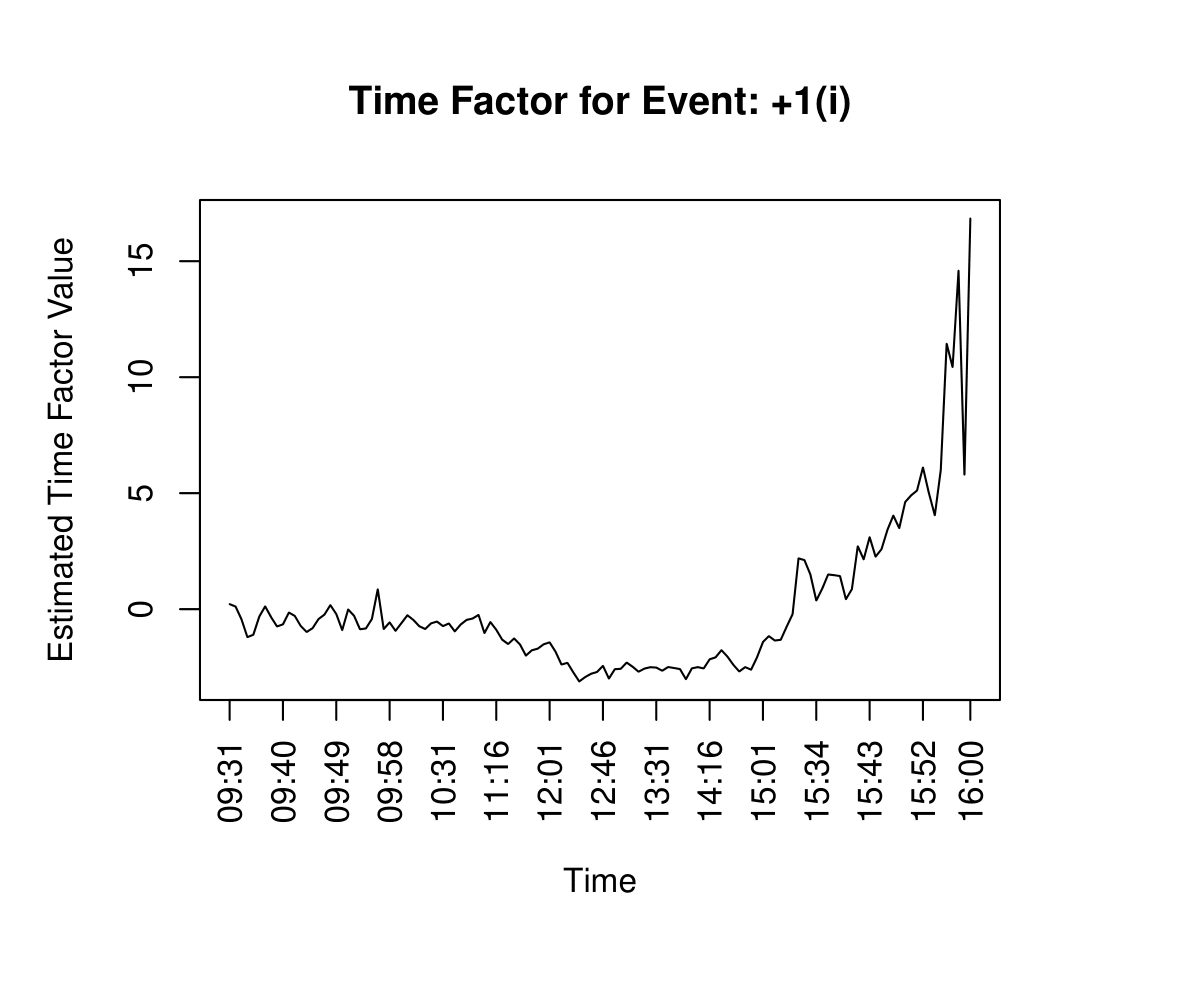} }}%
    \qquad
    \subfloat[Time factor for event \texttt{+3(t)}]{{\includegraphics[scale = 0.43]{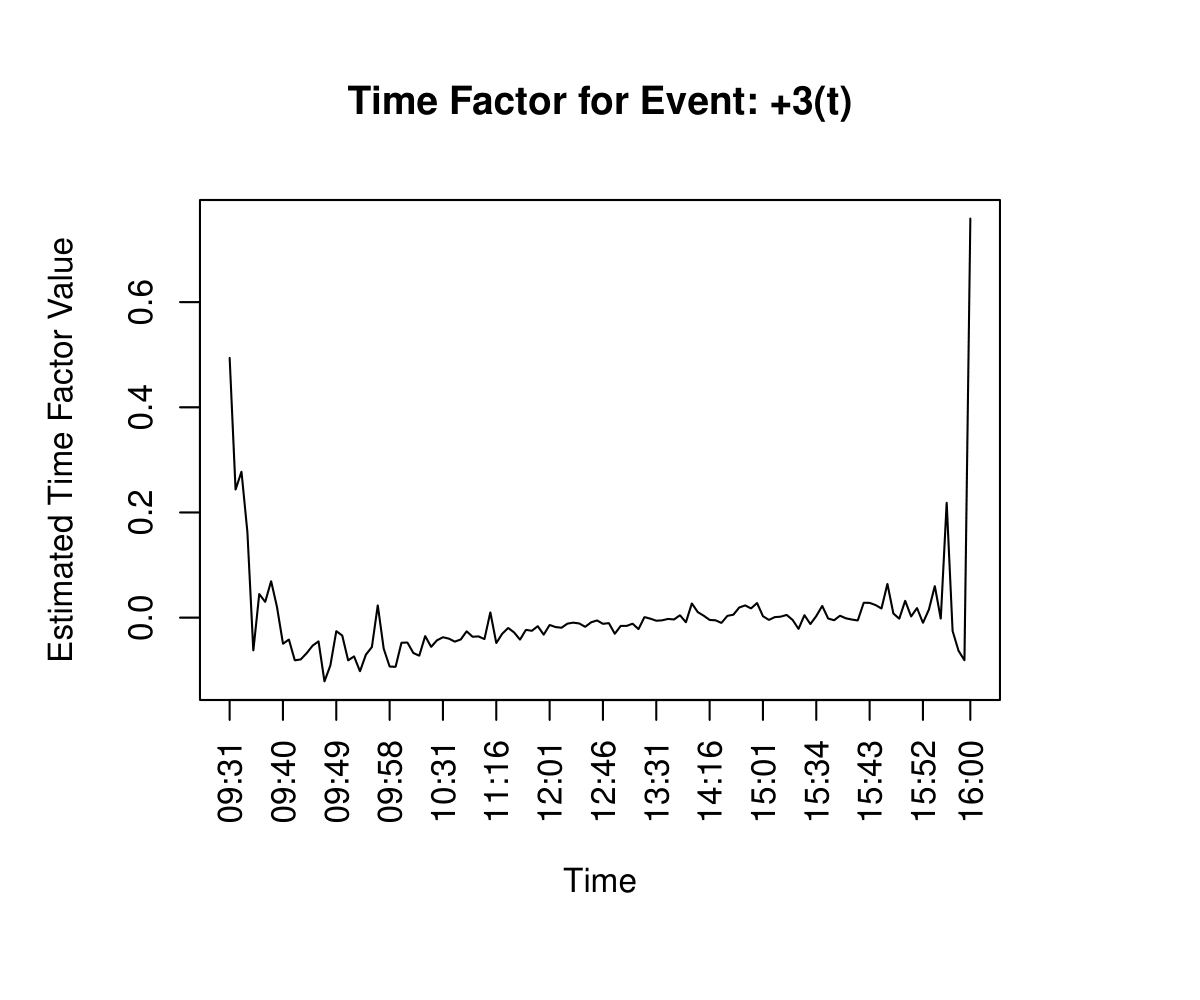} }}%
    \caption{Aggregated estimation result for time factor between 9:30 am and 4:00 pm under $(s=20 \text{ seconds}, \Delta=0.25 \text{ seconds})$. }
    \label{fig: time factor}
\end{figure}

For estimation results of time factor, the order arrival intensity tends to be larger at the beginning and end of the trading hours between 9:30 am and 4:00 pm (see Figure \ref{fig: time factor}). This feature matches our initial arrival rate estimation in Figure \ref{fig: rough arrival rate} and reinforces our model assumption that more time factor categories should be constructed in the beginning and ending 30 minutes.

Figure \ref{fig: time factor} demonstrates the time factor pattern mentioned above for event \texttt{+1(i)} and event \texttt{+3(t)}. From the estimation results we have also observed a significant intensity increase for many order types such as \texttt{p-(i)}, \texttt{p-(t)}, and \texttt{-1(t)} at 15:55 pm and 15:59 am. These patterns possibly stem from large algorithmic trader's execution rules or the policy of stock exchanges on last-minute order submission/cancellations.

\subsection{Sensitivity analysis}\label{sec: sensitivity analysis}
So far our results have been presented using the bin count construction method based on order sizes, a fixed bin-size $\Delta = 0.25$ seconds, and the LASSO loss function. This section provides a sensitivity analysis of these model assumptions to explore estimation robustness. The analysis consists of the following parts:
\begin{itemize}
    \item Estimation results when the size of order is ignored in bin count construction, assuming all orders to have size 1 in the estimation. (see Supplementary Material Section \ref{sec_support: result nosize})
    \item Estimation results when the LASSO regularization is removed. (see Supplementary Material Section \ref{sec_support: result noLASSO})
    \item Estimation results when the bin-size is increased from 0.25 seconds to 0.5 seconds. (see Supplementary Material Section \ref{sec_support: result large bin size})
\end{itemize}

Generally speaking, our estimation is robust for the above sensitivity tests. We have obtained qualitatively similar results as our original setting when we remove the LASSO regularization and enlarge the bin-size. After ignoring the order size, the estimation levels tend to decrease and the excitement function becomes smoother and less volatile. However, the observations given from Section \ref{sec:estimated excitement functions: ask} to Section \ref{sec: time factor} still hold.

\subsection{Model selection result} \label{sec: model selection result}

\begin{figure}[H]
\centering
\includegraphics[scale = 0.20]{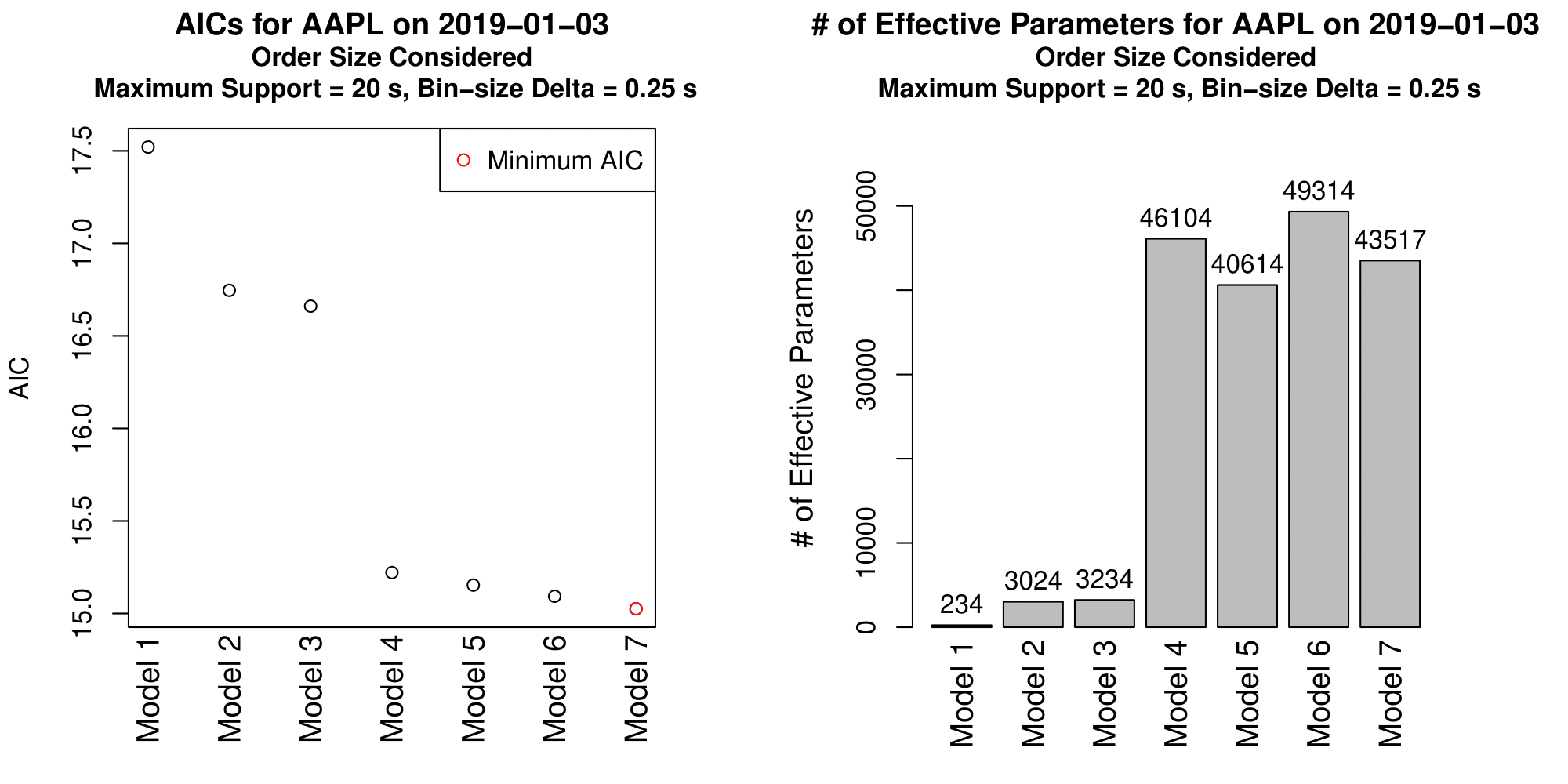}
\caption{AICs and number of effective parameters for seven model types for Apple.Inc on 2019-01-03 with maximum support $s = 20$ seconds and bin-size $\Delta=0.25$ seconds. Order size is considered to construct bin count sequence.}
\label{fig:aic_delta025_withsize}
\end{figure}

In this section, we validate the added explanatory power for different model parts using the AIC model selection method illustrated in Section \ref{sec: AIC model selection}. We consider the following model specifications for model selection: \textbf{Model \cirnum{1}}: liquidity state only; \textbf{Model \cirnum{2}}: time factor only; \textbf{Model \cirnum{3}}: liquidity state + time factor; \textbf{Model \cirnum{4}}: Hawkes only; \textbf{Model \cirnum{5}}: Hawkes + LASSO (LASSO parameter 0.0005); \textbf{Model \cirnum{6}}: liquidity state + time factor + Hawkes; \textbf{Model \cirnum{7}}: Liquidity state + time factor + Hawkes + LASSO (LASSO parameter 0.005). \textbf{Model \cirnum{7}} is the model we mainly proposed and discussed in previous sections (Section \ref{sec: final form model} and Section \ref{bigset:estimation procedure}). The "+LASSO" notation is used to illustrate that LASSO regularization is used in model estimations as discussed in Section \ref{sec: add LASSO}. Figure \ref{fig:aic_delta025_withsize} demonstrate AICs for the 7 models mentioned above for Apple.Inc on 2019-01-03.

 We conclude that for Apple.Inc on 2019-01-03, \textbf{Model \cirnum{1}/\cirnum{2}/\cirnum{3}} generally have much higher AICs with a much smaller number of effective parameters than the rest of the models. 

As expected, \textbf{Model \cirnum{4}} outperforms \textbf{Model \cirnum{1}/\cirnum{2}/\cirnum{3}}, indicating the Hawkes excitement functions have stronger explanatory power than the combination of liquidity state and time variables. Furthermore, the LASSO models (\textbf{Model \cirnum{5}/\cirnum{7}}) with a small regularization parameter $\lambda_{i}=0.0005$ generate even smaller AICs compared to \textbf{Model \cirnum{4}} and \textbf{Model \cirnum{6}}, respectively. \textbf{Model \cirnum{7}} apparently dwarfs all other models by including liquidity state, time factor, Hawkes kernels, and LASSO all together.

We have calculated the AICs for the 7 types of model across all the 20 trading days from 2019-01-02 to 2019-01-31. The AIC comparison and difference between different model types are demonstrated in the following Table \ref{table: aic diff}.

\begin{table}[H]
\centering

    \begin{threeparttable}
            \begin{tabular}{c|c|c|c|c|c|c|c} 
            % \hline\hline
            % \multicolumn{8}{c}{\begin{tabular}[c]{@{}c@{}}Summary Statistics of AIC Difference of Apple .Inc 2019-01-02 - 2019-01-31\\~Maximum Support s = 20s, bin-size $\Delta$ = 0.25s\\Order size is considered in bin-count construction\end{tabular}}                                                                    \\ 
            % \hline\hline
            \hline
            \begin{tabular}[c]{@{}c@{}}\textbf{AIC} \\\textbf{Difference}\end{tabular}  & \textbf{Min}   & \begin{tabular}[c]{@{}c@{}}\textbf{1st} \\\textbf{Quantile}\end{tabular} & \textbf{Median} & \textbf{Mean}  & \begin{tabular}[c]{@{}c@{}}\textbf{3rd} \\\textbf{Quantile}\end{tabular} & \textbf{Max}          & \begin{tabular}[c]{@{}c@{}} \textbf{\# of} \\\textbf{days with} \\ \textbf{decreased }\\\textbf{AIC}\end{tabular}  \\ 
            \hline
            \begin{tabular}[c]{@{}c@{}}\textbf{\cirnum{4}} - \textbf{\cirnum{3}}\tnote{1}\end{tabular} & -4.09 & -1.19 & -0.58  & -0.89 & -0.16  & 0.09 & 18 out of 20  \\ 
            \hline
            \begin{tabular}[c]{@{}c@{}}\textbf{\cirnum{6}} - \textbf{\cirnum{4}}\tnote{2}\end{tabular}         & -0.31 & -0.24 & -0.23  & -0.22 & -0.18  & -0.13 & 20 out of 20  \\ 
            \hline
            \begin{tabular}[c]{@{}c@{}}\textbf{\cirnum{5}} - \textbf{\cirnum{4}}\tnote{3}\end{tabular}              & -0.12 & -0.09 & -0.08  & -0.08 & -0.07  & -0.03 & 20 out of 20  \\ 
            \hline
            \begin{tabular}[c]{@{}c@{}}\textbf{\cirnum{5}} - \textbf{\cirnum{3}}\tnote{4}\end{tabular}    & -4.15 & -1.28 & -0.63  & -0.97 & -0.23  & -0.01 & 20 out of 20  \\ 
            \hline
            \begin{tabular}[c]{@{}c@{}}\textbf{\cirnum{7}} - \textbf{\cirnum{6}}\tnote{5}\end{tabular}            & -0.13 & -0.09 & -0.08  & -0.08 & -0.07  & -0.03 & 20 out of 20                                                              \\
            \hline
            \end{tabular}
    
    \begin{tablenotes}
    {\setlength\itemindent{-10pt} \item Interpretations:}
        \item[1] The Hawkes part has stronger explanatory power than the liquidity state and time factor part.
        \item[2] Adding the liquidity state and time factor to the Hawkes part further improves explanatory power.
        \item[3] Adding LASSO (LASSO parameter 0.0005) to the Hawkes part further improves explanatory power.
        \item[4] The Hawkes part with LASSO (LASSO parameter 0.0005) generates stronger explanatory power than the liquidity state and time factor.
        \item[5] Adding LASSO (LASSO parameter 0.0005) further improves the explanatory power of the model with the liquidity state, time factor, and Hawkes.
    \end{tablenotes}
\end{threeparttable}
        \caption{AIC difference summary statistics of Apple. Inc from 2019-01-02 to 2019-01-31. Maximum Support s = 20s, bin-size $\Delta$ = 0.25s. AIC has been adjusted for sample size so that it reflects the AIC per single sample.}
    \label{table: aic diff}
\end{table}

Following the sensitivity analysis provided in Section \ref{sec: sensitivity analysis}, we also present the model selection results when the size of order is ignored and when the bin-size is enlarged from 0.25s to 0.5s. The results are demonstrated in Supplementary Material Section \ref{sec_support: additional model selection}.

In summary, Figure \ref{fig:aic_delta025_withsize}, Table \ref{table: aic diff}, and Supplementary Material Section \ref{sec_support: additional model selection} demonstrate the following conclusions:
\begin{itemize}
    \item In event arrival intensity modeling, the inclusion of the Hawkes stimulating function part, the liquidity state part, and the time factor part all contribute to decreased AIC. Moreover, the Hawkes part is more powerful than the liquidity state and time factor parts in terms of reducing AIC.
    \item Adding LASSO regularization with a small regularization parameter can effectively eliminate redundant parameters for the Hawkes stimulating function and thereby further reduce AIC.
    \item Under maximum support $s = 20$ seconds, smaller AIC is achieved when the bin-size $\Delta$ decreases to a shorter period.
\end{itemize}

The model selection results are consistent whether we consider the order sizes or not. This is intuitive since the two methods are just different ways to account for the size of orders based on the same LOB dataset.
 %because we hope to make Model VI and VII consistent, we don not intend to set a large proportion of the Hawkes estimators to zeros. 

A very small $\lambda_{i} = 0.0005$ is chosen throughout our implementations. We have also implemented the same model AIC selections under $\lambda_{i} = 0.001$ and achieved similar results. Therefore, we believe a choice of $\lambda_{i}$ that regularizes 10\% to 20\% of parameters to zeros given a small choice of $\Delta$ (around $0.5$ seconds) is enough to reduce estimation AIC. Any more advanced method for selecting the appropriate $\lambda_{i}$ is beyond the scope of this paper and is left for future works.

For the choice of bin-size $\Delta$, Kirchner \cite{kirchner2017estimation} has discussed that the choice of $\Delta$ is a bias/variance trade-off as well as a bias/computational-issue trade-off when $\Delta$ is extremely small. Our estimation demonstrates that for all else being equal, smaller AICs are achieved when we reduce $\Delta$ from 0.5 seconds to 0.25 seconds for all 20 days discussed, taking the AIC difference between Figure \ref{fig:aic_delta025_withsize} and Figure \ref{fig:aic_delta05_withsize} as an example.  In addition, when $\Delta = 0.25$ seconds, \textbf{Model \cirnum{5}} (Hawkes+LASSO) achieves smaller AIC than \textbf{Model \cirnum{3}} (liquidity state+time factor) for all 20 days discussed (see Table \ref{table: aic diff}) while the number of days reduces to 19 when $\Delta = 0.5$ seconds with all else equal (see Table \ref{table: aic diff largeDelta}). This result implies that choosing $\Delta = 0.25s$ is always better than choosing $\Delta = 0.5s$ throughout our estimation. Also, we anticipate smaller AIC will be achieved if $\Delta$ is reduced to a further smaller value until the model reaches a threshold when the increasing number of parameters brought by the decreasing $\Delta$ generates too much penalty. However, the computational budget required for model estimation increases quadratically as $\Delta$ decreases. Due to this computational barrier, any experiment with $\Delta$ smaller than 0.25s is left for future research.

\section{Discussion and conclusion}

Though we have demonstrated our choice of maximum support $s$ and bin-size $\Delta$ is appropriate enough to derive expected results, research on the choice of $(s,\Delta)$ and their dynamics is a natural direction of further study. Besides, as an extension to \cite{kirchner2017estimation}, \cite{kirchner2018nonparametric} has introduced the comparison between the Hawkes model non-parametric estimation and maximum likelihood estimation (MLE), and concluded that the non-parametric method outperforms MLE under some circumstances. Whether this result holds under our proposed framework is also worth investigating. Another possible direction for future research would be to explore more LOB state variables that could reduce the estimation AIC. For example, it is worthwhile to test queue imbalance as proposed by \cite{morariu2018state}. Moreover, based on previous research \cite{huang2017ergodicity,cinlar1981representation} on the mathematical properties for Markov models, further research on the stationary and ergodicity properties for our proposed model is also a direction of investigation.

To conclude, in this work we have proposed a comprehensive method in high-frequency limit order book modeling, integrated Markovian state factors into plain-vanilla Hawkes process, and applied a flexible non-parametric method for high-dimensional estimation. Our model provides more careful classification rules for LOB event types and does not require strict parametric assumptions on event stimulating kernels. The mathematical property of our model enables us to implement the estimation on large scale under a parallel and distributed computing framework. We believe our proposed model will bring valuable insights for researchers, financial institutions, and policymakers who attempt to understand the distribution of the order book, the stimulating effects between orders, and more topics related to market microstructure.

\bibliographystyle{siamplain}
\bibliography{ms.bib}

\clearpage
\renewcommand{\appendixtocname}{Supplementary Material}
\renewcommand\appendixname{Supplementary Material}
\renewcommand\appendixpagename{Supplementary Material}
\newcommand{\beginsupplement}{%
        \setcounter{table}{0}
        \renewcommand{\thetable}{S\arabic{table}}%
        \setcounter{figure}{0}
        \renewcommand{\thefigure}{S\arabic{figure}}%
     }

\begin{appendices}
\beginsupplement
This Supplementary Material is organized as following: Section \ref{sec_support: reference price} illustrates the construction of the reference price in the order book representation mentioned in Section \ref{bigsec:order book representation}; Section \ref{sec_support:estimation math} illustrates the  mathematical representation details of the estimation procedure following Section \ref{sec:nonparametric estimation}; Section \ref{sec_support: LASSO math} demonstrates the mathematical representation for the LASSO regularization discussed in Section \ref{sec: add LASSO}; Section \ref{sec_support: estimated excitement functions bid} demonstrates the estimated Hawkes excitement functions of insertion event at the bid side following Section \ref{sec:estimated excitement functions: ask}; Section \ref{sec_support: more non_exponential} shows more examples on the estimated non-exponential shaped excitement functions following Section \ref{sec: exponential and Non-exponential shape of excitement function}; Section \ref{sec_support: more liquidity state} presents more examples on the liquidity state estimation following Section \ref{sec: liquidity state}; Section \ref{sec_support: result nosize} demonstrates the estimation results when the size of order is ignored as mentioned in Section \ref{sec: sensitivity analysis}; Section \ref{sec_support: result noLASSO} demonstrates the estimation results when the LASSO regularization is removed as mentioned in Section \ref{sec: sensitivity analysis}; Section \ref{sec_support: result large bin size} demonstrates the estimation results when the bin-size is enlarged as mentioned in section \ref{sec: sensitivity analysis}; Section \ref{sec_support: additional model selection} presents additional model selection results as mentioned in Section \ref{sec: model selection result}.

\section{Specifications on reference price} \label{sec_support: reference price}
Following Section \ref{bigsec:order book representation}, this supporting section demonstrates the construction of the reference price $p_{ref}$. The construction method presented here is mainly adopted from \cite{huang2015simulating}. 

When the bid-ask spread is odd in tick unit, it is intuitive to use the mid price $p_{mid}$ to approximate the reference price $p_{ref}$. Though one can still use $p_{mid}$ as a proxy of $p_{ref}$ when the spread is even in tick unit, it is no longer appropriate enough since $p_{mid}$ itself can be a position for order arrivals. To be more strict, when $(Q_\text{best-ask} - Q_\text{best-bid}) = 2n+1, n\in\mathbb{Z}$, we have $p_{ref} = p_{mid} = (Q_\text{best-bid} + Q_\text{best-ask})/2$; When $(Q_\text{best-ask} - Q_\text{best-bid}) = 2n, n\in\mathbb{Z}$, we have $p_{ref} = (Q_{best-bid} + Q_{best-ask})/2+\alpha/2$ or $p_{ref} = (Q_{best-bid} + Q_{best-ask})/2-\alpha/2$, whichever is closer to the previous value of $p_{ref}$.

\section{Mathematical details on non-parametric estimation} \label{sec_support:estimation math}
Following Section \ref{sec:nonparametric estimation}, this supporting section illustrates the mathematical representations of the non-parametric estimation over the $(6K+6)$-dimensional LOB data. 

Precisely, based on Definition 3.3, Theorem 3.5, and Definition 3.6 from \cite{kirchner2017estimation}, as well as \cite{lutkepohl2005new}(page70-75), the mathematical details of the estimation procedure is given in Definition \ref{def: nonparam estimation} and Definition \ref{def: design matrix}. 

According to our model specification discussed in Section \ref{bigsec:model specification}, a total number of $6K+6$ event types is considered for a level-$K$ order book and the number of event types serves as the dimension of the multivariate Hawkes process. For notation simplicity, we denote $6K+6$ as $d$ in the following discussions (Definition \ref{def: nonparam estimation}, Definition \ref{def: design matrix}, Remark \ref{remark: OLS-implement}, and Definition \ref{def: LASSO math} in Supplementary Materials \ref{sec_support: LASSO math}).

\begin{definition} \label{def: nonparam estimation}
Let $X_{i}$ ($\forall i = 1,2,\ldots,d$) be a $d$-variate Hawkes process derived from LOB data with varying baseline intensities controlled by liquidity state sequence $l_{i}(t)$ and time factor $t$. Let $T>0$ and consider the time interval $(0,T]$. For some bin-size $\Delta>0$, construct the following bin-count sequences according to Section \ref{sec: bin construction}:
\[
 \left(B_{k}^{(\Delta)}, l_{k}^{(\Delta)}, t_{k}^{(\Delta)}\right), \forall k = 1,2,...,n:= \floor{T/\Delta}
\], where $B_{k}^{(\Delta)}$, $l_{k}^{(\Delta)}$ and $t_{k}^{(\Delta)}$ are $d\times1$ column vectors defined on $\mathbb{R}^{d}$.\\
Then assume $l_{k}^{(\Delta)}$ can be bucketized into 10 categories $[L_1,L_2,\ldots,L_{10}]$ and $t_{k}^{(\Delta)}$ can be bucketized into 126 categories $[T_1,T_2,\ldots,T_{126}]$ according to Section \ref{sec:estimation-baseline}. Given some maximum support $s$ such that $\Delta < s < T$,  The $d$-variate estimator for the proposed model is defined as:
\begin{align}
& \left(  \hat{\Phi}^{(\Delta,s)}_1   ,\hat{\Phi}^{(\Delta,s)}_2 ,..., \hat{\Phi}^{(\Delta,s)}_p, \hat{\mu}_{1}^{(\Delta,s)}, \ldots,\hat{\mu}_{10}^{(\Delta,s)},\hat{\theta}_{1}^{(\Delta,s)}, \ldots,\hat{\theta}_{126}^{(\Delta,s)}  \right):=\hat{\Phi}^{(\Delta,s)} \in \mathbb{R}^{d\times(dp+10+126)}, \nonumber\\
& \text{with } p :=\floor{s/\Delta}  
\label{eq3}
\end{align} Specifically, 
\[
\hat{\Phi}^{(\Delta,s)}_r := \begin{bmatrix}
\hat{\phi}^{(\Delta,s)}_{11,r} & \hat{\phi}^{(\Delta,s)}_{12,r} & \hat{\phi}^{(\Delta,s)}_{1d,r}\\
\hat{\phi}^{(\Delta,s)}_{21,r} & \hat{\phi}^{(\Delta,s)}_{22,r} & \hat{\phi}^{(\Delta,s)}_{2d,r}\\
\dots\\
\hat{\phi}^{(\Delta,s)}_{d1,r} & \hat{\phi}^{(\Delta,s)}_{d2,r} & \hat{\phi}^{(\Delta,s)}_{dd,r}\\
\end{bmatrix} \in \mathbb{R}^{d \times d}, \text{with } \forall r = 1,2,\ldots,p
\], where the matrix element 
\[
\hat{\phi}^{(\Delta,s)}_{ji,r}, \forall i,j = 1,2,\ldots,d; \forall r = 1,2,\ldots,p
\] are weakly consistent estimators for the Multivaraite Hawkes excitement function for event $j$ stimulating event $i$ at the $r$-th function discretizated short period. Substitute $r$ with $\floor{t/\Delta}$ yields that $\hat{\phi}^{(\Delta,s)}_{ji,\floor{t/\Delta}}$ are weakly consistent estimator (for $T \rightarrow \infty, \Delta \rightarrow 0$ and $s = \Delta p \rightarrow \infty$) for $\phi_{ji}(t)$ as shown in Eq.(\ref{equ: final form model}).

Also, $\left( \hat{\mu}_{1}^{(\Delta,s)}, \ldots,\hat{\mu}_{10}^{(\Delta,s)}  \right) \in\mathbb{R}^{d\times 10}$ and $\left( \hat{\theta}_{1}^{(\Delta,s)}, \ldots,\hat{\theta}_{126}^{(\Delta,s)}  \right) \in\mathbb{R}^{d\times 126}$ are weakly consistent estimators for function $M(\cdot)$ and $\Theta(\cdot)$ (for $T \rightarrow \infty, \Delta \rightarrow 0$ and $s = \Delta p \rightarrow \infty$).
\end{definition}

Definition \ref{def: nonparam estimation} gives the detailed description on the structure of the estimator. Then we elucidate the estimation formulas for estimator $\hat{\Phi}^{(\Delta,s)}$ in Definition \ref{def: design matrix}:

\begin{definition} \label{def: design matrix}
Followed from Definition \ref{def: nonparam estimation}, $\hat{\Phi}^{(\Delta,s)}$ can be obtained by applying the following multivariate conditional least-squares (CLS) estimator:
\[
   \hat{\Phi}^{(\Delta,s)} := \frac{1}{\Delta}\hat{\theta}_{CLS}^{(p,n)} \left( B_k^{(\Delta)},  l_k^{(\Delta)}, t_k^{(\Delta)}      \right)_{k = 1,\ldots,n}
\] 
The CLS estimator is defined as \[
\hat{\theta}_{CLS}^{(p,n)} : \mathbb{R}^{d\times (n-p)} \rightarrow \mathbb{R}^{d\times (dp + 10 + 126 )}
\]

\begin{align*}
    &\left( B_{1}^{(\Delta)},\ldots, B_{n}^{(\Delta)}; l_{1}^{(\Delta)},\ldots, l_{n}^{(\Delta)}; t_{1}^{(\Delta)},\ldots, t_{n}^{(\Delta)} \right) \rightarrow\\ &\hat{\theta}_{CLS}^{(p,n)}\left( B_{1}^{(\Delta)},\ldots, B_{n}^{(\Delta)}; l_{1}^{(\Delta)},\ldots, l_{n}^{(\Delta)}; t_{1}^{(\Delta)},\ldots, t_{n}^{(\Delta)} \right)
    := YZ^\top \left( ZZ^\top \right)^{-1}
\end{align*},where

\begin{align*}
& Z\left( B_{1}^{(\Delta)},\ldots, B_{n}^{(\Delta)}; l_{1}^{(\Delta)},\ldots, l_{n}^{(\Delta)}; t_{1}^{(\Delta)},\ldots, t_{n}^{(\Delta)} \right):=\\ 
& \begin{bmatrix}
 B_{p}^{(\Delta)} & B_{p+1}^{(\Delta)} & \dots & B_{n-1}^{(\Delta)}\\
B_{p-1}^{(\Delta)} & B_{p}^{(\Delta)} & \dots & B_{n-2}^{(\Delta)}\\
\dots\\
B_{1}^{(\Delta)} & B_{2}^{(\Delta)} & \dots & B_{n-p}^{(\Delta)}\\
\mathbbm{1}_{l_{p+1}^{(\Delta)}\in L_{1}} & \mathbbm{1}_{l_{p+2}^{(\Delta)}\in L_{1}} & \dots & \mathbbm{1}_{l_{n}^{(\Delta)}\in L_{1}}\\
\mathbbm{1}_{l_{p+1}^{(\Delta)}\in L_{2}} & \mathbbm{1}_{l_{p+2}^{(\Delta)}\in L_{2}} & \dots & \mathbbm{1}_{l_{n}^{(\Delta)}\in L_{2}}\\
\dots\\
\mathbbm{1}_{l_{p+1}^{(\Delta)}\in L_{10}} & \mathbbm{1}_{l_{p+2}^{(\Delta)}\in L_{10}} & \dots & \mathbbm{1}_{l_{n}^{(\Delta)}\in L_{10}}\\
[0]_{d\times1} & [0]_{d\times1} & \dots & [0]_{d\times1}\\
\mathbbm{1}_{t_{p+1}^{(\Delta)}\in T_{2}} & \mathbbm{1}_{t_{p+2}^{(\Delta)}\in T_{2}} & \dots & \mathbbm{1}_{t_{n}^{(\Delta)}\in T_{2}}\\
\mathbbm{1}_{t_{p+1}^{(\Delta)}\in T_{3}} & \mathbbm{1}_{t_{p+2}^{(\Delta)}\in T_{3}} & \dots & \mathbbm{1}_{t_{n}^{(\Delta)}\in T_{3}}\\
\dots\\
\mathbbm{1}_{t_{p+1}^{(\Delta)}\in T_{126}} & \mathbbm{1}_{t_{p+2}^{(\Delta)}\in T_{126}} & \dots & \mathbbm{1}_{t_{n}^{(\Delta)}\in T_{126}}\\
\end{bmatrix} \in \mathbb{R}^{(dp+10+126)\times (n-p)}
\end{align*} is the design matrix and 
\[
Y\left( B_{1}^{(\Delta)},\ldots, B_{n}^{(\Delta)}\right) := \left(  B_{p+1}^{(\Delta)}, B_{p+2}^{(\Delta)},\ldots,B_{n}^{(\Delta)} \right) \in \mathbb{R}^{d \times (n-p)}
\] is the response. 

Within the design matrix $Z$, $\left( \mathbbm{1}_{l_{k}^{(\Delta)}\in L_{1}}, \ldots, \mathbbm{1}_{l_{k}^{(\Delta)}\in L_{10}}   \right)$ and $\left( \mathbbm{1}_{t_{k}^{(\Delta)}\in L_{1}}, \ldots, \mathbbm{1}_{l_{k}^{(\Delta)}\in L_{126}}   \right)$ with $\forall k = p+1,\ldots,n$ are $d \times 1$ column indicator functions that returns $1$ (returns $0$ otherwise) when the $i$-th dimension element falling into the corresponding category; $[0]_{d \times 1}$ is $d$-dimensional column vector consisting of zeros. 
\end{definition}

The design matrix $Z$ in Definition \ref{def: design matrix} contains a $(n-p)$ dimensional row consisting of zeros. This row references the part of design matrix such that the time factor sequence $t_{k}^{(\Delta)}$ belongs to the $T_{1}$ category. The $T_{1}$ category is treated as the ``reference group" in the presence of the two categorical variables $[l_{k}^{(\Delta)},t_{k}^{(\Delta)}]$ and the regular (non-categorical) variables derived from $B_{k}^{(\Delta)}$. The \texttt{R} programming language we use is built with its default ``contrast coding" system that requires the existence of at least one ``reference group" when implementing linear regression models with more than one categorical variable. The CLS estimators for the "reference group" are automatically set to zeros according to the ``contrast coding" rule. Therefore, the CLS estimator in Definition \ref{def: design matrix} returns a zero vector of dimensional $d$ as the estimator for category $T_{1}$. The ``contrast coding" system may be different for other programming languages and their statistical packages. The choice of categorical variable coding rules and the reference group can be considered as adding/subtracting a constant on the estimators for one categorical variable and subtracting/adding it back on another, leading to no change on model fit.

Followed Definition \ref{def: design matrix}, the following definition gives an equivalent but easier way for model implementation:
\begin{remark}\label{remark: OLS-implement}
Given Definition \ref{def: nonparam estimation} and \ref{def: design matrix}, \cite{lutkepohl2005new}(page72) illustrates that the multivariate CLS-estimation is equivalent to $d$ individual Ordinary-Least-Squared(OLS) estimations, in which $d$ is the dimension of the CLS-estimation. Using design matrix $Z$, response $Y$, and estimator $\hat{\Phi}^{(\Delta,s)}$ given in Definition \ref{def: nonparam estimation} and \ref{def: design matrix}, and let $y_{i}$ be the transpose of the $i$-th row vector of response Y:
\[
 y_{i} = \left(  B_{i,p+1}^{(\Delta)}, B_{i,p+2}^{(\Delta)}, \ldots, B_{i,n}^{(\Delta)}  \right)^\top \in \mathbb{R}^{(n-p) \times 1}, \forall i = 1,2,\ldots,d
\]; Let $\hat{\phi_{i}}^{(\Delta,s)}$ be the transpose of the $i$-th row vector of $\hat{\Phi}^{(\Delta,s)}$:
\begin{align*}
    \hat{\phi_{i}}^{(\Delta,s)} = & \Big[  \left( \hat{\phi}^{(\Delta,s)}_{i1,1}, \ldots, \hat{\phi}^{(\Delta,s)}_{id,1} \right), \ldots, \left( \hat{\phi}^{(\Delta,s)}_{i1,p}, \ldots, \hat{\phi}^{(\Delta,s)}_{id,p} \right), \\
    & \left(  \hat{\mu}_{i,1}^{(\Delta,s)} ,\ldots, \hat{\mu}_{i,10}^{(\Delta,s)} \right), \left( \hat{\theta}_{i,1}^{(\Delta,s)} ,\ldots, \hat{\theta}_{i,126}^{(\Delta,s)} \right) \Big]^\top \in \mathbb{R}^{(dp+10+126) \times 1}
\end{align*} We have that $\hat{\phi_{i}}^{(\Delta,s)} := \left( ZZ^\top \right)^{-1}Zy_{i}$ is the OLS estimator for the model:
\[
y_{i} = Z^\top \phi_{i}^{(\Delta,s)} + u_{i}, \forall i = 1,2,\ldots,d
\], where $u_{i}$ is $(n-p)\times1$ white-noise column vector $\left( u_{i,p+1}, u_{i,p+2}, \ldots, u_{i,n} \right)^\top$ with $i = 1,\ldots,d$.
\end{remark}

In model implementation, we prefer conducting the OLS estimations based on $y_{i} = Z^\top \phi_{i}^{(\Delta,s)} + u_{i}$ over all dimensions $1,2,\ldots,d$, over the one single CLS-estimation shown in Definition \ref{def: design matrix}, since the OLS estimations involve less dimensions and thereby more computationally efficient under a parallel computing setting.

\section{Non-parametric estimation with LASSO}\label{sec_support: LASSO math}
The following definition illustrates LASSO regularization for our proposed model discussed in Section \ref{sec: add LASSO}. Consistent with Supplementary Materials \ref{sec_support:estimation math}, we denote the dimension of our estimation (the number of event types considered) $6\times(K+1)$ as $d$ for simplicity. Note $K$ represents the level of the order book we consider.

\begin{definition}\label{def: LASSO math}
Consider the design matrix $Z$, the response $Y$, and the OLS-estimators proposed in Definition \ref{def: design matrix} and Remark \ref{remark: OLS-implement}. The OLS-estimators minimize the loss function:
\[
  \left( y_{i} - Z^\top \phi_{i}^{(\Delta,s)}   \right)^\top\left( y_{i} - Z^\top \phi_{i}^{(\Delta,s)}   \right)
\]

Then consider adding a LASSO regularization term that only applies to the Hawkes excitement function $\phi_{i}^{(\Delta,s),excitements} := \Big[  \left( \hat{\phi}^{(\Delta,s)}_{i1,1}, \ldots, \hat{\phi}^{(\Delta,s)}_{id,1} \right), \ldots, \left( \hat{\phi}^{(\Delta,s)}_{i1,p}, \ldots, \hat{\phi}^{(\Delta,s)}_{id,p} \right)\Big]^\top \in \mathbb{R}^{dp\times1}$
to the loss function, the LASSO loss function becomes:
\begin{align*}
  \left( y_{i} - Z^\top \phi_{i}^{(\Delta,s)}   \right)^\top\left( y_{i} - Z^\top \phi_{i}^{(\Delta,s)}   \right) + \lambda_{i} \rVert  \phi_{i}^{(\Delta,s),excitements}       \rVert_{1}
\end{align*}
where $\lambda_{i}$ denotes the regularization penalty and $\rVert \cdot \rVert_{1}$ denotes the $\ell_{1}$-norm for estimators.
\end{definition}

\section{Estimated excitement functions: bid orders}\label{sec_support: estimated excitement functions bid}
Following Section \ref{sec:estimated excitement functions: ask}, this section illustrates excitement functions of insertion event at 1st bid (event \texttt{-1(i)}) stimulating insertion and cancellation at the 1st bid (event \texttt{-1(i)} and \texttt{-1(c)}). The shapes of the estimated excitement functions exhibit similar time-decaying patterns as shown in Section \ref{sec:estimated excitement functions: ask} for ask orders.

\begin{figure}[H]
    \centering
    \subfloat[\texttt{-1(i)} stimulate \texttt{-1(i)}]{{\includegraphics[scale = 0.16]{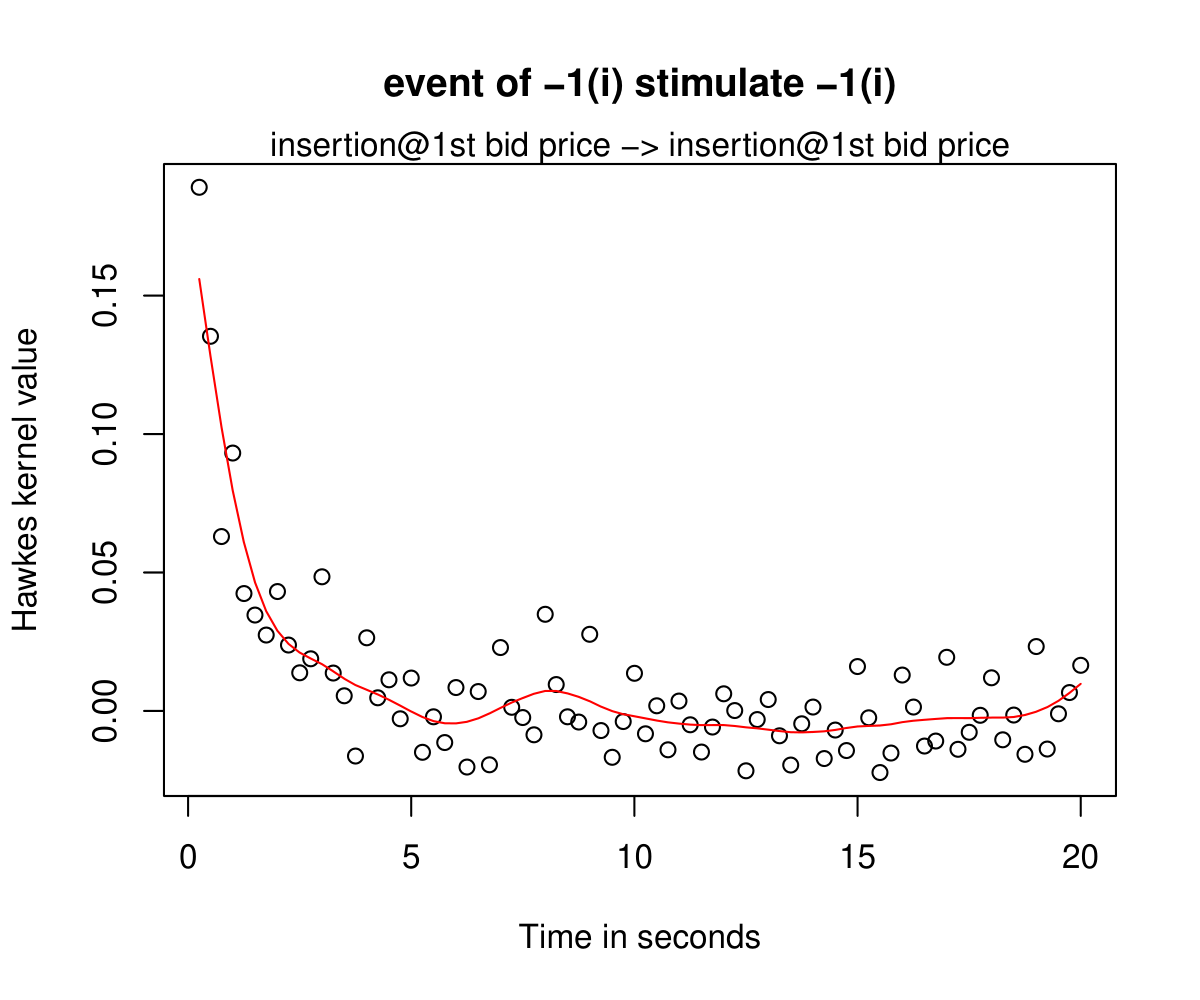} }}%
    \qquad
    \subfloat[\texttt{-1(i)} stimulate \texttt{-1(c)}]{{\includegraphics[scale =0.16]{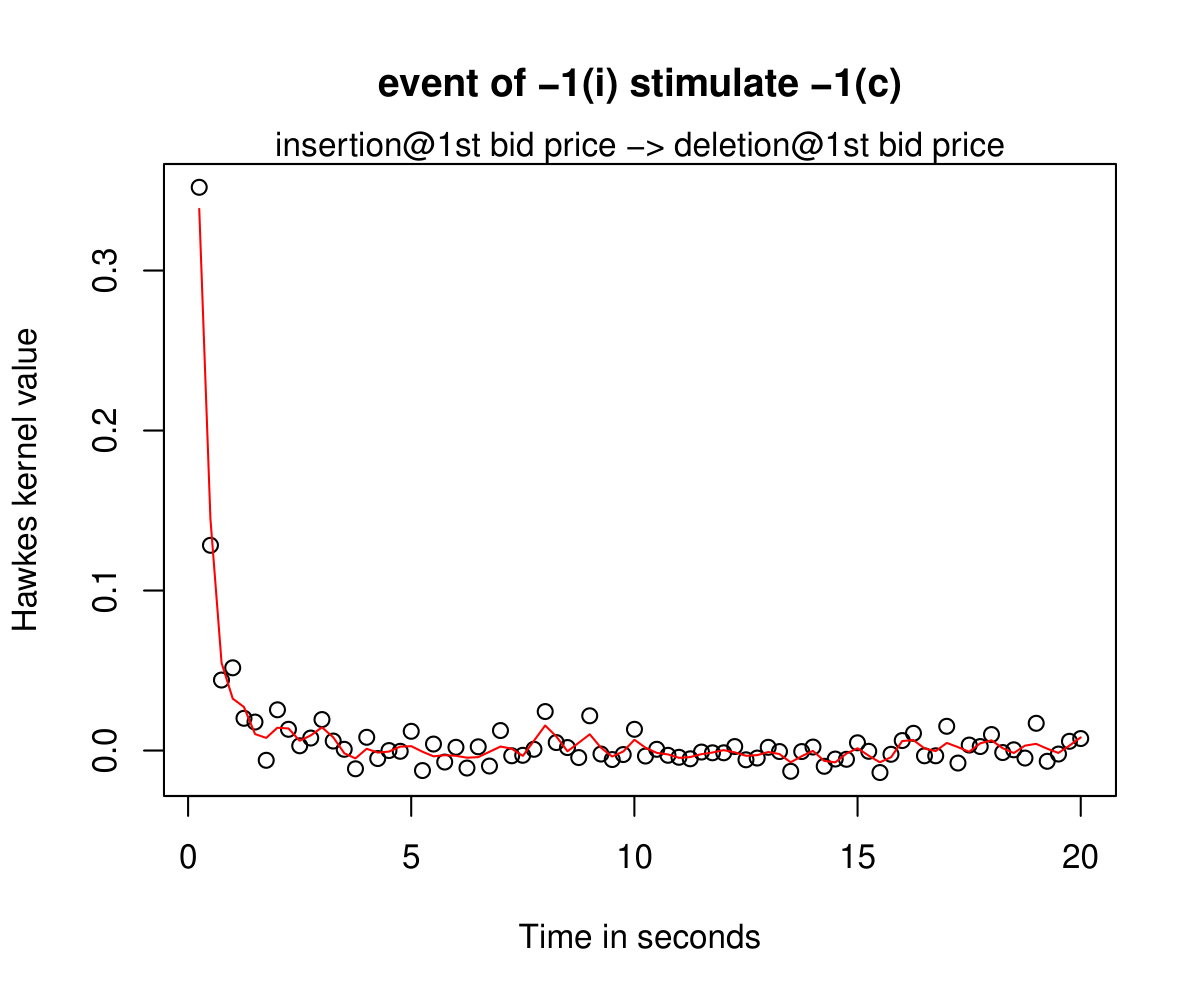} }}%
    \caption{Aggregated Hawkes excitement function estimation result under $(s=20 \text{ seconds}, \Delta=0.25 \text{ seconds})$ with LASSO regularization. The points illustrate the discrete function estimators. The red line illustrates the cubic smoothing spline for the points. }
    \label{fig: first-level similar bid}
\end{figure}

Additionally, we observed similarities between Fig.\ref{fig: first-level similar bid}(a) and Fig.\ref{fig: first-level similar bid}(b), which show the stimulation of \texttt{-1(i)} to \texttt{-1(i)} and \texttt{-1(c)}. Both excitement functions spike at around 8 seconds and 20 seconds. This observation suggests that the estimated Hawkes excitement functions are similar for the effect towards the insertion and cancellation at the 1st bid (effect towards \texttt{-1(i)} and \texttt{-1(c)}).

\section{More examples on non-exponential shaped excitement functions} \label{sec_support: more non_exponential}

Following Section \ref{sec: exponential and Non-exponential shape of excitement function} and Figure \ref{fig: non-exponential kernels}, this section demonstrates more examples on the estimated excitement functions with non-exponential shapes in the following Figure \ref{fig: more non-exponential kernels}.

\begin{figure}[H]
    \centering
    \subfloat[\texttt{-1(c)} stimulate \texttt{+2(t)}]{{\includegraphics[scale =0.135]{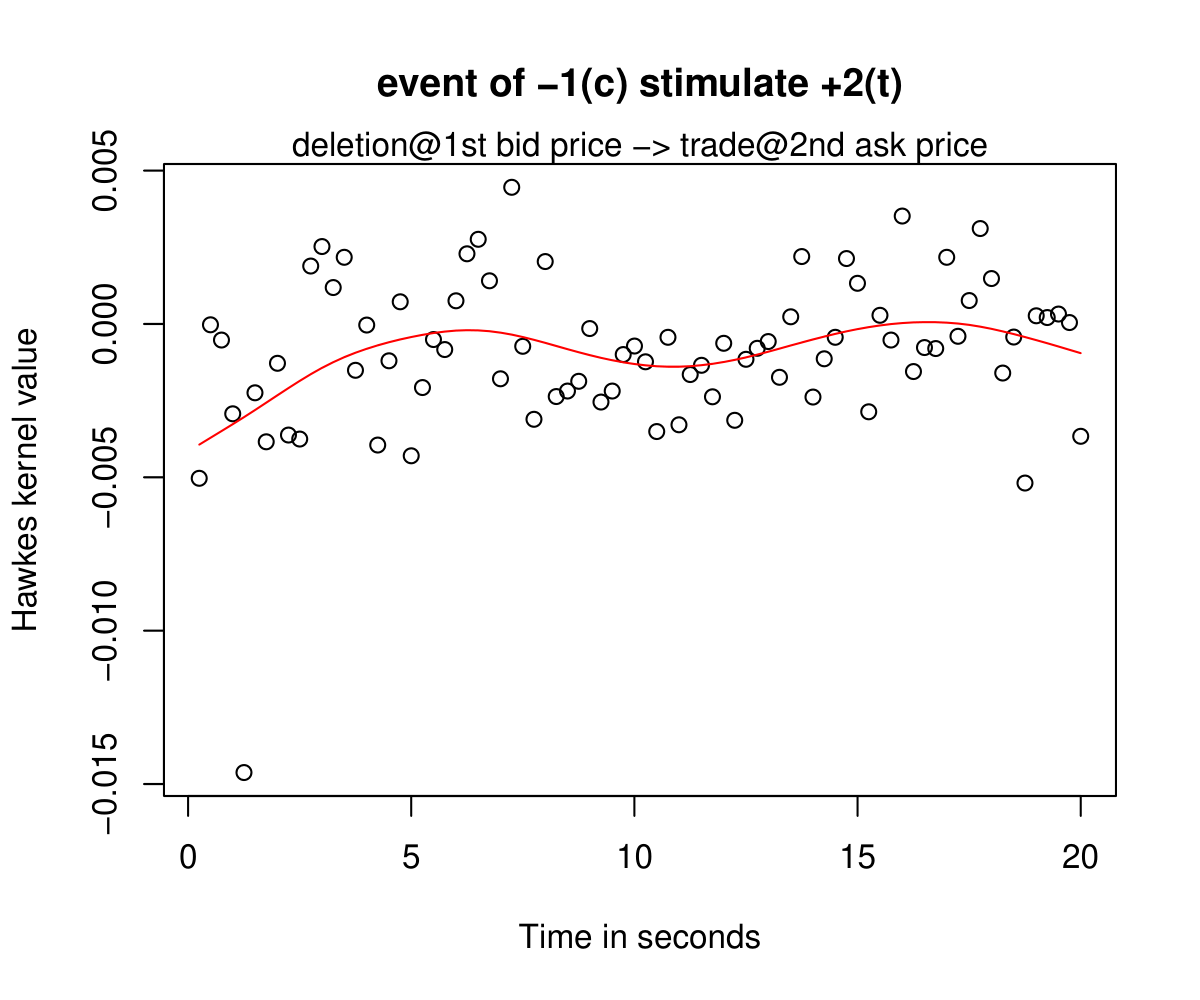} }}%
    \qquad
    \subfloat[\texttt{+2(t)} stimulate \texttt{-2(t)}]{{\includegraphics[scale =0.135]{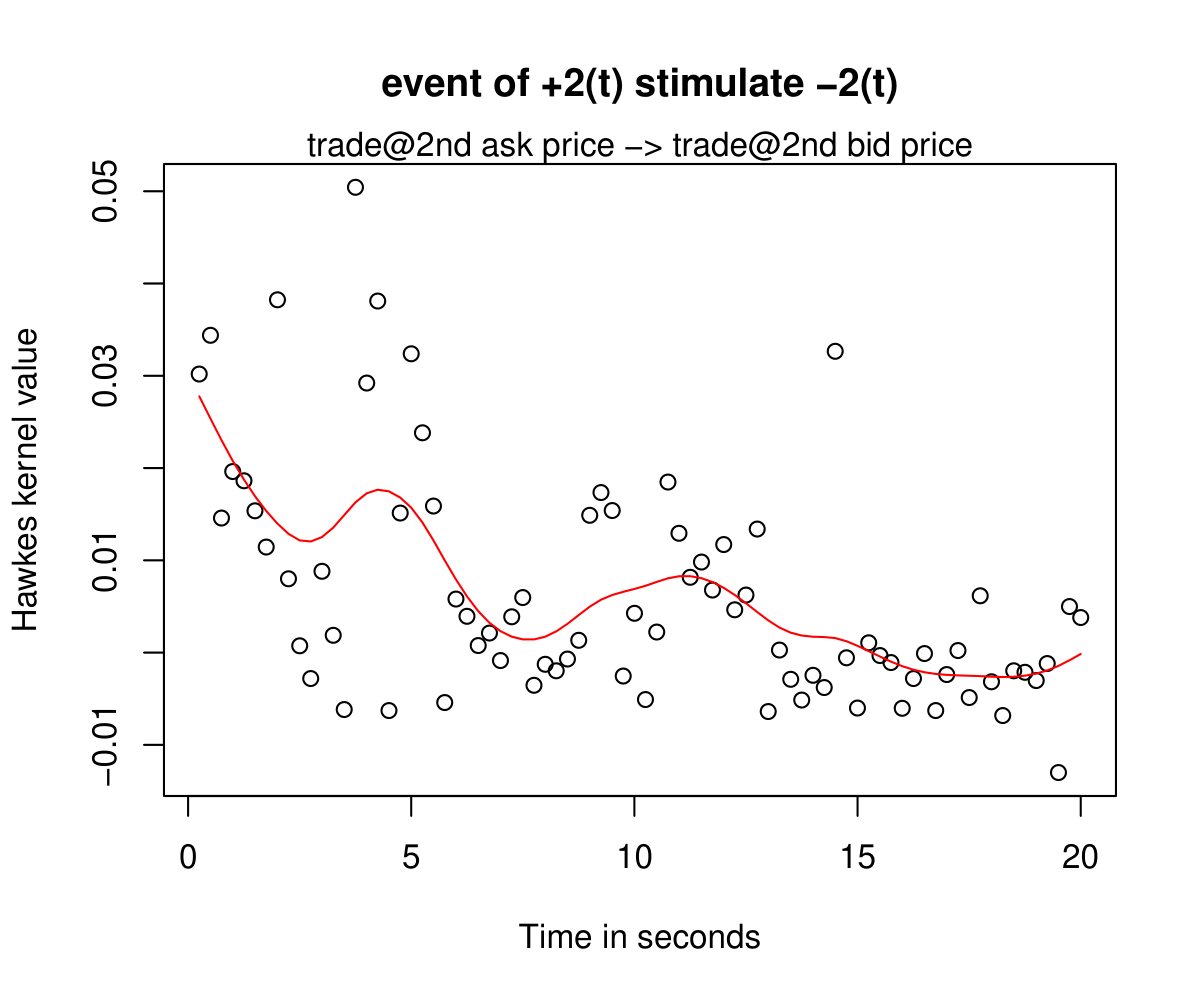} }}%
    \caption{Aggregated Hawkes excitement function estimation under $(s=20 \text{ seconds}, \Delta =0.25 \text{ seconds})$ with LASSO regularization. The points illustrate the discrete function estimators. The red line illustrates the cubic smoothing spline for the points. } 
    \label{fig: more non-exponential kernels}
\end{figure}

\section{More examples on liquidity state}\label{sec_support: more liquidity state}

\begin{figure}[H]
    \centering
    \subfloat[Liquidity state for \texttt{+1(i)}]{{\includegraphics[scale = 0.43]{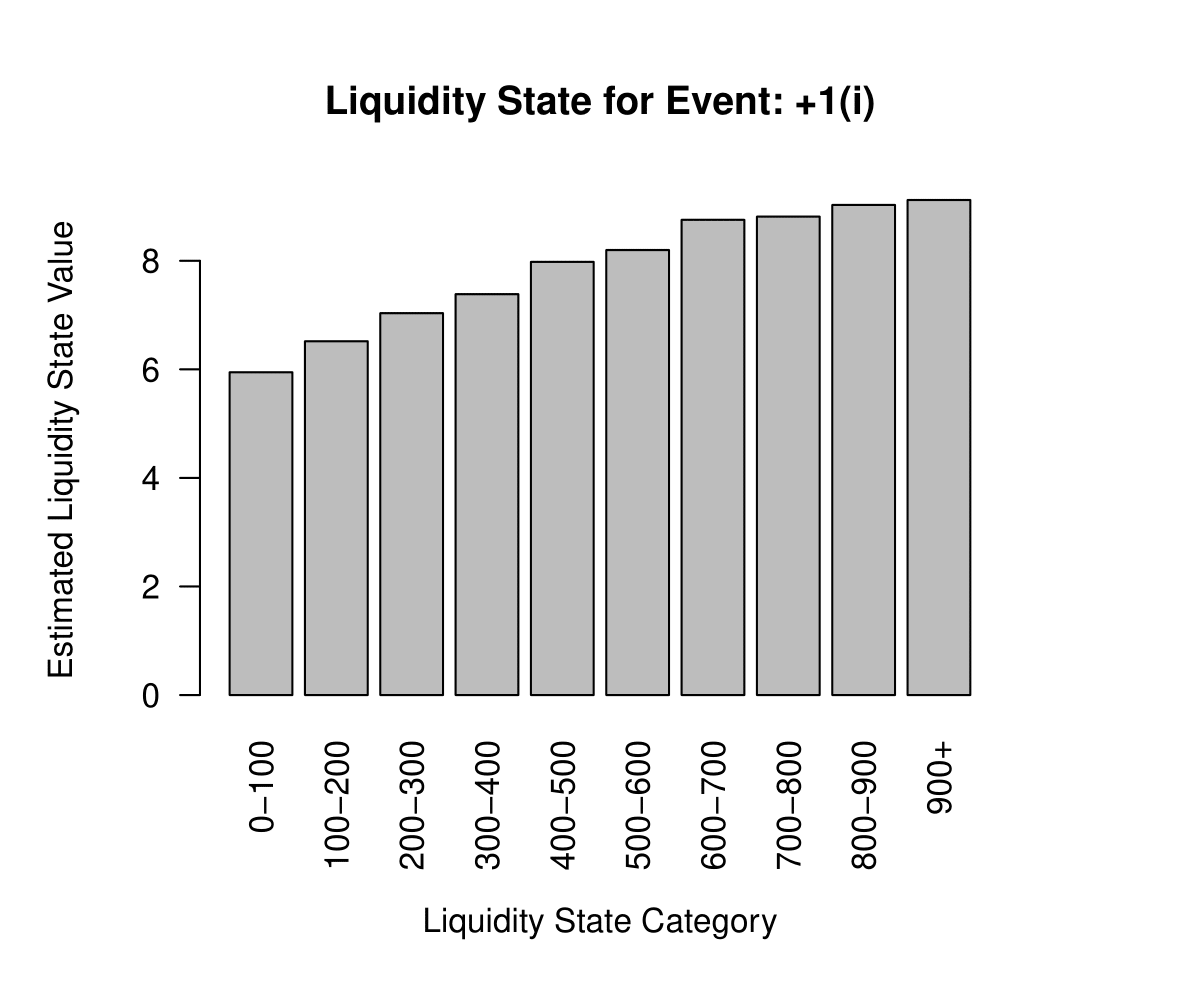} }}%
    \qquad
    \subfloat[Liquidity state for \texttt{+1(t)} ]{{\includegraphics[scale = 0.43]{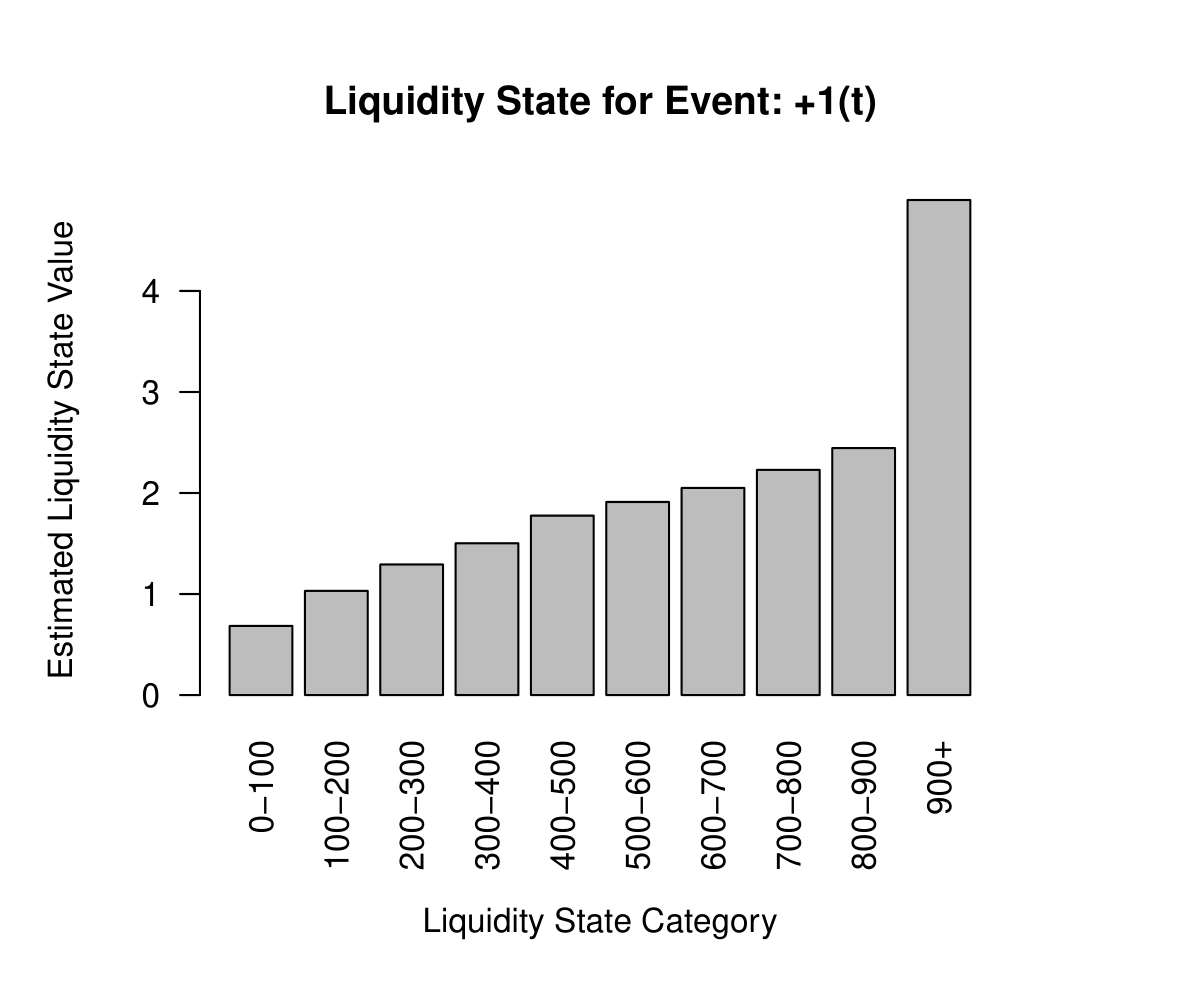} }}%
    \qquad
    \subfloat[Liquidity state for \texttt{-1(c)} ]{{\includegraphics[scale = 0.43]{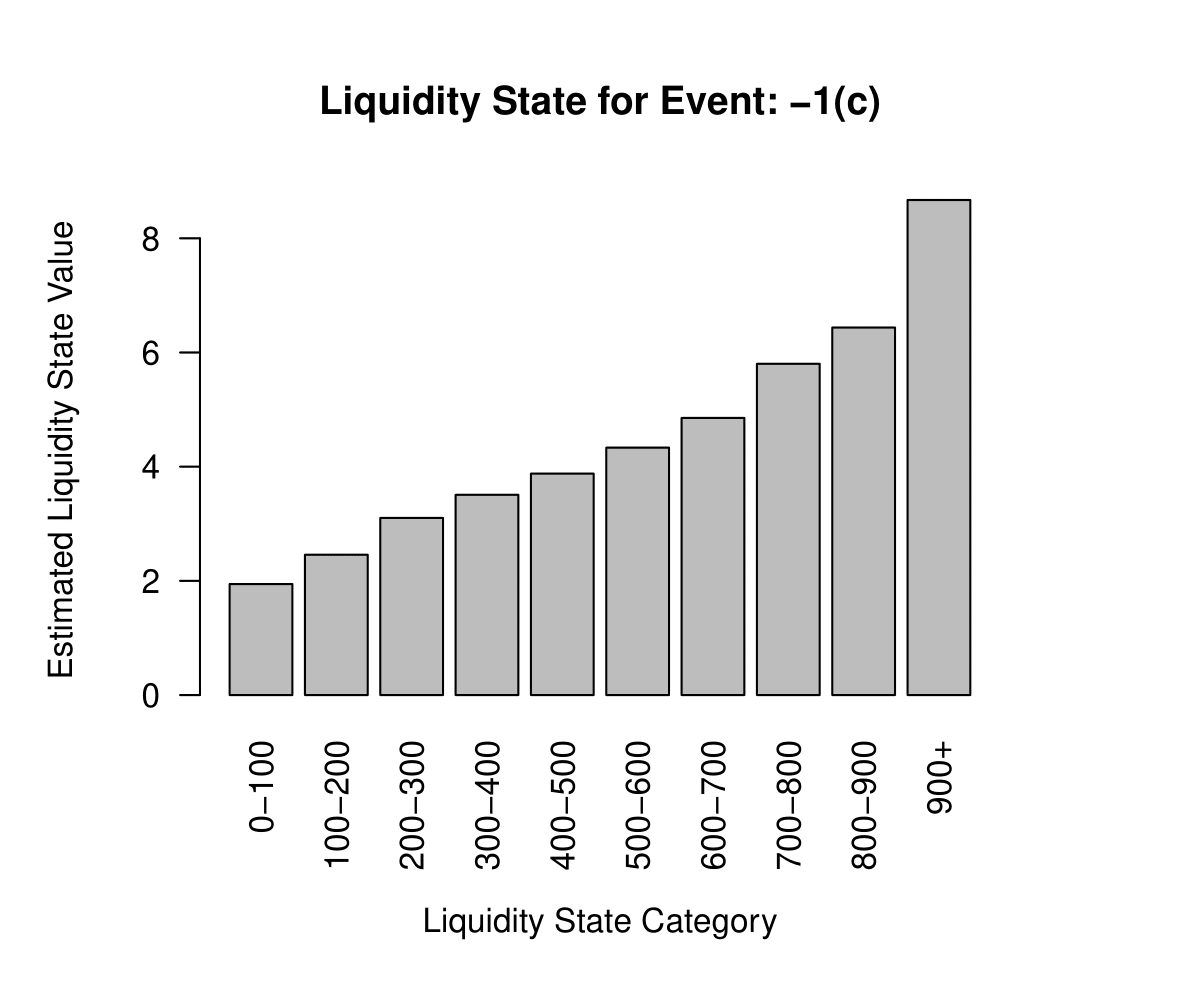} }}%
    \caption{Aggregated estimation results for liquidity state for selected events under $(s=20 \text{ seconds}, \Delta=0.25 \text{ seconds})$.}
    \label{fig: state_increase}
\end{figure}

\begin{figure}[H]
    \centering
    \subfloat[Liquidity state for \texttt{+3(i)}]{{\includegraphics[scale = 0.43]{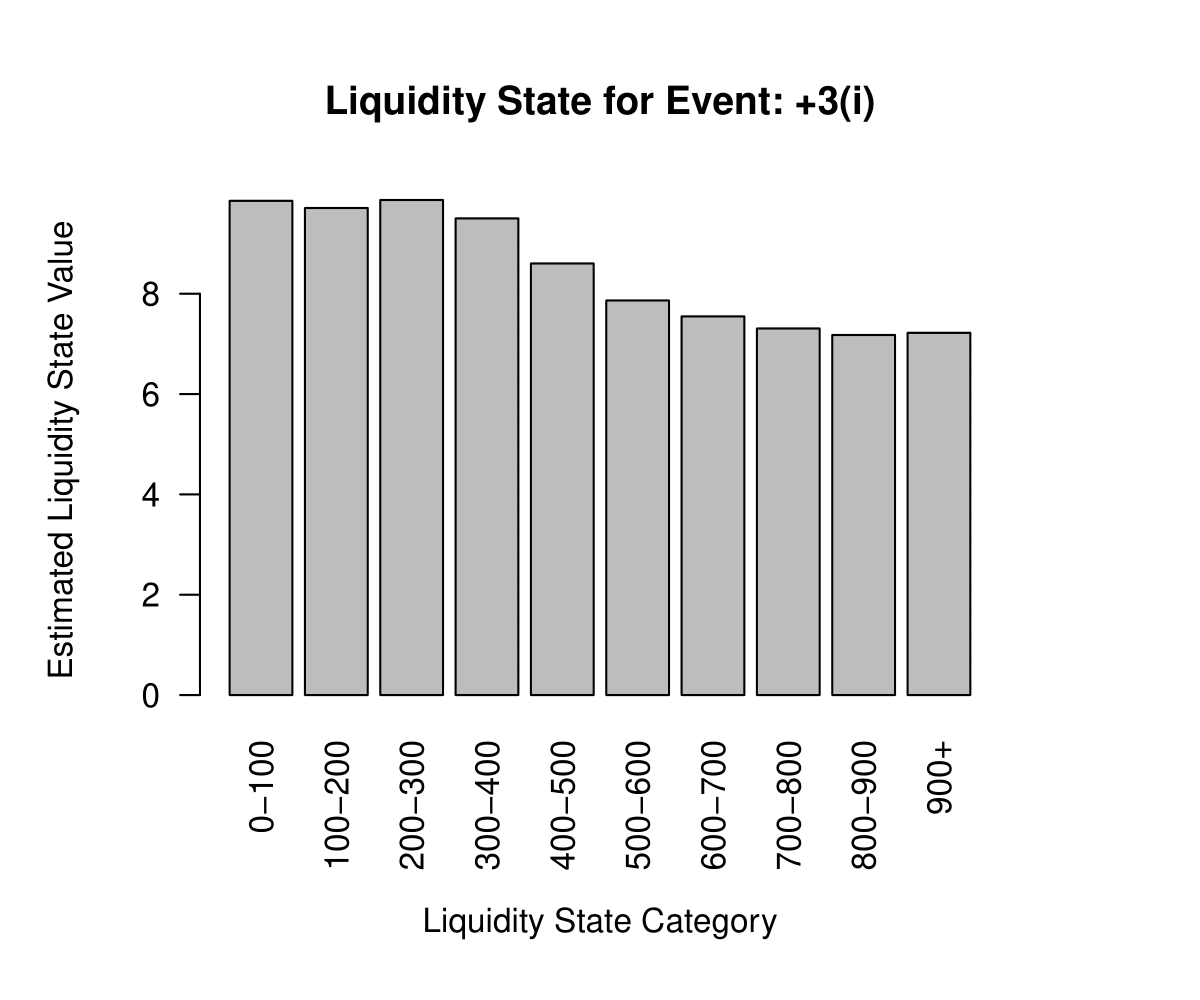} }}%
    \qquad
    \subfloat[Liquidity state for \texttt{+2(i)} ]{{\includegraphics[scale = 0.43]{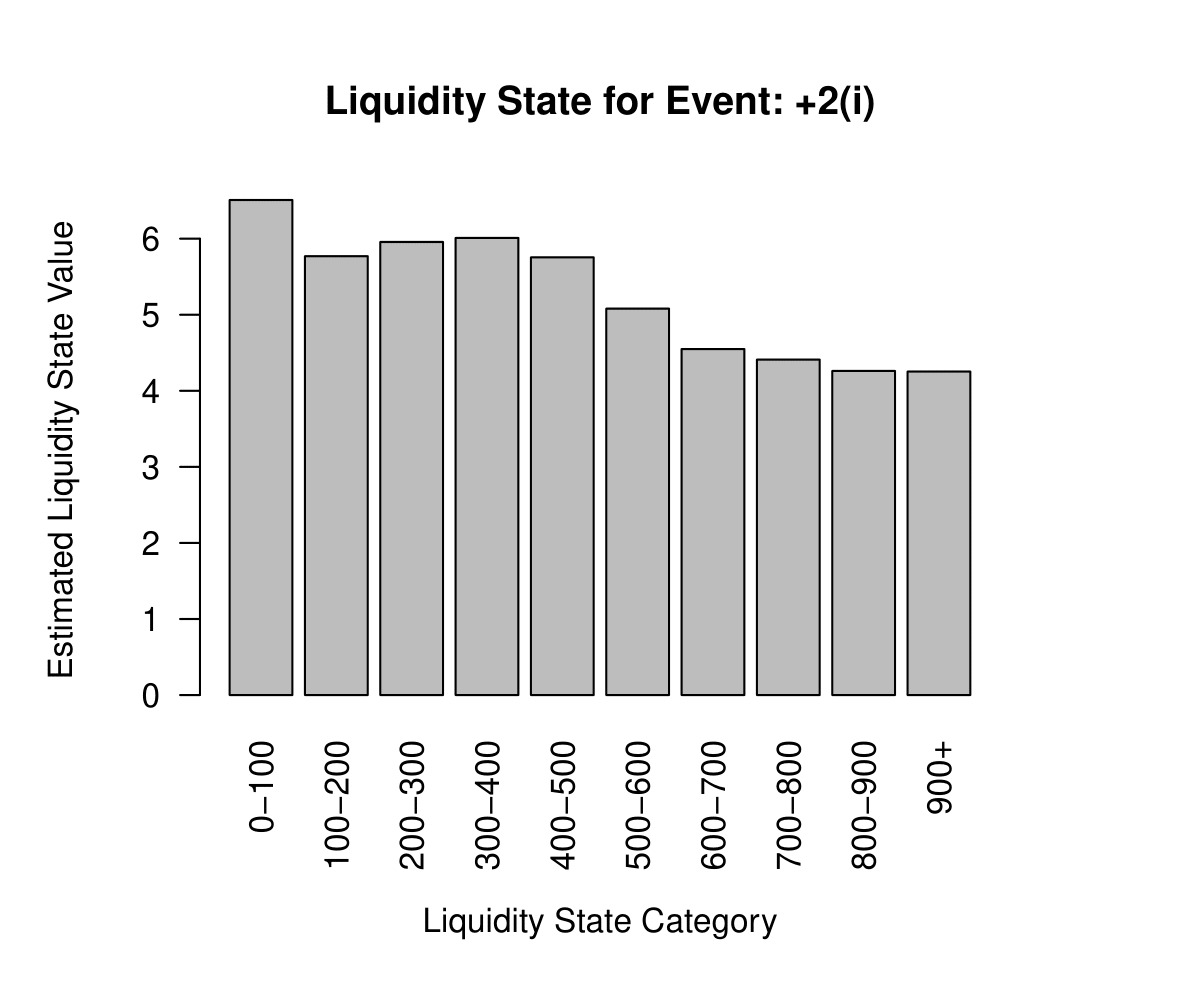} }}%
    \qquad
    \subfloat[Liquidity state for \texttt{-2(i)} ]{{\includegraphics[scale = 0.43]{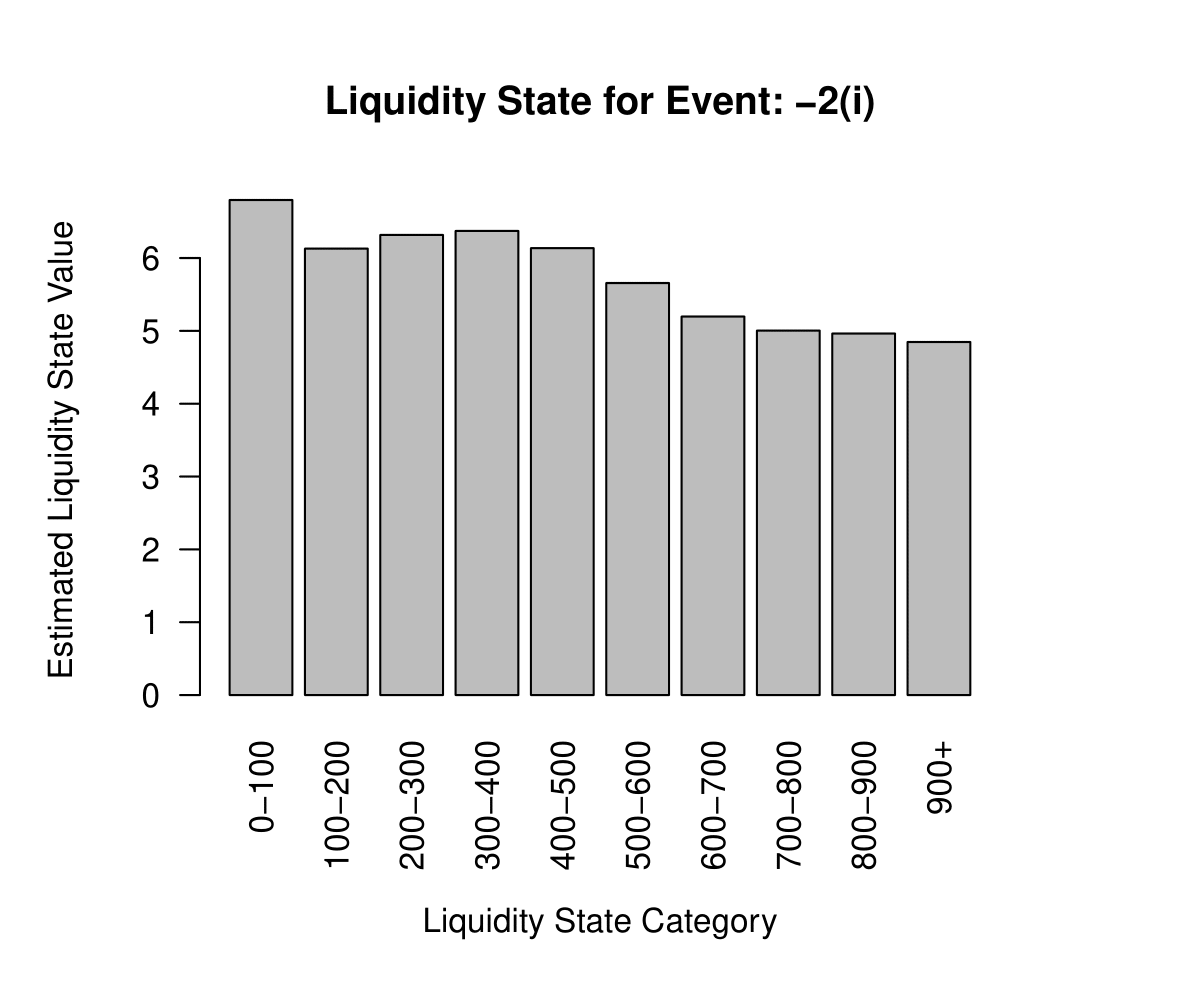} }}%
    \caption{Aggregated estimation results for liquidity state for selected events under $(s=20 \text{ seconds}, \Delta=0.25 \text{ seconds})$.}
    \label{fig: state_decrease}
\end{figure}

Following Section \ref{sec: liquidity state} and Figure \ref{fig: state_example}, this section presents more examples of the estimated result for liquidity state. The examples are presented in Figure \ref{fig: state_increase} and Figure \ref{fig: state_decrease}. In general, the arrival intensity increases as the liquidity state increases for trade/cancellation events and insertion events on the 1st level (i.e., \texttt{-3(c)}, \texttt{-3(t)}, \texttt{-2(c)}, \texttt{-2(t)},\texttt{-1(i)}, \texttt{-1(c)}, \texttt{-1(t)},\texttt{+1(i)}, \texttt{+1(c)}, \texttt{+1(t)},\texttt{+2(c)}, \texttt{+2(t)}, \texttt{+3(c)}, \texttt{+3(t)}); the arrival intensity decreases as liquidity state increases for the insertion events on the 2nd and 3rd level (i.e., \texttt{+3(i)}, \-\texttt{-3(i)}, \-\texttt{+2(i)}, \-\texttt{-2(i)}).

\section{Empirical results when order size is ignored}\label{sec_support: result nosize}
This supporting section demonstrates the empirical estimation results when the order size is ignored, as mentioned in Section \ref{sec: sensitivity analysis}. The demonstrations will be presented in the same format as the demonstrations from Section \ref{sec:estimated excitement functions: ask} to Section \ref{sec: time factor}.  In general, the results on excitement function, liquidity state, and time factor still hold qualitatively in the sense that most estimated functions have similar shapes. 

However, different order size considerations tend to give the estimated functions in different levels, where the estimations in general have lower intensity levels for many events when the order size is ignored. Also, the estimated Hawkes excitement functions tend to be less volatile if we ignore the order size. This behavior is expected since we aggregate order sizes in the original model setting while all orders are considered to have size 1 if we ignore order size. Therefore, when order size is ignored, it is natural for the estimates to have relatively lower intensity and volatility, especially during peak trading hours.

\subsection{Estimated excitement functions}
\begin{figure}[H]
    \centering
    \subfloat[\texttt{-1(i)} stimulate \texttt{-1(i)}]{{\includegraphics[scale = 0.16]{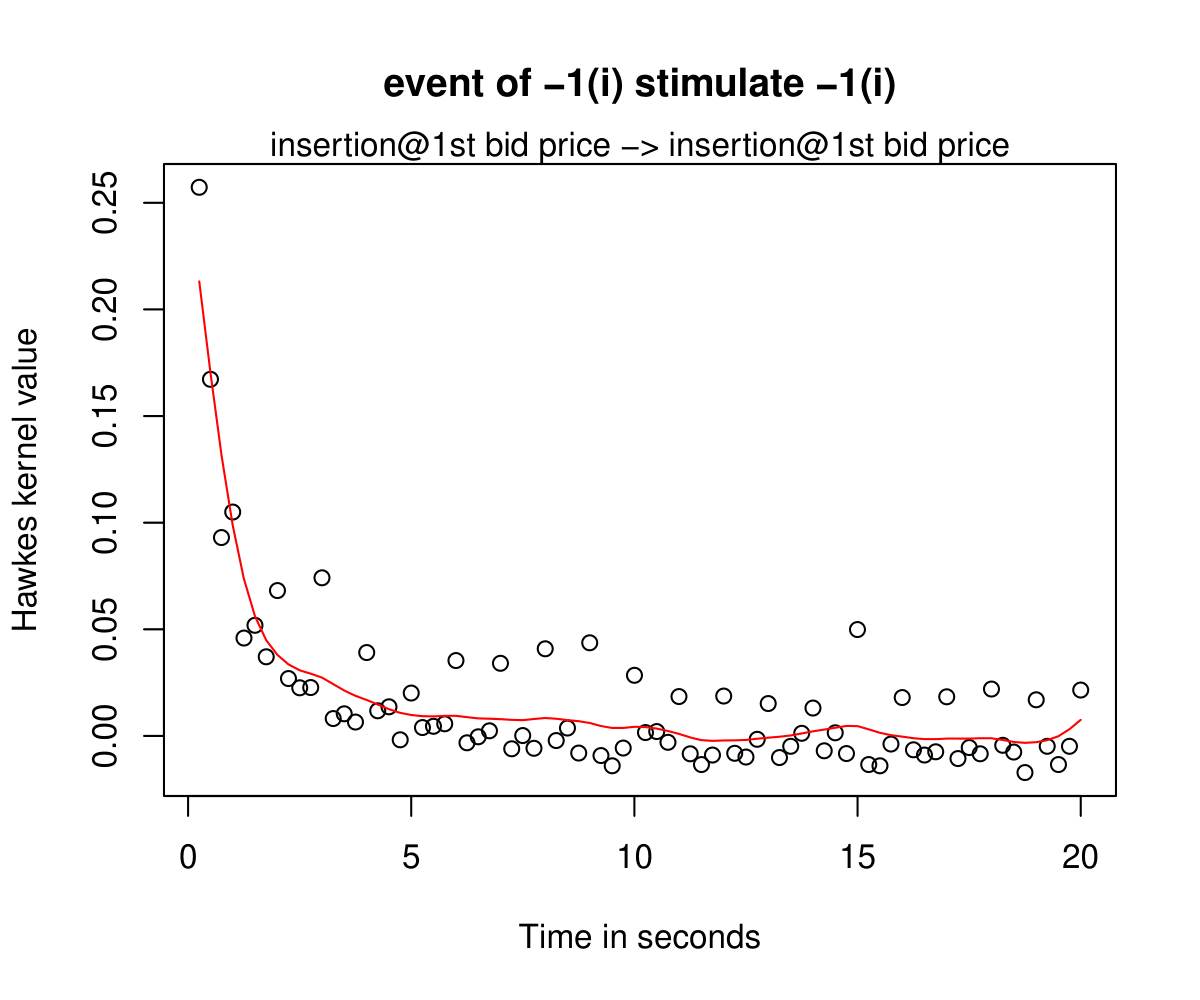} }}%
    \qquad
    \subfloat[\texttt{-1(i)} stimulate \texttt{-1(c)}]{{\includegraphics[scale =0.16]{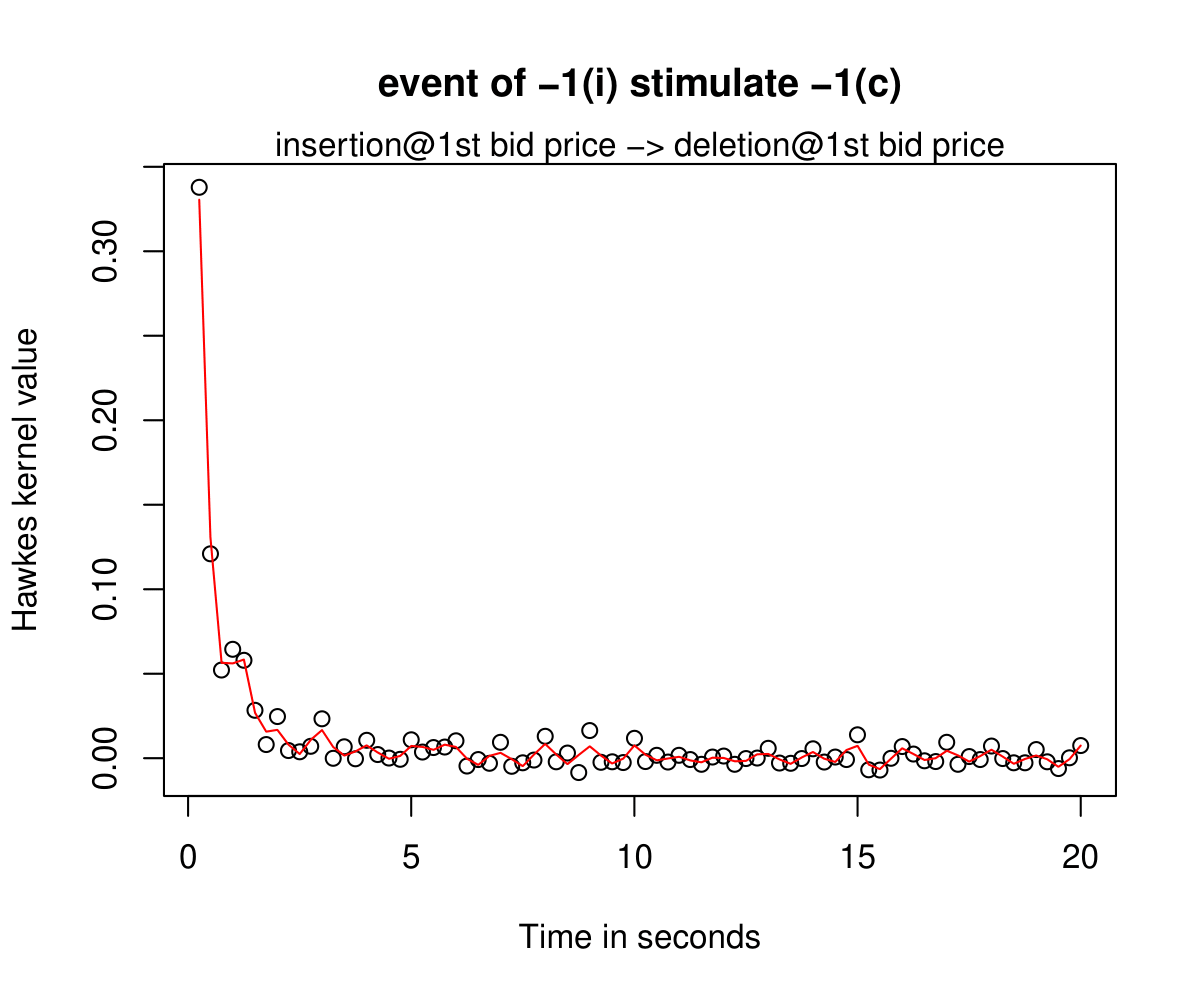} }}%
    \caption{Aggregated Hawkes excitement function estimation under $(s=20 \text{ seconds}, \Delta=0.25 \text{ seconds})$ with LASSO regularization. All orders are considered to have size 1. The points illustrate the discrete function estimators. The red line illustrates the cubic smoothing spline for the points.}
    \label{fig: first-level similar bid nosize}
\end{figure}

Based on Figure \ref{fig: first-level similar} and Figure \ref{fig: first-level similar bid}, the following Figure \ref{fig: first-level similar nosize} and Figure \ref{fig: first-level similar bid nosize} demonstrate the estimated Hawkes excitement functions when the order size is ignored.

\begin{figure}[H]
    \centering
    \subfloat[\texttt{+1(i)} stimulate \texttt{+1(i)}]{{\includegraphics[scale = 0.16]{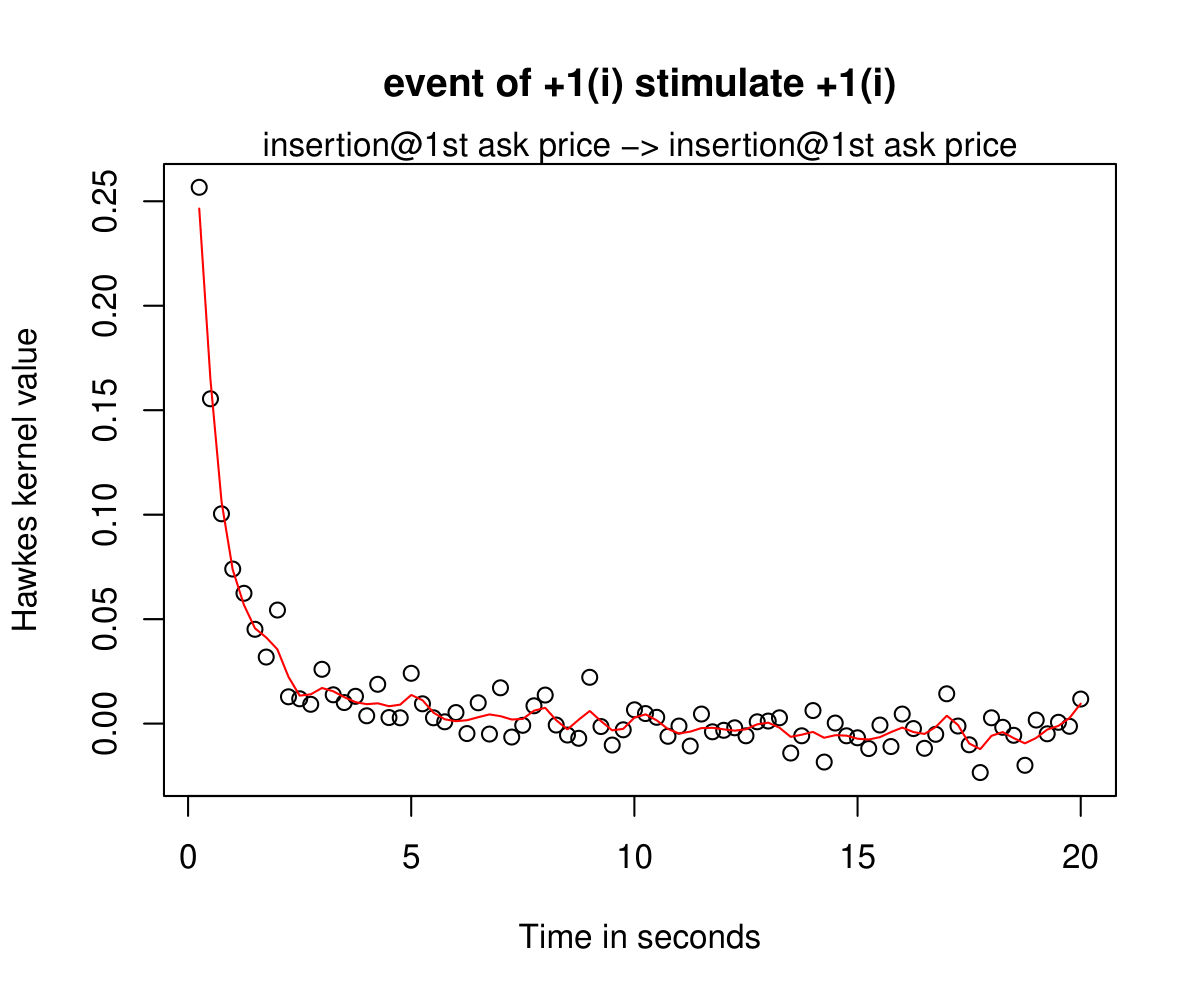} }}%
    \qquad
    \subfloat[\texttt{+1(i)} stimulate \texttt{+1(c)}]{{\includegraphics[scale =0.16]{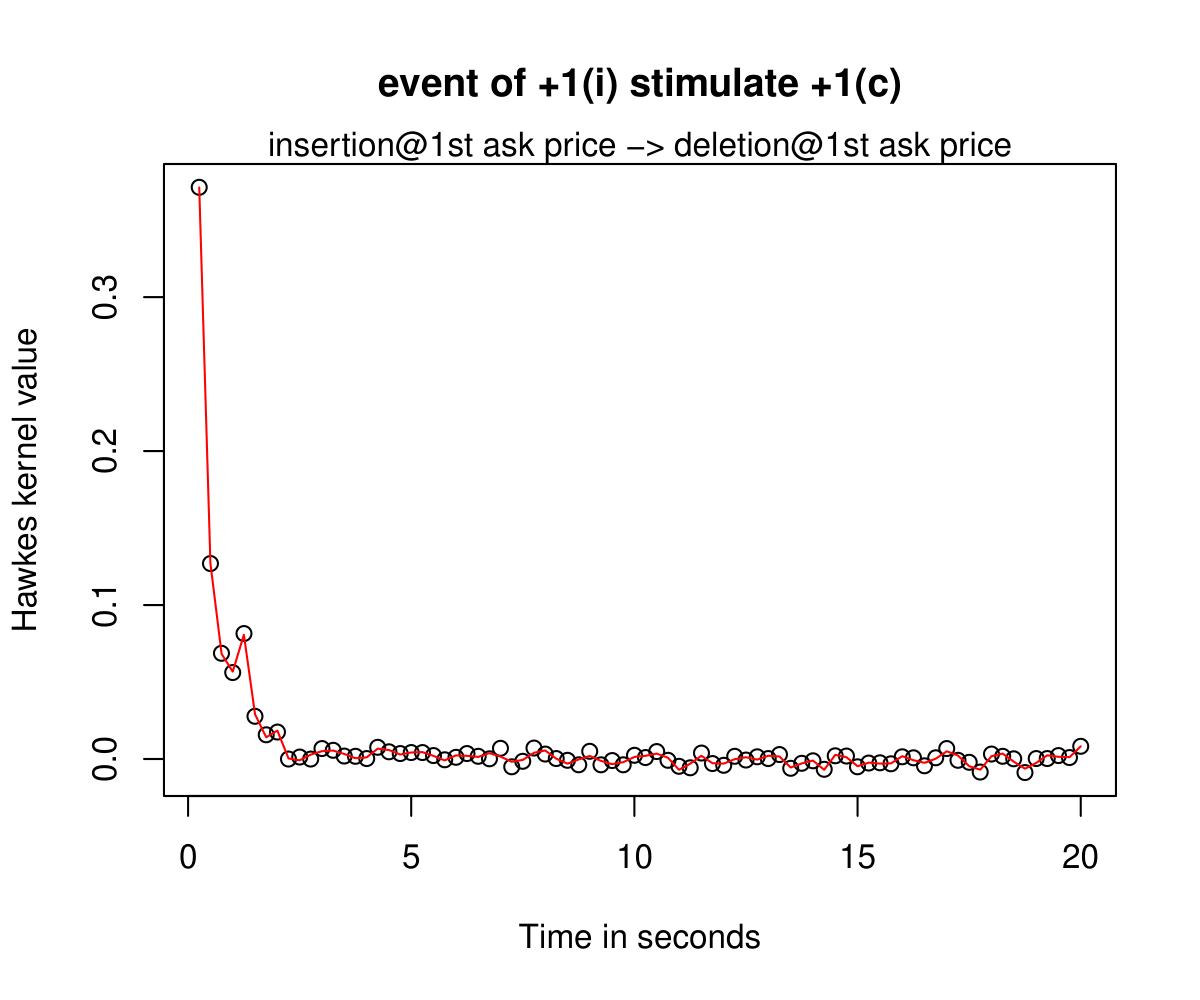} }}%
    \caption{Aggregated Hawkes excitement function estimation under $(s=20 \text{ seconds}, \Delta=0.25 \text{ seconds})$ with LASSO regularization. All orders are considered to have size 1. The points illustrate the discrete function estimators. The red line illustrates the cubic smoothing spline for the points.}
    \label{fig: first-level similar nosize}
\end{figure}

As we can observe, the above estimated functions are consistent with the 1st-ask and 1st-bid similarity patterns discussed in Section \ref{sec:estimated excitement functions: ask} and Section \ref{sec_support: estimated excitement functions bid}.

\subsection{Exponential and non-exponential shape of excitement functions}

Based on Figure \ref{fig: non-exponential kernels} in Section \ref{sec: exponential and Non-exponential shape of excitement function}, the following Figure \ref{fig: non-exponential kernels nosize} demonstrates the estimated Hawkes excitement functions with non-exponential shapes when the order size is ignored.

\begin{figure}[H]
    \centering
    \subfloat[\texttt{-1(t)} stimulate \texttt{+3(t)}]{{\includegraphics[scale = 0.16]{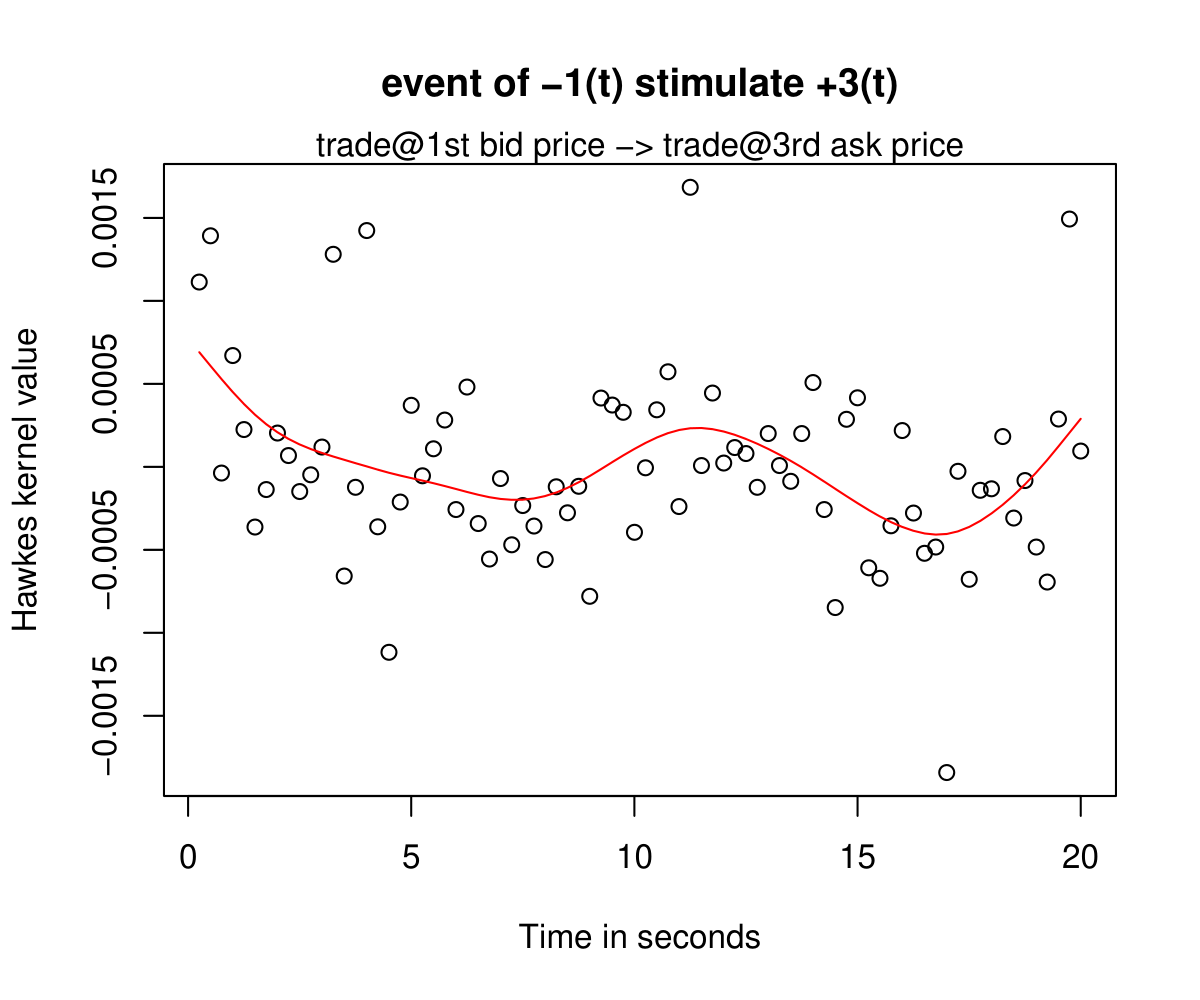} }}%
    \qquad
    \subfloat[\texttt{p+(t)} stimulate \texttt{-3(t)}]{{\includegraphics[scale =0.16]{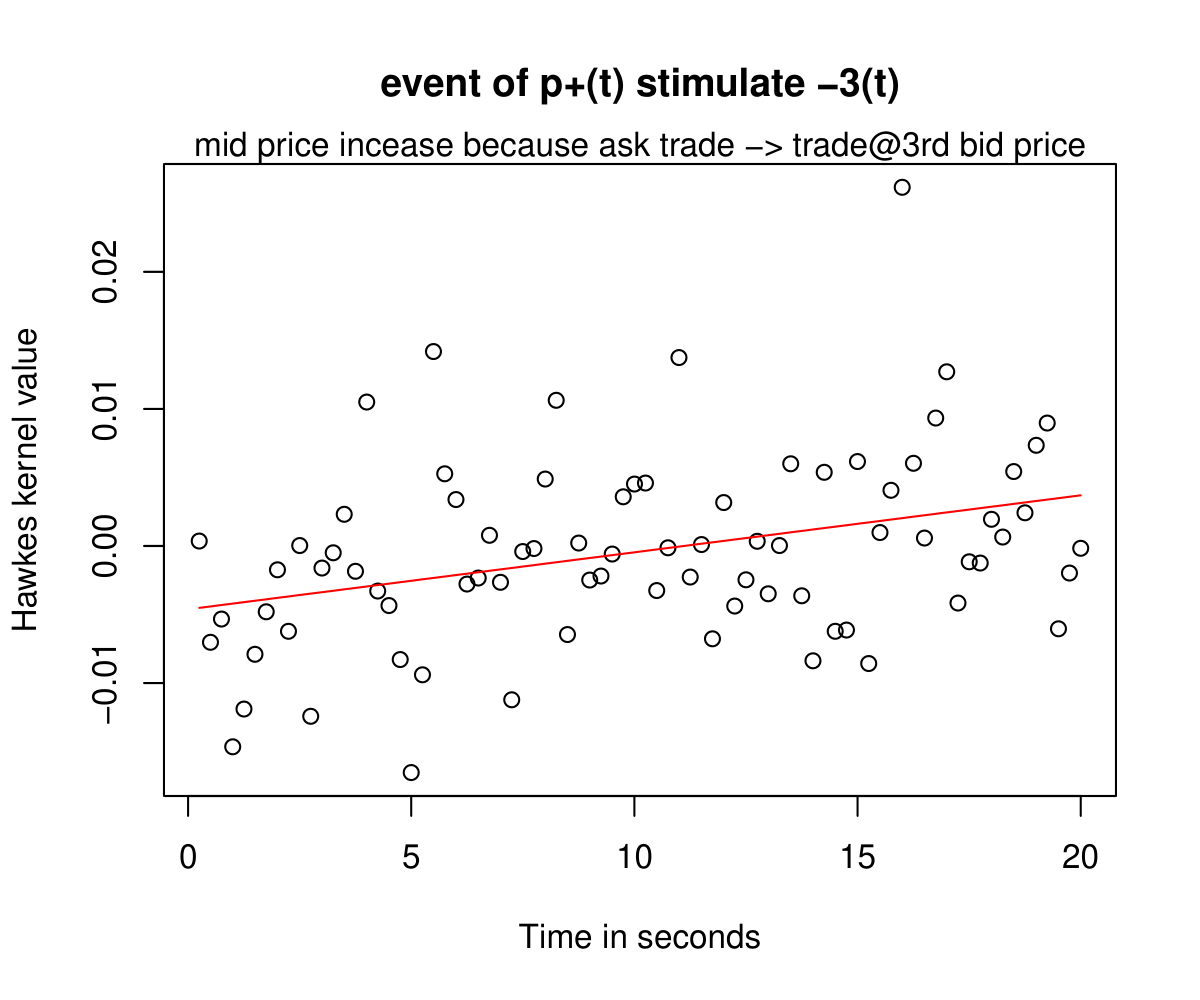} }}%
    \qquad
    \subfloat[\texttt{-1(c)} stimulate \texttt{+2(t)}]{{\includegraphics[scale =0.16]{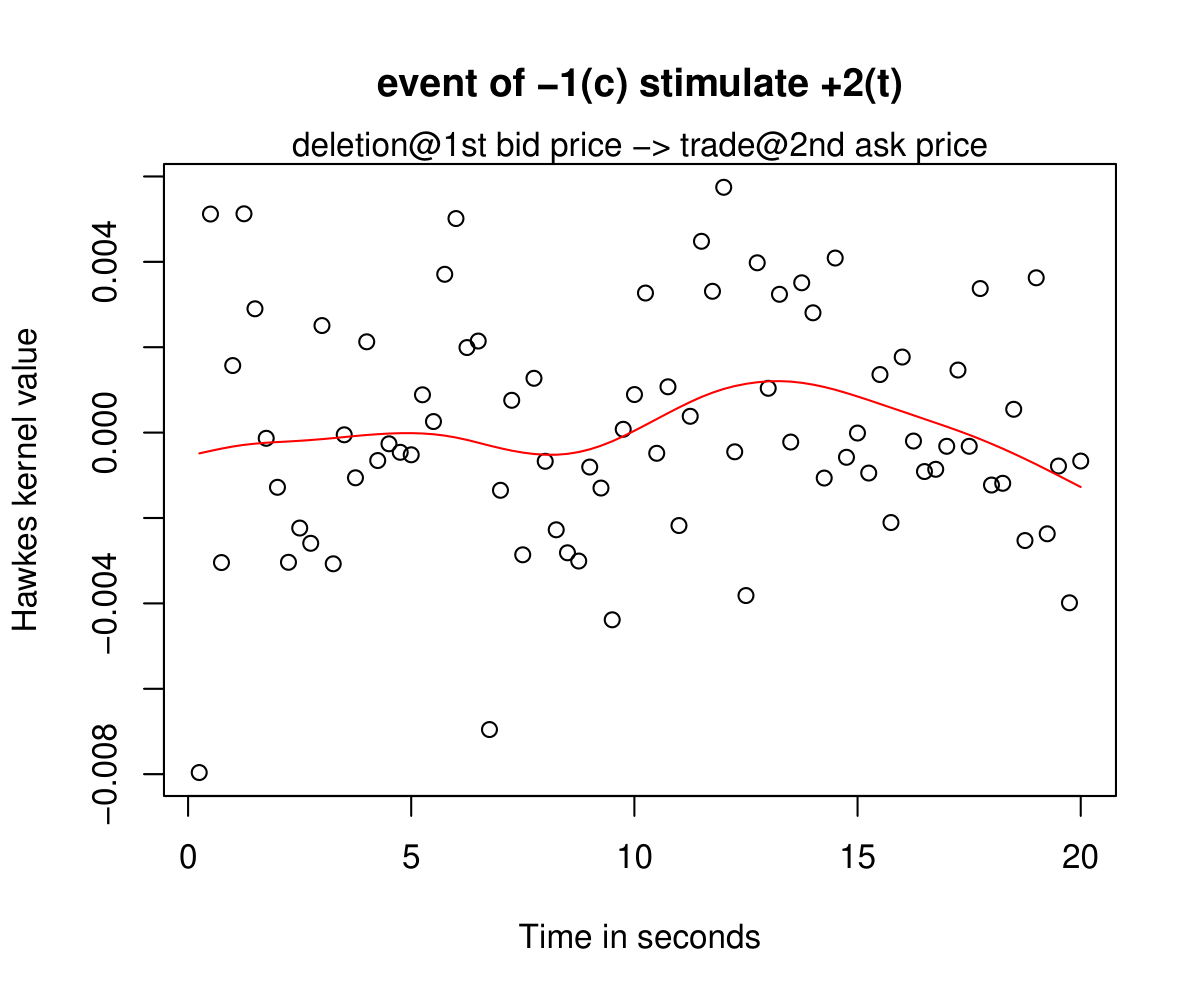} }}%
    \qquad
    \subfloat[\texttt{+2(t)} stimulate \texttt{-2(t)}]{{\includegraphics[scale =0.16]{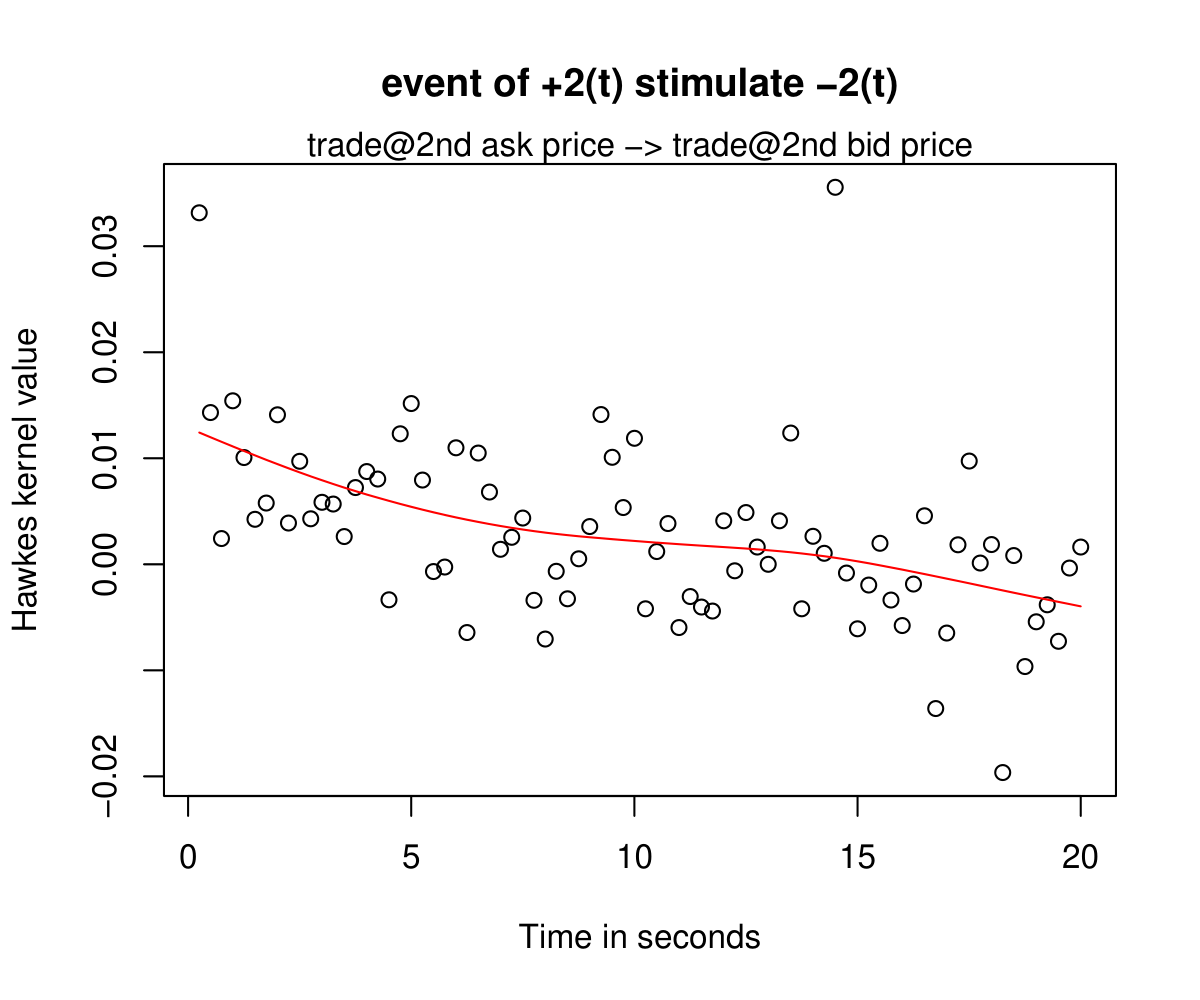} }}%
    
    \caption{Aggregated Hawkes excitement function estimation under $(s=20 \text{ seconds}, \Delta=0.25 \text{ seconds})$ with LASSO regularization. All orders are considered to have size 1. The points illustrate the discrete function estimators. The red line illustrates the cubic smoothing sline for the points.}
    \label{fig: non-exponential kernels nosize}
\end{figure}
The demonstrated results for the model with LASSO regularization are consistent with the results discussed in Section \ref{sec: exponential and Non-exponential shape of excitement function}: while most of the estimated Hawkes functions exhibit exponential time-decaying shapes, some estimated results do exhibit non-exponential shapes.

\subsection{Liquidity state}

\begin{figure}[H]
    \centering
    \subfloat[Liquidity state for \texttt{+3(i)}]{{\includegraphics[scale = 0.43]{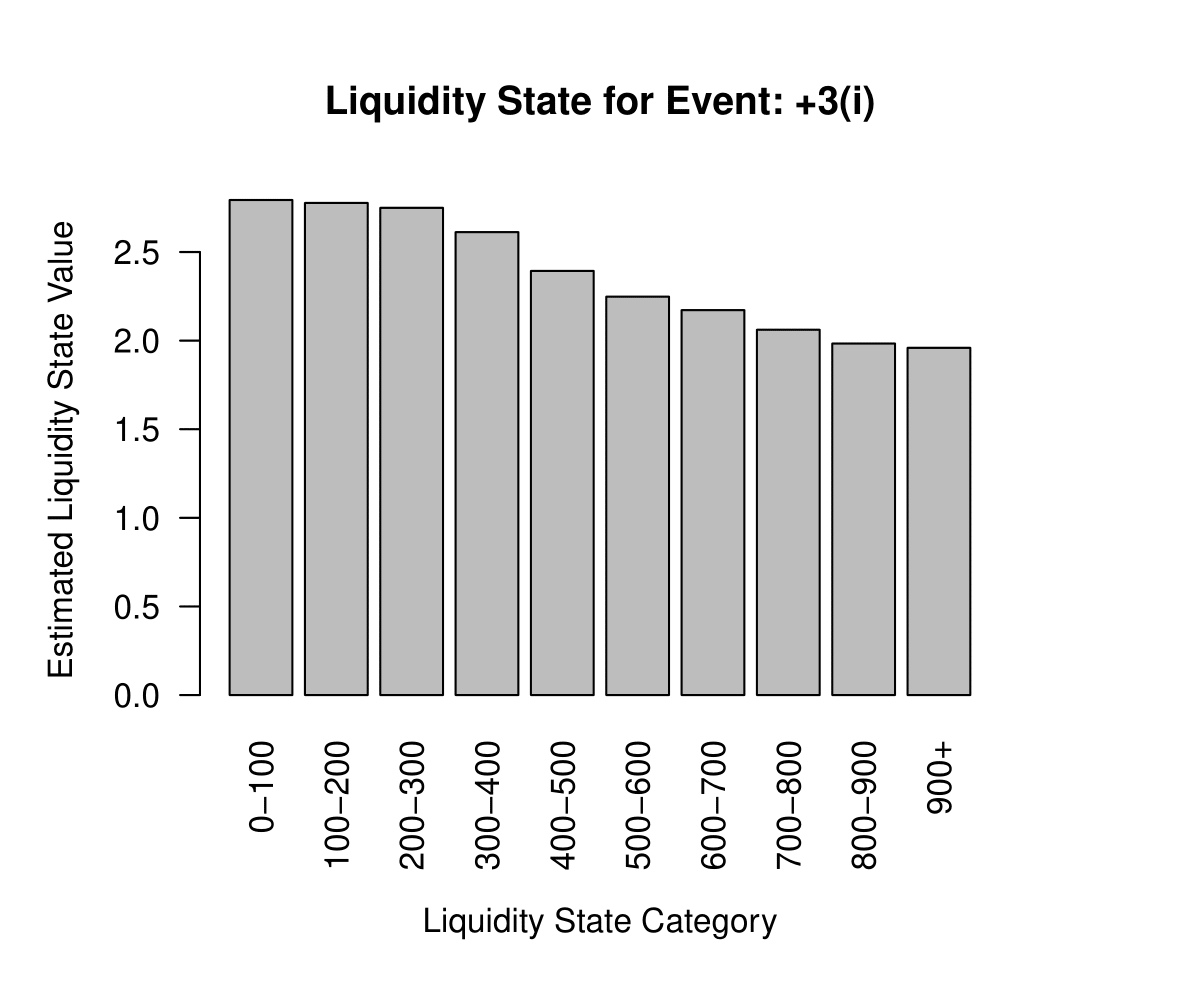} }}%
    \qquad
    \subfloat[Liquidity state for \texttt{-3(i)} ]{{\includegraphics[scale = 0.43]{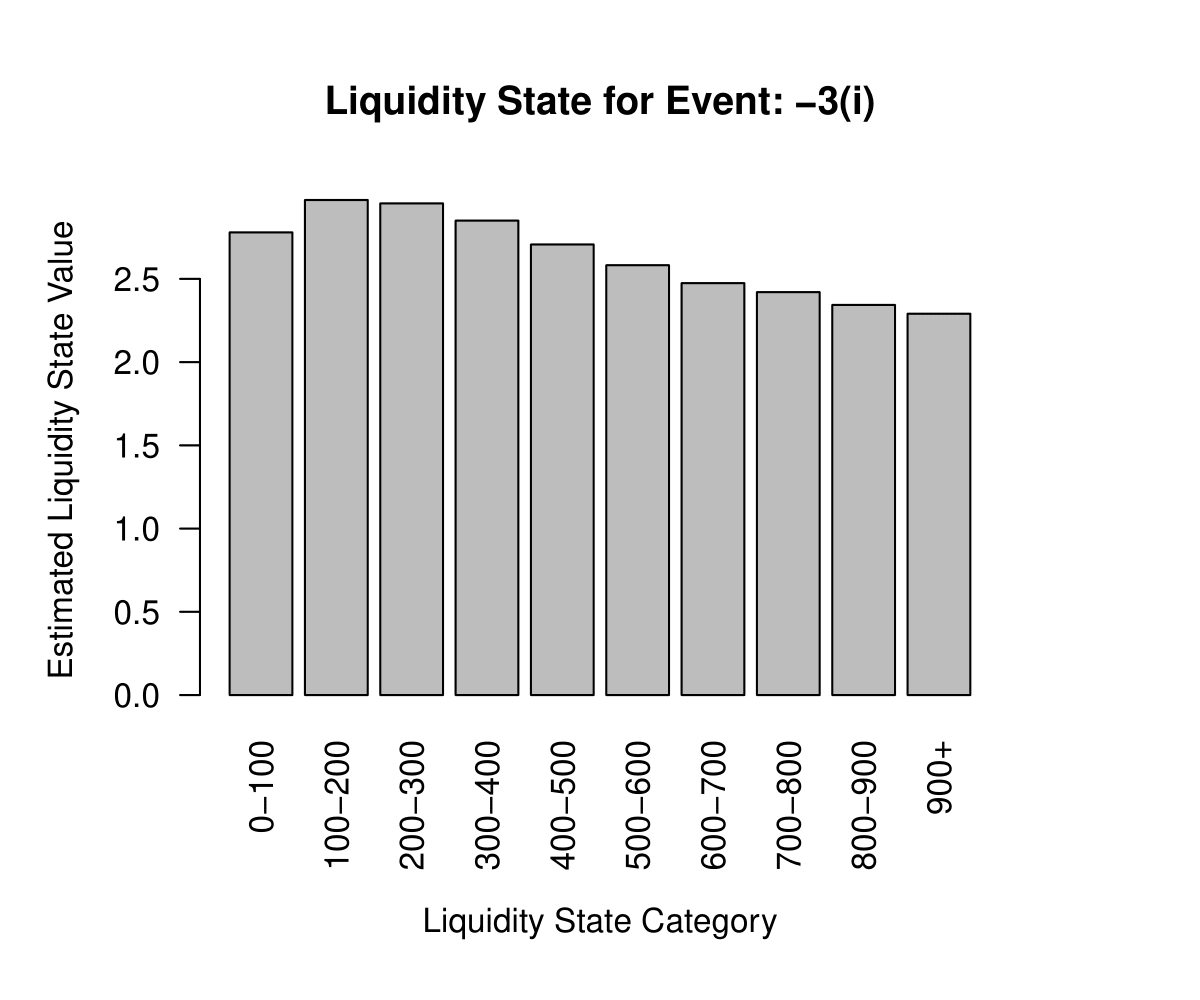} }}%
    \qquad
    \subfloat[Liquidity state for \texttt{+2(i)} ]{{\includegraphics[scale = 0.43]{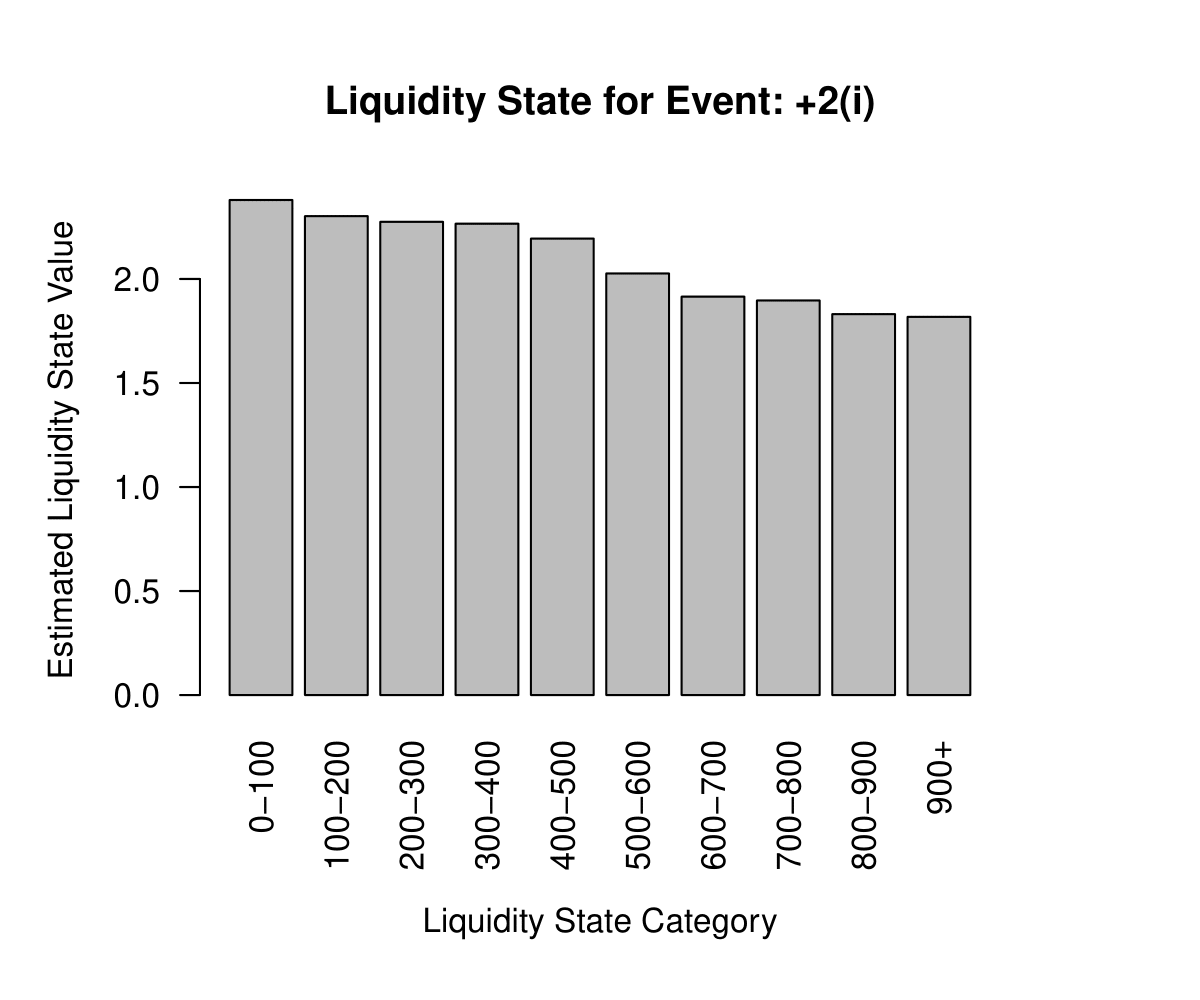} }}%
    \qquad
    \subfloat[Liquidity state for \texttt{-2(i)} ]{{\includegraphics[scale = 0.43]{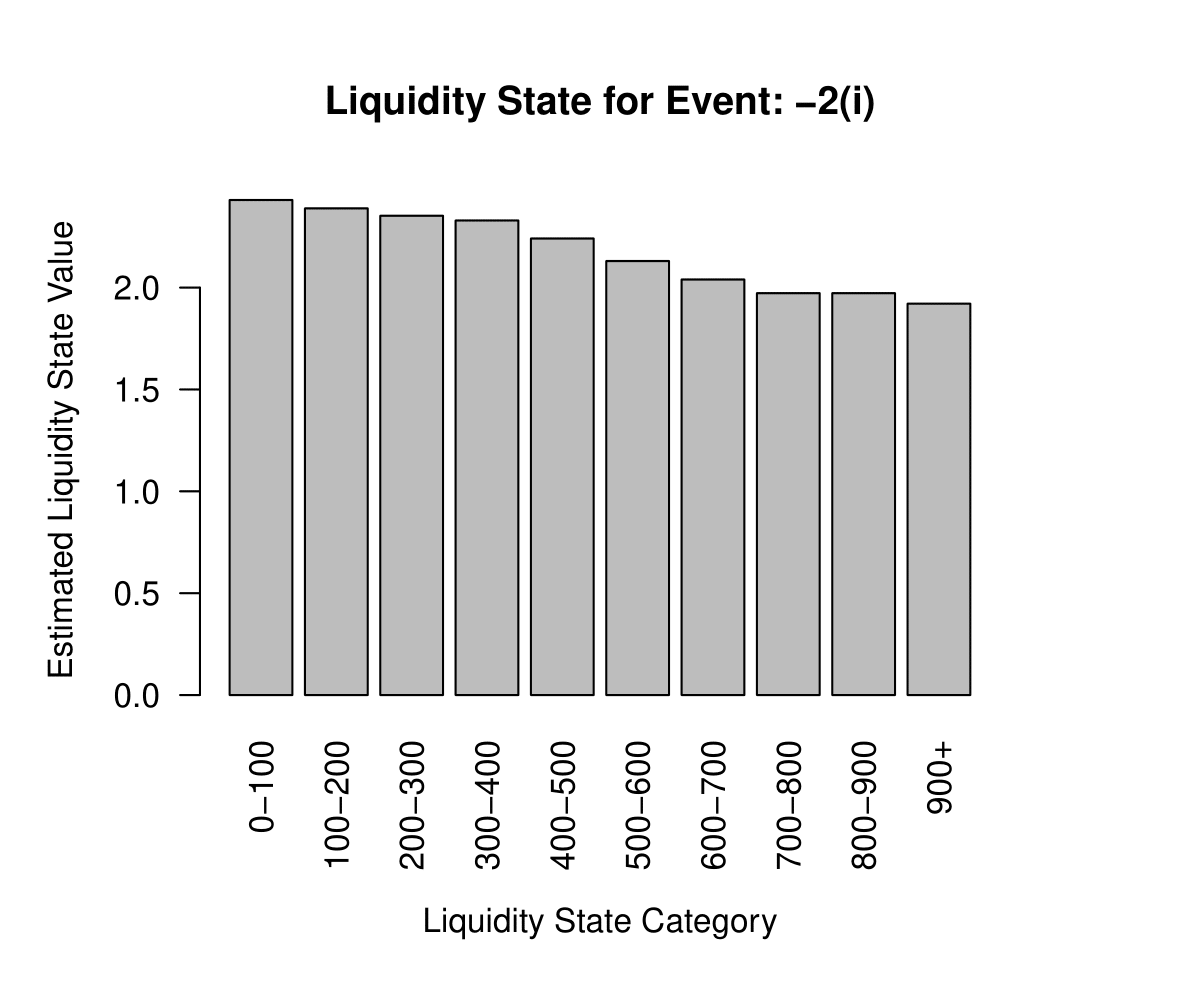} }}%
    \caption{Aggregated estimation result for liquidity state for event \texttt{+3(i)}, \-\texttt{-3(i)}, \-\texttt{+2(i)}, \-\texttt{-2(i)} under $(s=20 \text{ seconds}, \Delta=0.25 \text{ seconds})$. All orders are considered to have size 1. For these events the event arrival intensity decreases as liquidity state increases.}
    \label{fig: state_decrease nosize}
\end{figure}

Based on Figure \ref{fig: state_increase} and Figure \ref{fig: state_decrease} in Section \ref{sec: liquidity state}, the following Figure \ref{fig: state_increase nosize} and Figure \ref{fig: state_decrease nosize} demonstrate the liquidity state estimations of the model with LASSO regularization.

\begin{figure}[H]
    \centering
    \subfloat[Liquidity state for \texttt{+1(i)}]{{\includegraphics[scale = 0.41]{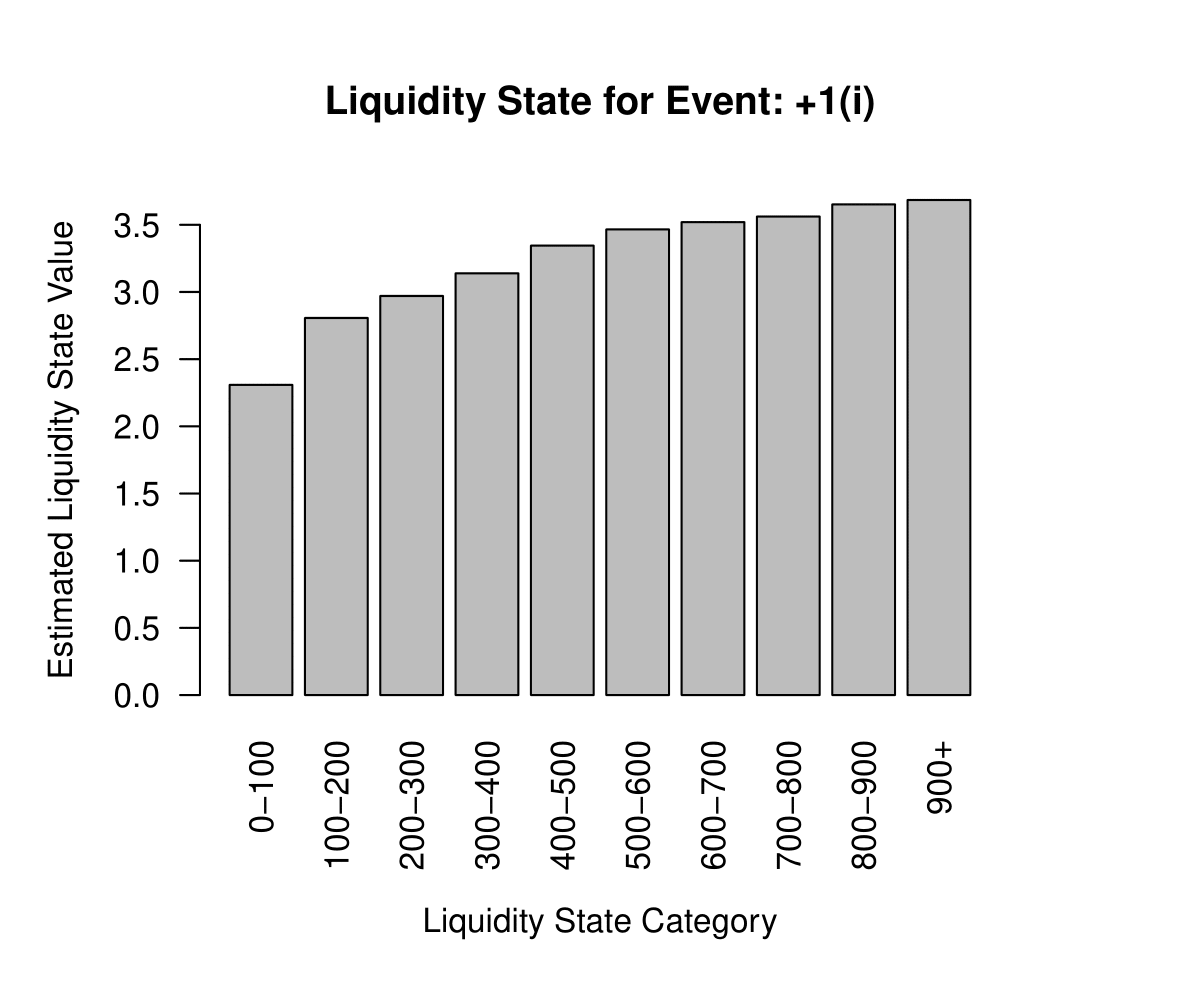} }}%
    \qquad
    \subfloat[Liquidity state for \texttt{+1(c)} ]{{\includegraphics[scale = 0.41]{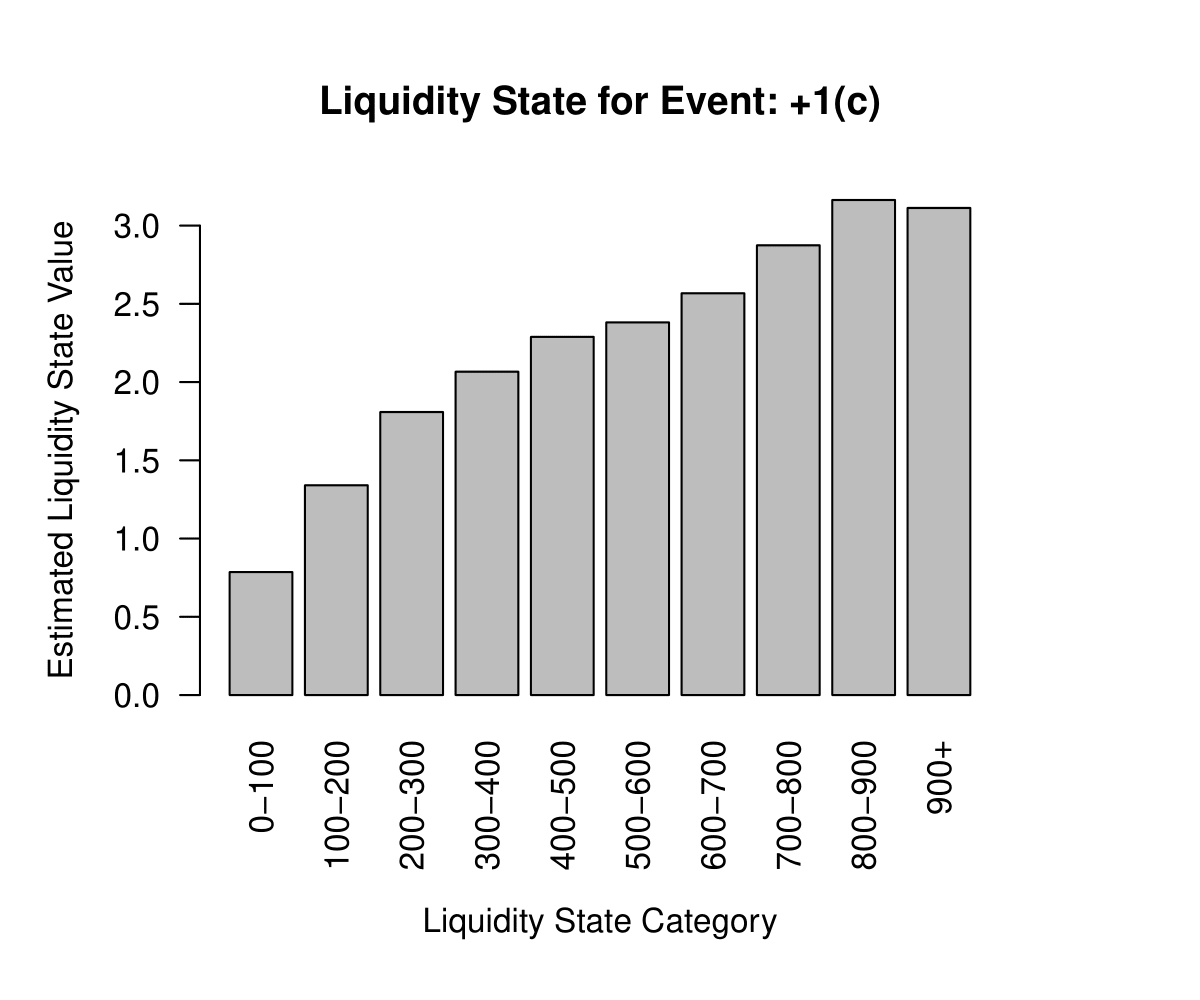} }}%
    \qquad
    \subfloat[Liquidity state for \texttt{+1(t)} ]{{\includegraphics[scale = 0.41]{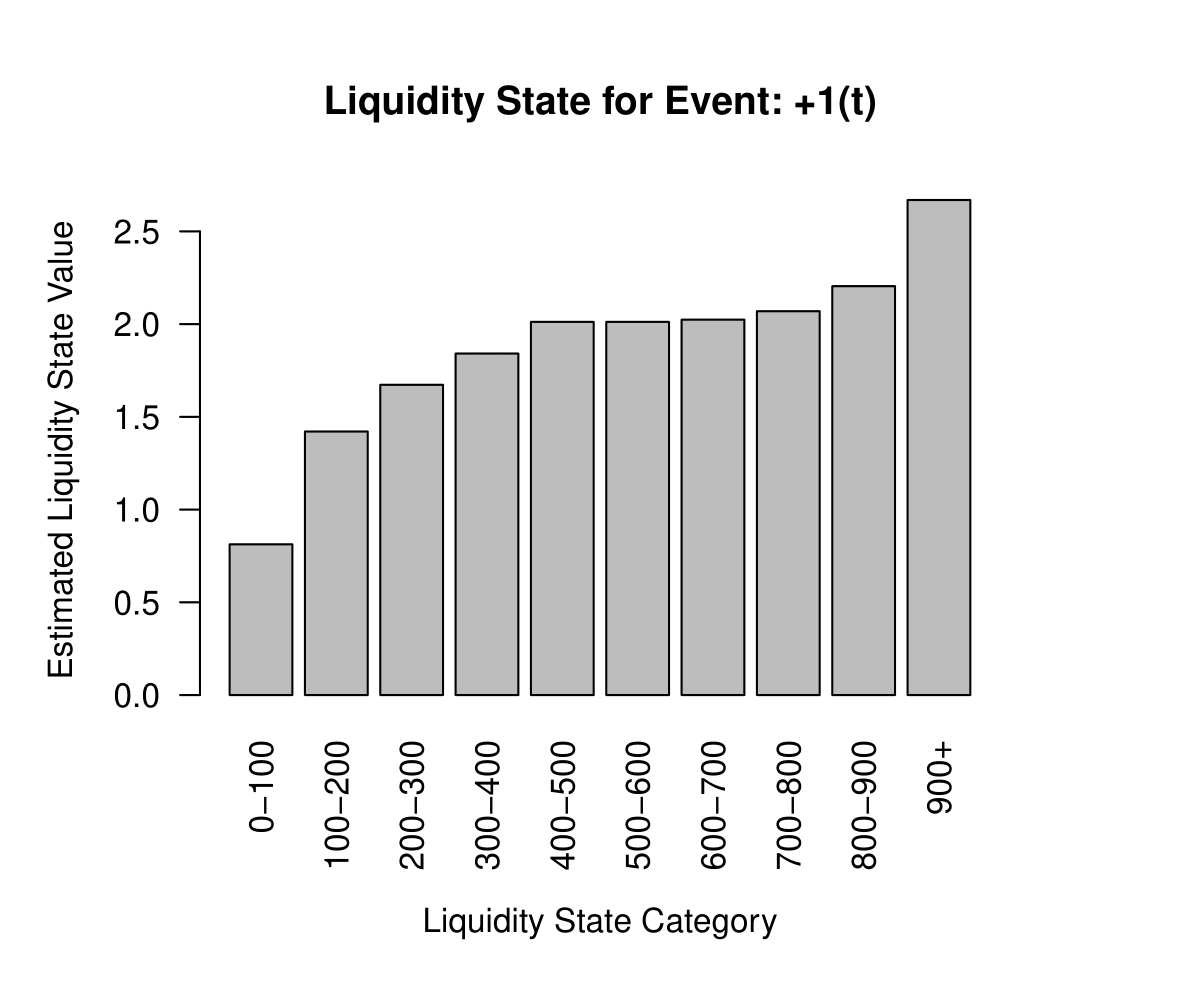} }}%
    \qquad
    \subfloat[Liquidity state for \texttt{-1(c)} ]{{\includegraphics[scale = 0.41]{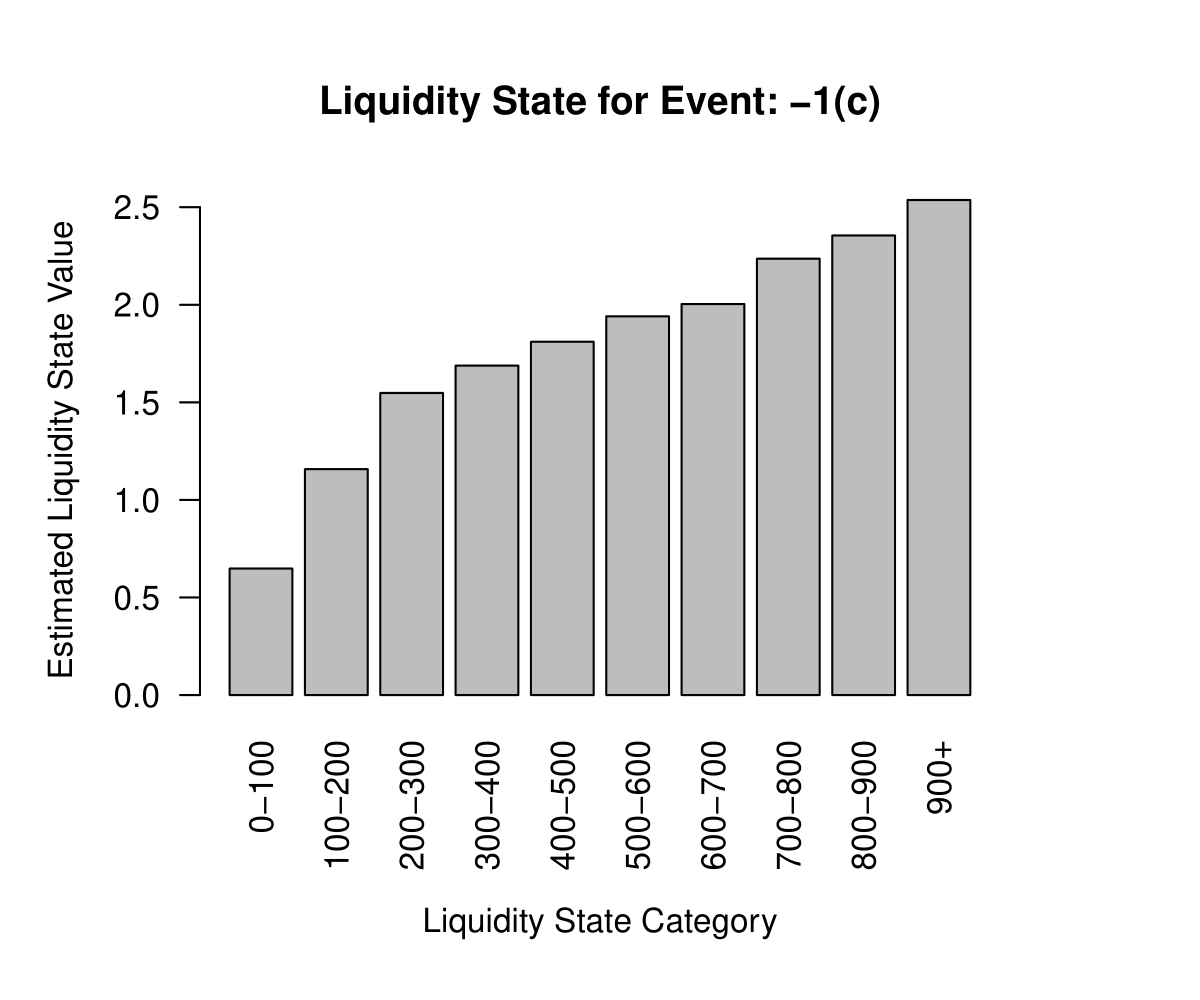} }}%
    \caption{Aggregated estimation result for liquidity state for event \texttt{+1(i)}, \-\texttt{+1(c)}, \-\texttt{+1(t)}, \-\texttt{-1(c)} under $(s=20 \text{ seconds}, \Delta=0.25 \text{ seconds})$. All orders are considered to have size 1. For these events the event arrival intensity increases as liquidity state increases.}
    \label{fig: state_increase nosize}
\end{figure}

The demonstrated liquidity state estimation results of the model with LASSO regularization is consistent with the results discussed in Section \ref{sec: liquidity state}.

\subsection{Time factor}

Based on Figure \ref{fig: time factor} in Section \ref{sec: time factor}, the following Figure \ref{fig: time factor nosize} demonstrates the time factor estimations of the model with LASSO regularization.

\begin{figure}[H]
    \centering
    \subfloat[Time factor for event \texttt{+1(i)}]{{\includegraphics[scale = 0.43]{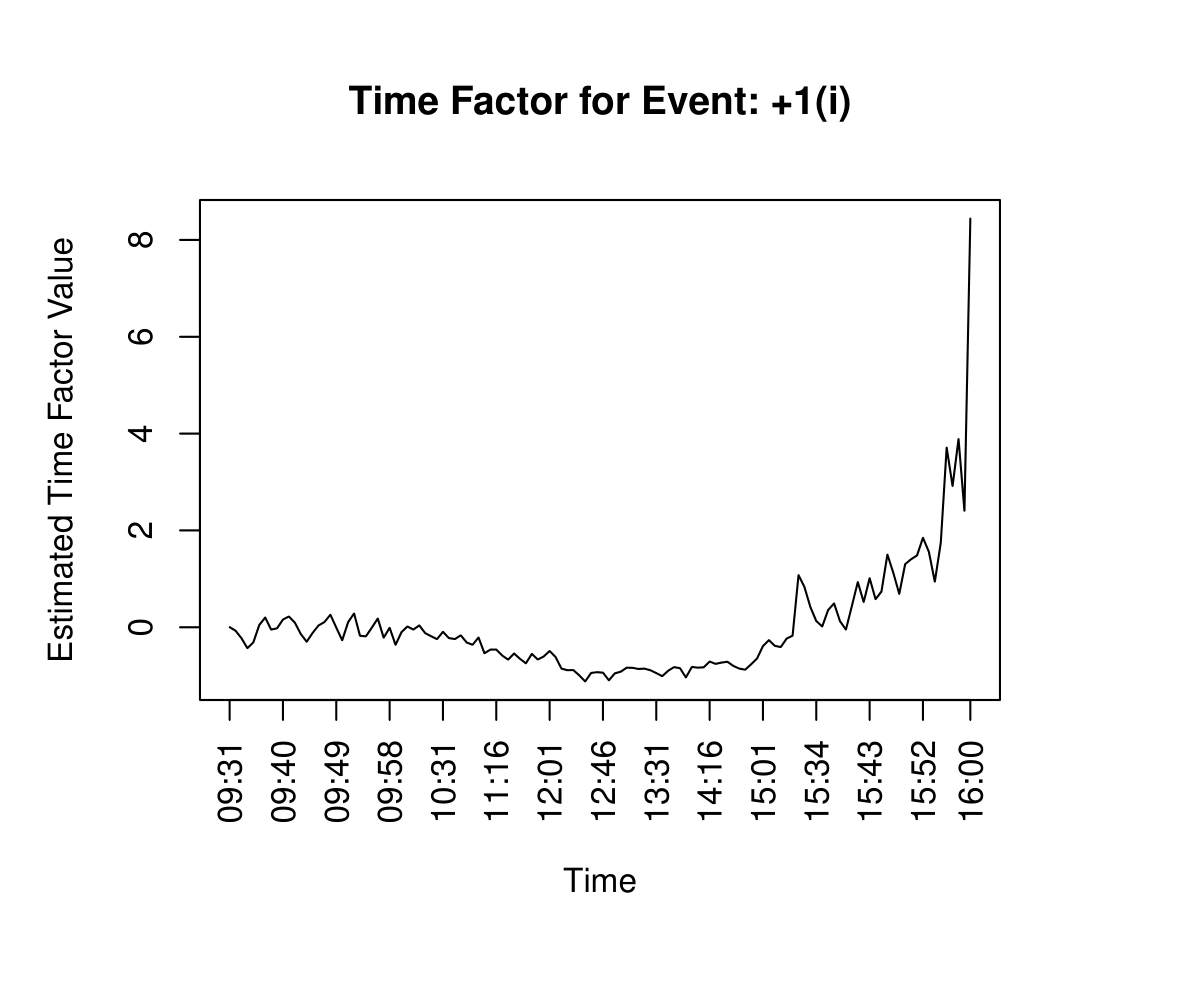} }}%
    \qquad
    \subfloat[Time factor for event \texttt{+3(t)}]{{\includegraphics[scale = 0.43]{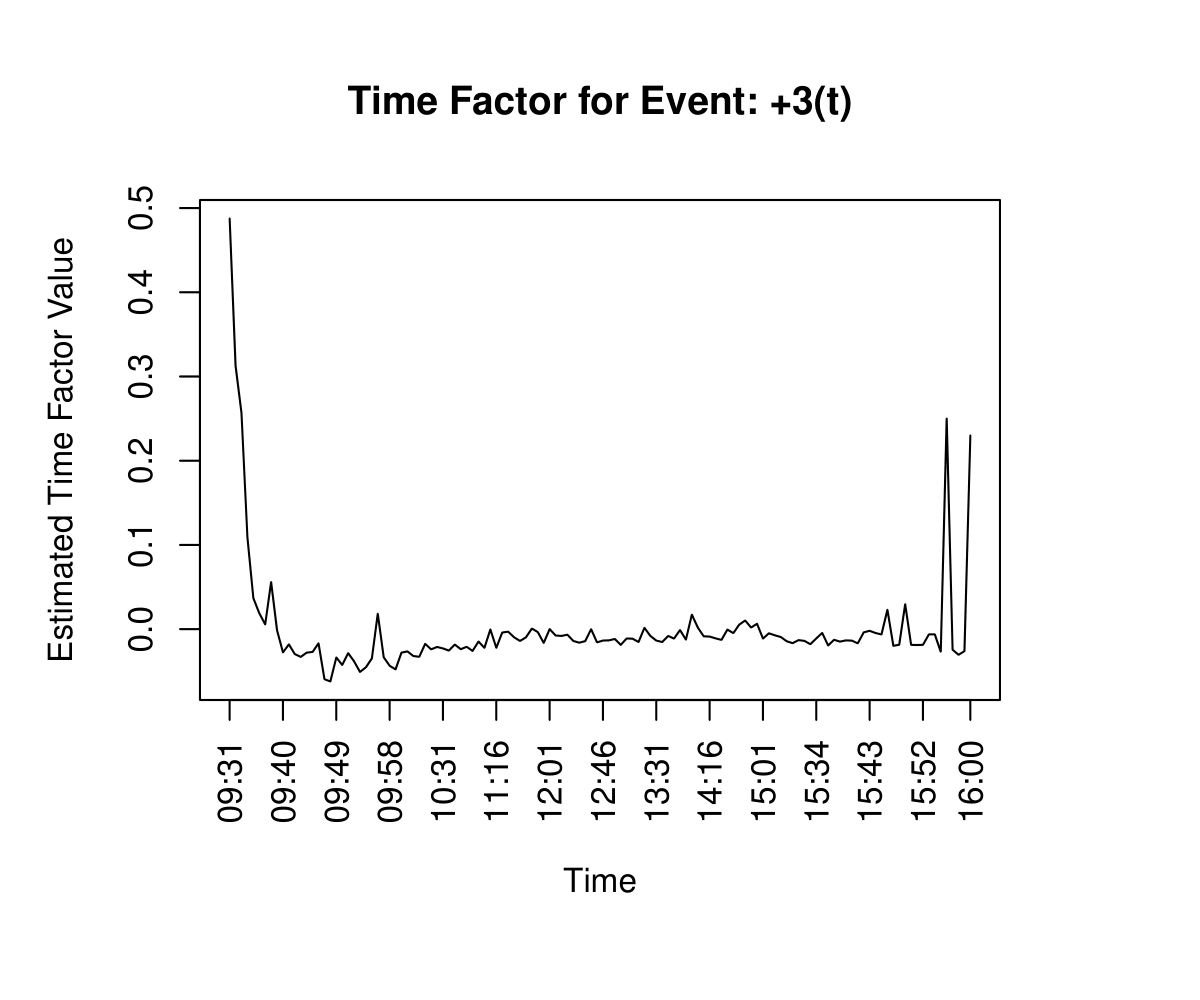} }}%
    \caption{Aggregated estimation result for time factor between 9:30 am and 4:00 pm under $(s=20 \text{ seconds}, \Delta=0.25 \text{ seconds})$. All orders are considered to have size 1.}
    \label{fig: time factor nosize}
\end{figure}

The demonstrated time factor estimation results of the model with LASSO regularization is consistent with the results discussed in Section \ref{sec: time factor}.

\section{Empirical results without LASSO}\label{sec_support: result noLASSO}
This supporting section demonstrates the empirical estimation results when the LASSO regularization is removed, as mentioned in Section \ref{sec: sensitivity analysis}. The demonstrations will be presented in the same format as the demonstrations from Section \ref{sec:estimated excitement functions: ask} to Section \ref{sec: time factor}.  As a whole, the results on excitement function, liquidity state, and time factor still hold qualitatively. The estimation result with or without LASSO (small regularization $\lambda_{i} = 0.0005$) are very similar visually. Furthermore, the cubic smoothing spline for the LASSO model is smoother than the model without LASSO since the estimator distribution is more concentrated to zero after adding LASSO.

\subsection{Estimated excitement functions}
\begin{figure}[H]
    \centering
    \subfloat[\texttt{-1(i)} stimulate \texttt{-1(i)}]{{\includegraphics[scale = 0.16]{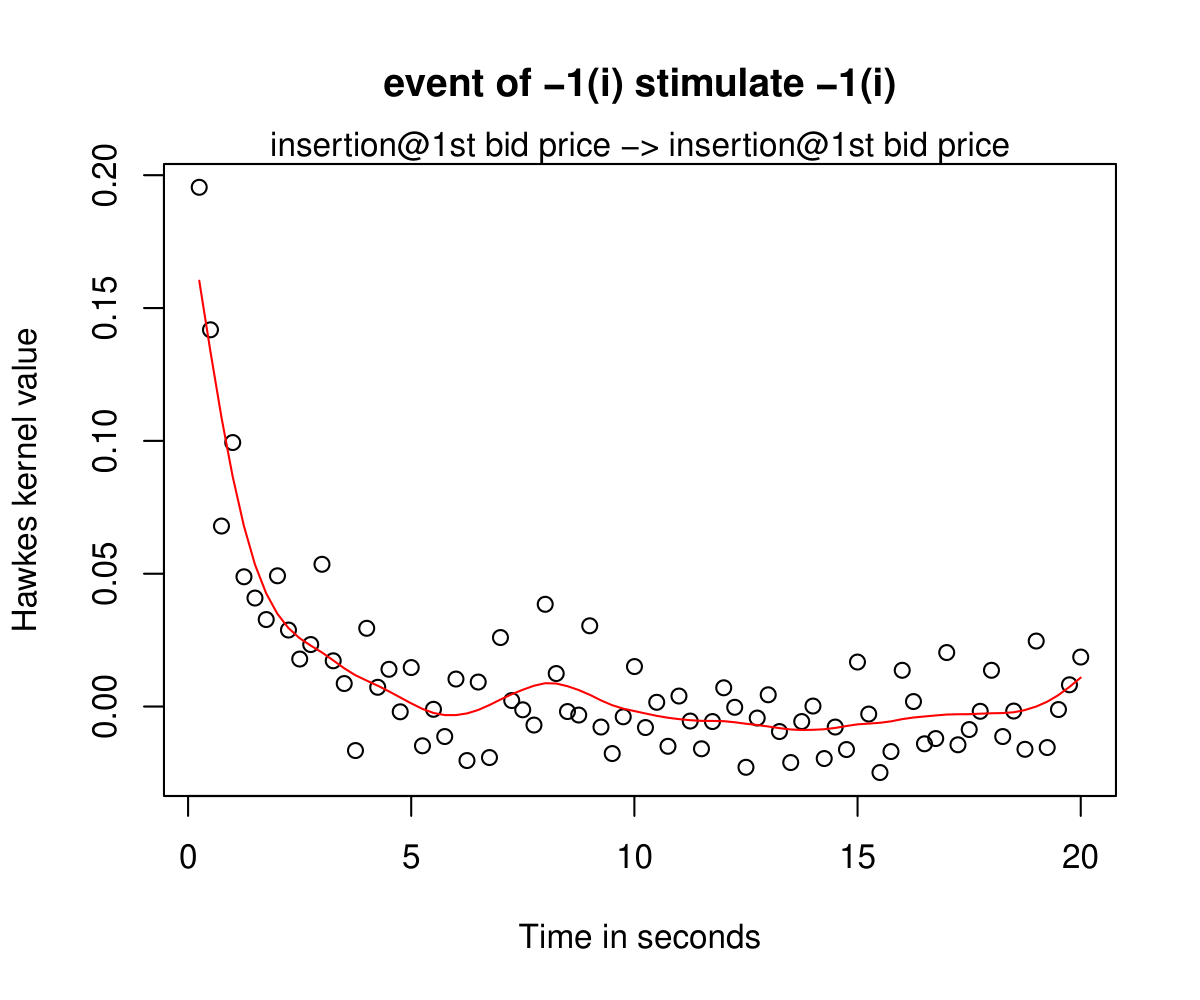} }}%
    \qquad
    \subfloat[\texttt{-1(i)} stimulate \texttt{-1(c)}]{{\includegraphics[scale =0.16]{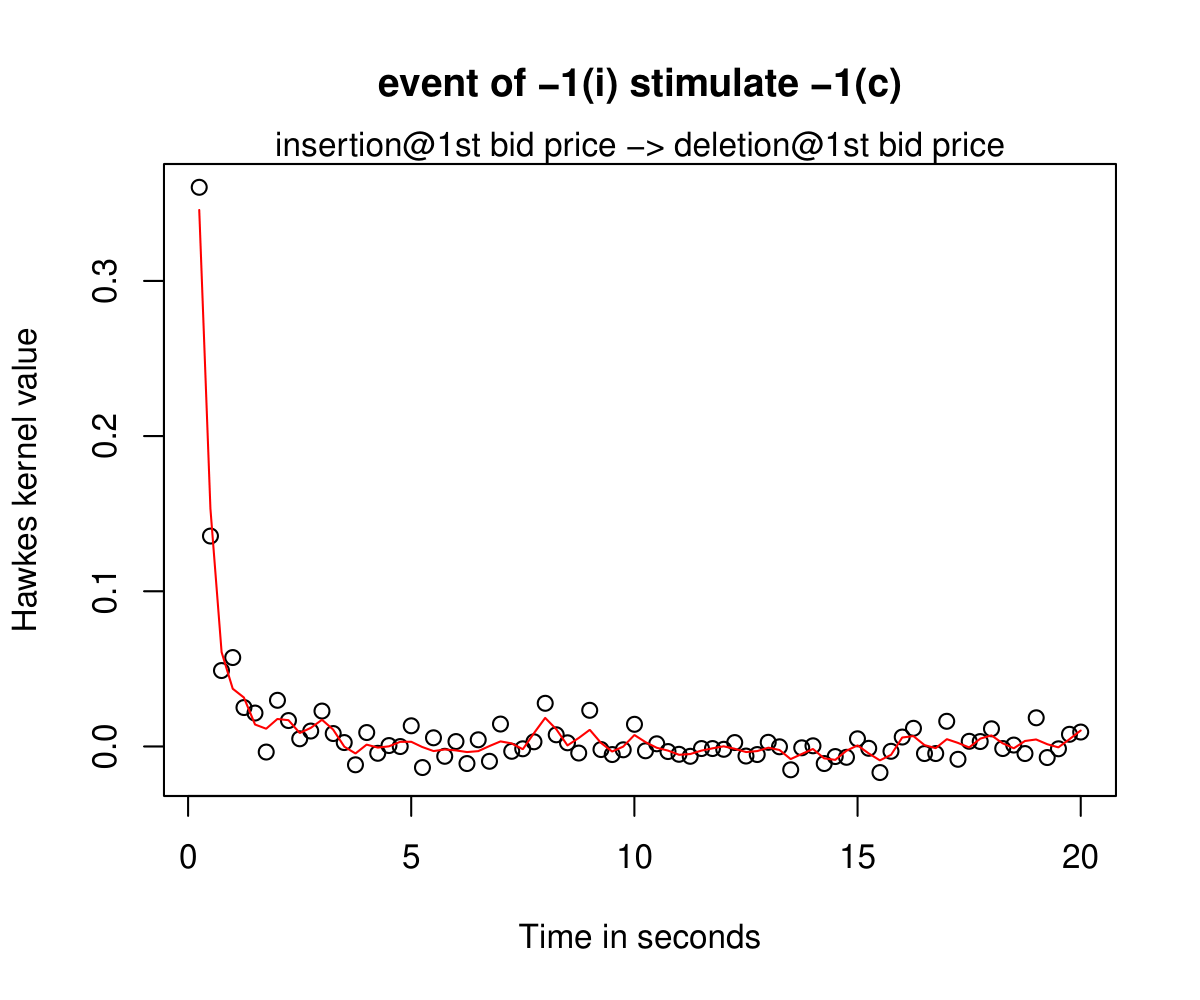} }}%
    \caption{Aggregated Hawkes excitement function estimation under $(s=\\20 \text{ seconds}, \Delta=0.25 \text{ seconds})$ when LASSO regularization is removed. The points illustrate the discrete function estimators. The red line illustrates the cubic smoothing spline for the points.}
    \label{fig: first-level similar bid noLASSO}
\end{figure}

Based on Figure \ref{fig: first-level similar} and Figure \ref{fig: first-level similar bid}, the following Figure \ref{fig: first-level similar noLASSO} and Figure \ref{fig: first-level similar bid noLASSO} demonstrate the estimated Hawkes excitement functions when the LASSO regularization is removed.

\begin{figure}[H]
    \centering
    \subfloat[\texttt{+1(i)} stimulate \texttt{+1(i)}]{{\includegraphics[scale = 0.16]{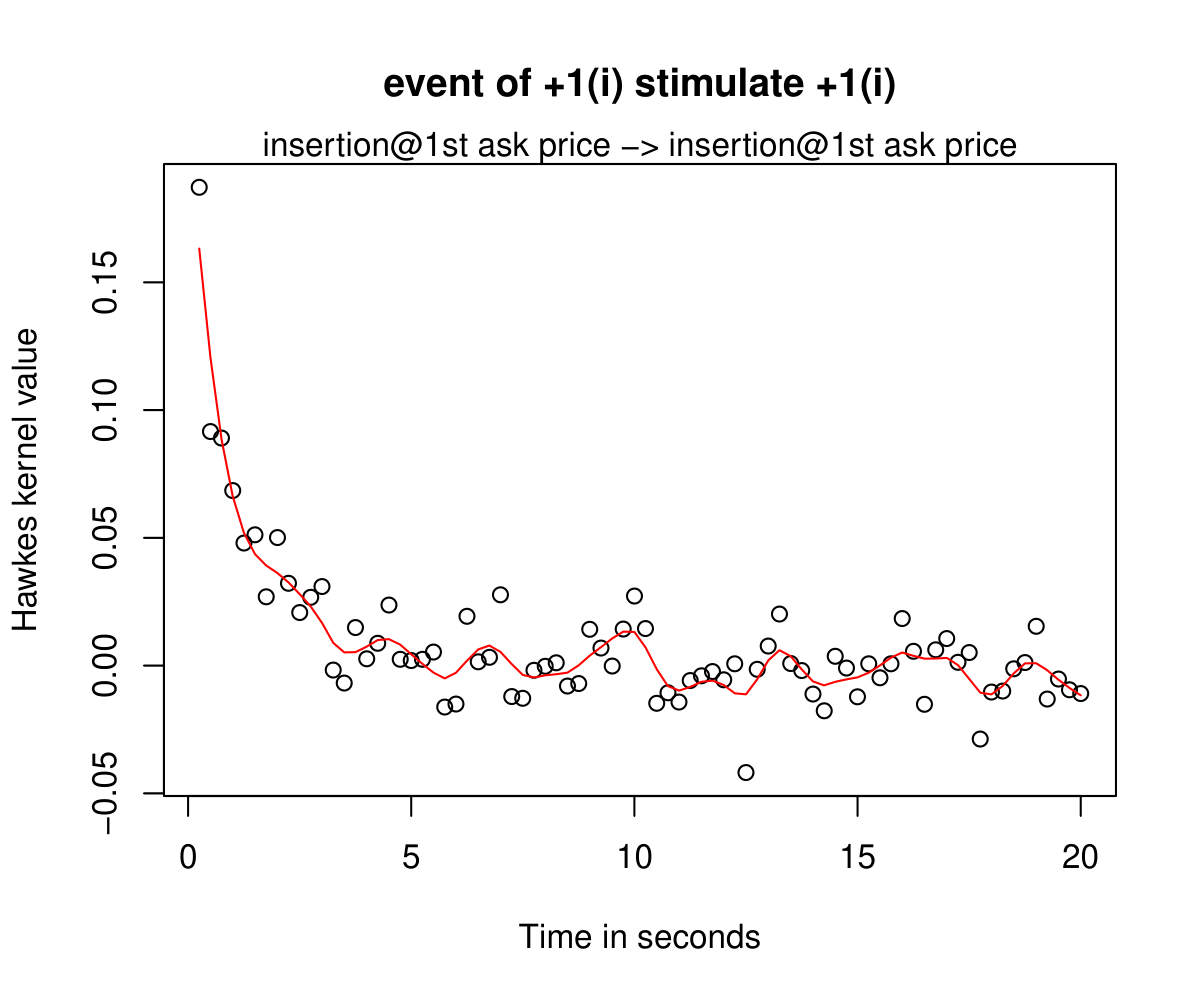} }}%
    \qquad
    \subfloat[\texttt{+1(i)} stimulate \texttt{+1(c)}]{{\includegraphics[scale =0.16]{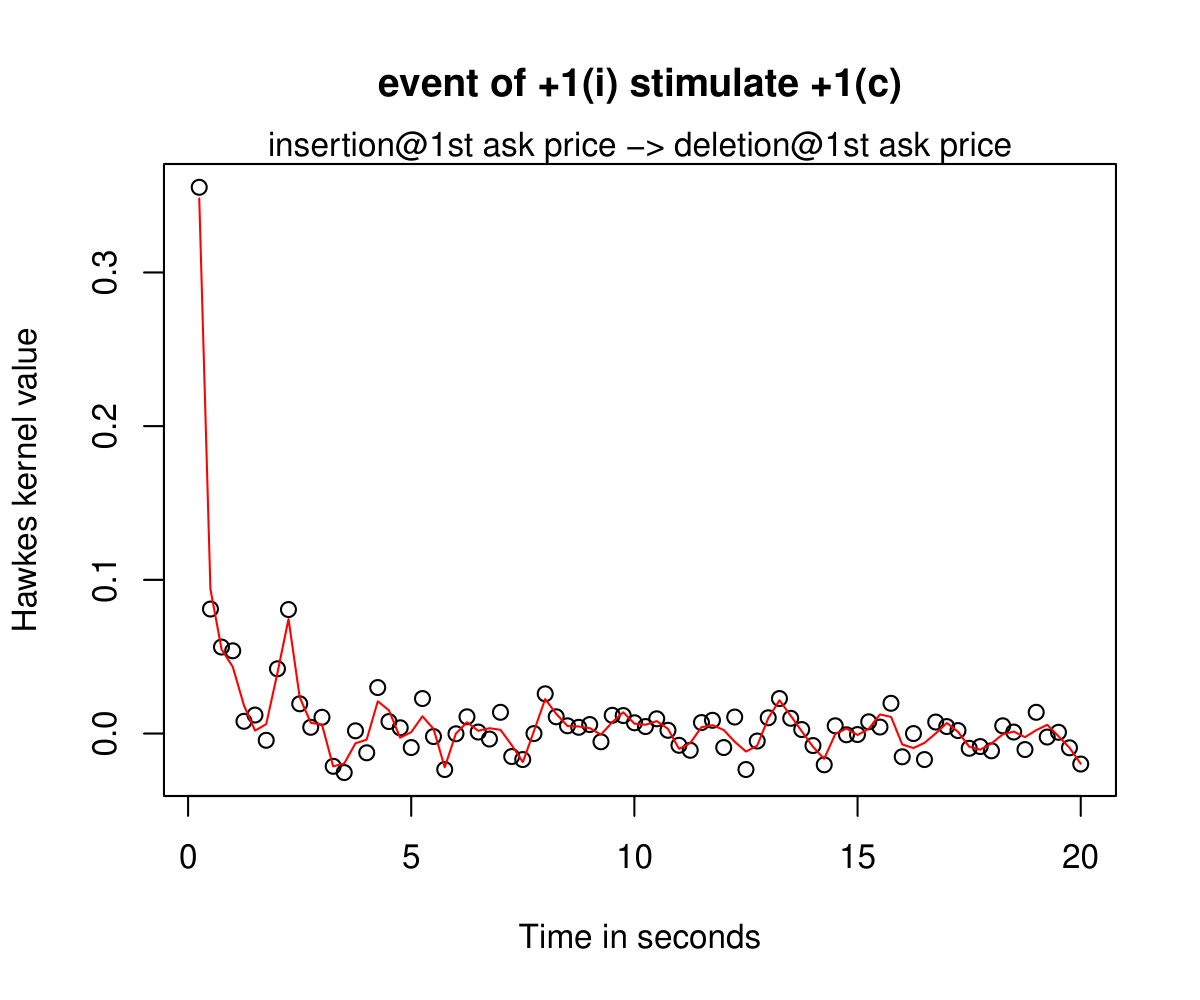} }}%
    \caption{Aggregated Hawkes excitement function estimation under $(s=\\20 \text{ seconds}, \Delta=0.25 \text{ seconds})$ when LASSO regularization is removed.  the discrete function estimators. The red line illustrates the cubic smoothing spline for the points.}
    \label{fig: first-level similar noLASSO}
\end{figure}

As we can observe, the above estimated functions are consistent with the 1st-ask and 1st-bid similarity patterns discussed in Section \ref{sec:estimated excitement functions: ask} and Section \ref{sec_support: estimated excitement functions bid}, when the LASSO regularization is removed.

\subsection{Exponential and non-exponential shape of excitement functions}

Based on Figure \ref{fig: non-exponential kernels} in Section \ref{sec: exponential and Non-exponential shape of excitement function}, the following Figure \ref{fig: non-exponential kernels noLASSO} demonstrates the estimated Hawkes excitement functions with non-exponential shapes when the LASSO regularization is removed.

\begin{figure}[H]
    \centering
    \subfloat[\texttt{-1(t)} stimulate \texttt{+3(t)}]{{\includegraphics[scale = 0.16]{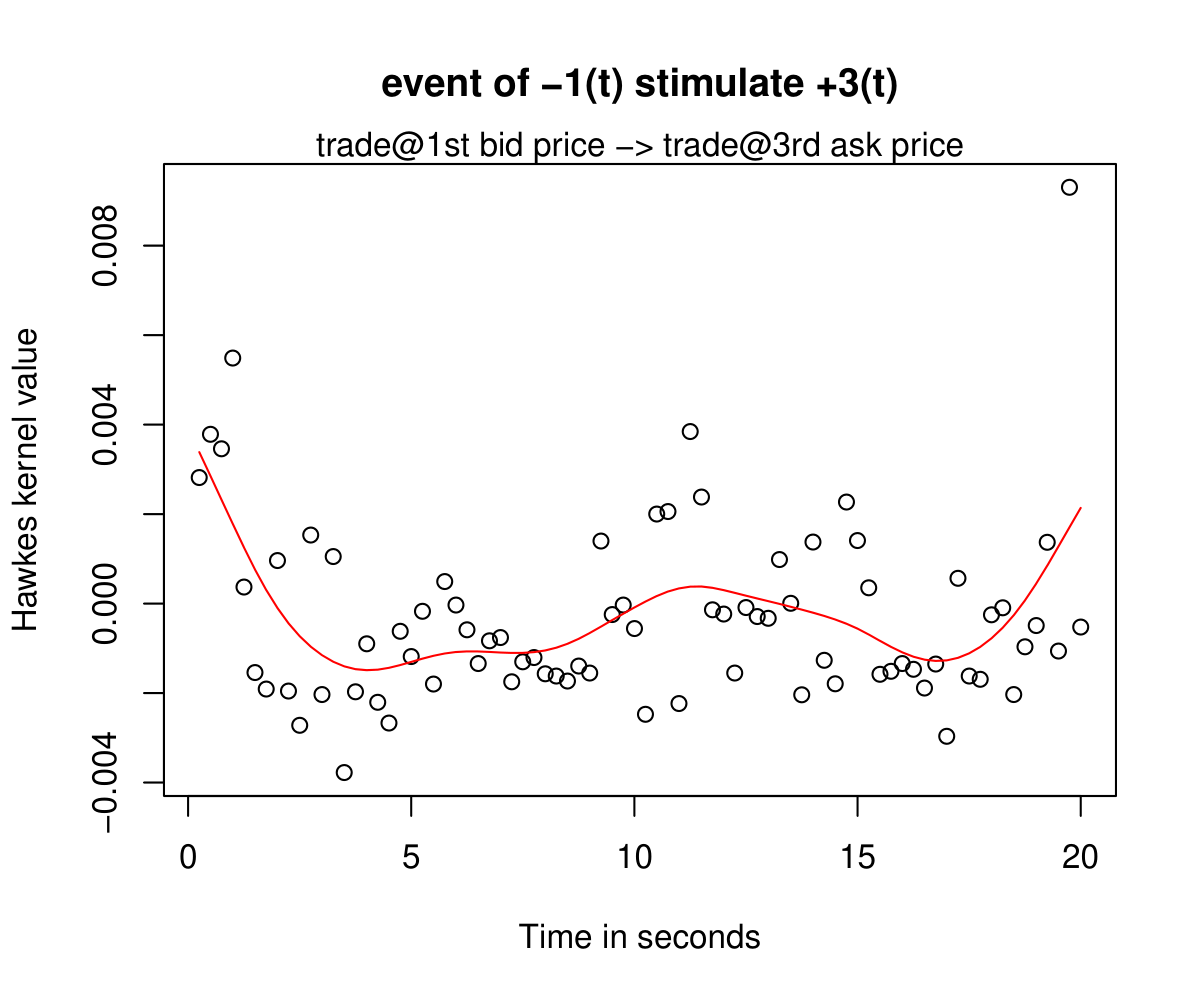} }}%
    \qquad
    \subfloat[\texttt{p+(t)} stimulate \texttt{-3(t)}]{{\includegraphics[scale =0.16]{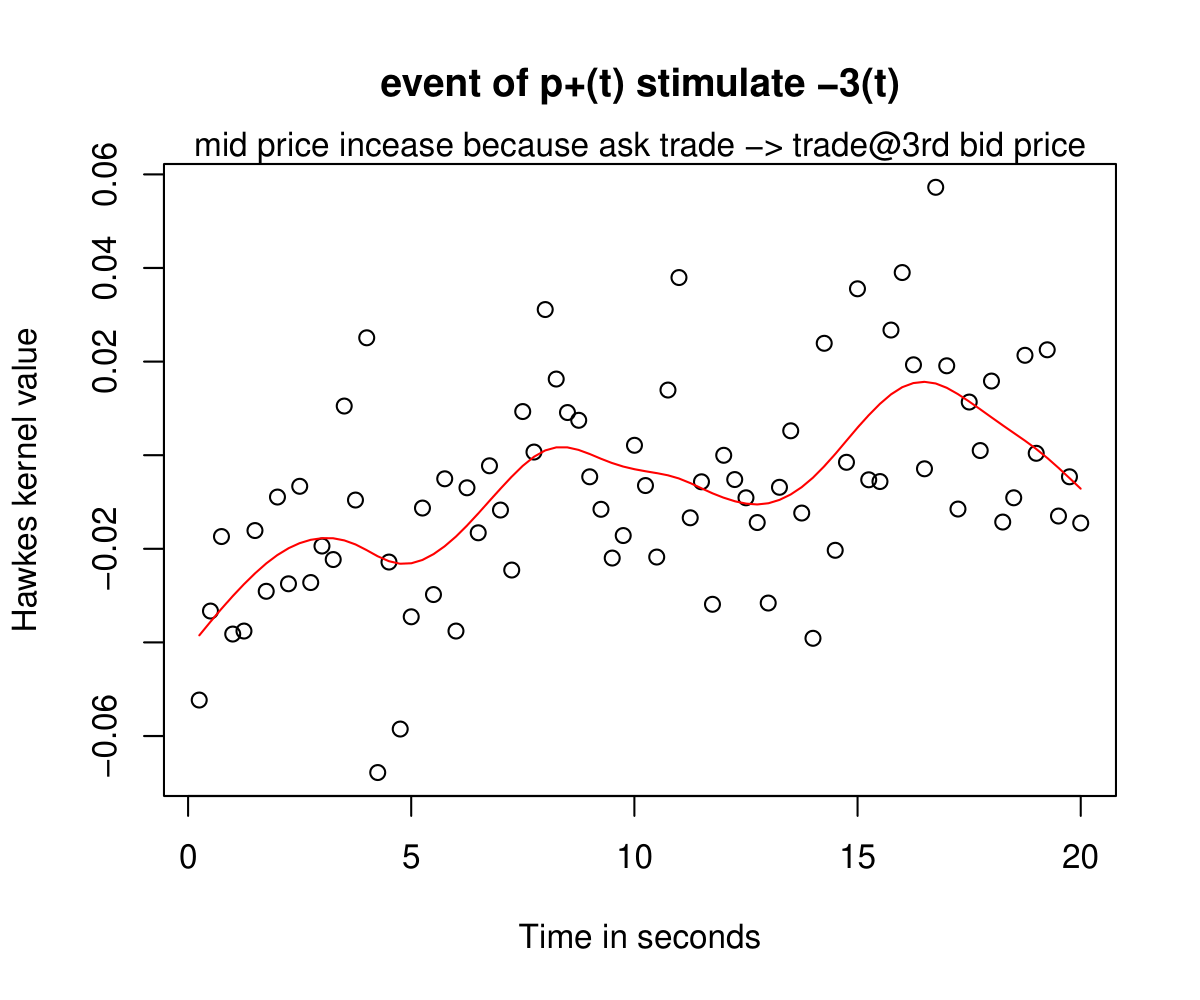} }}%
    \qquad
    \subfloat[\texttt{-1(c)} stimulate \texttt{+2(t)}]{{\includegraphics[scale =0.16]{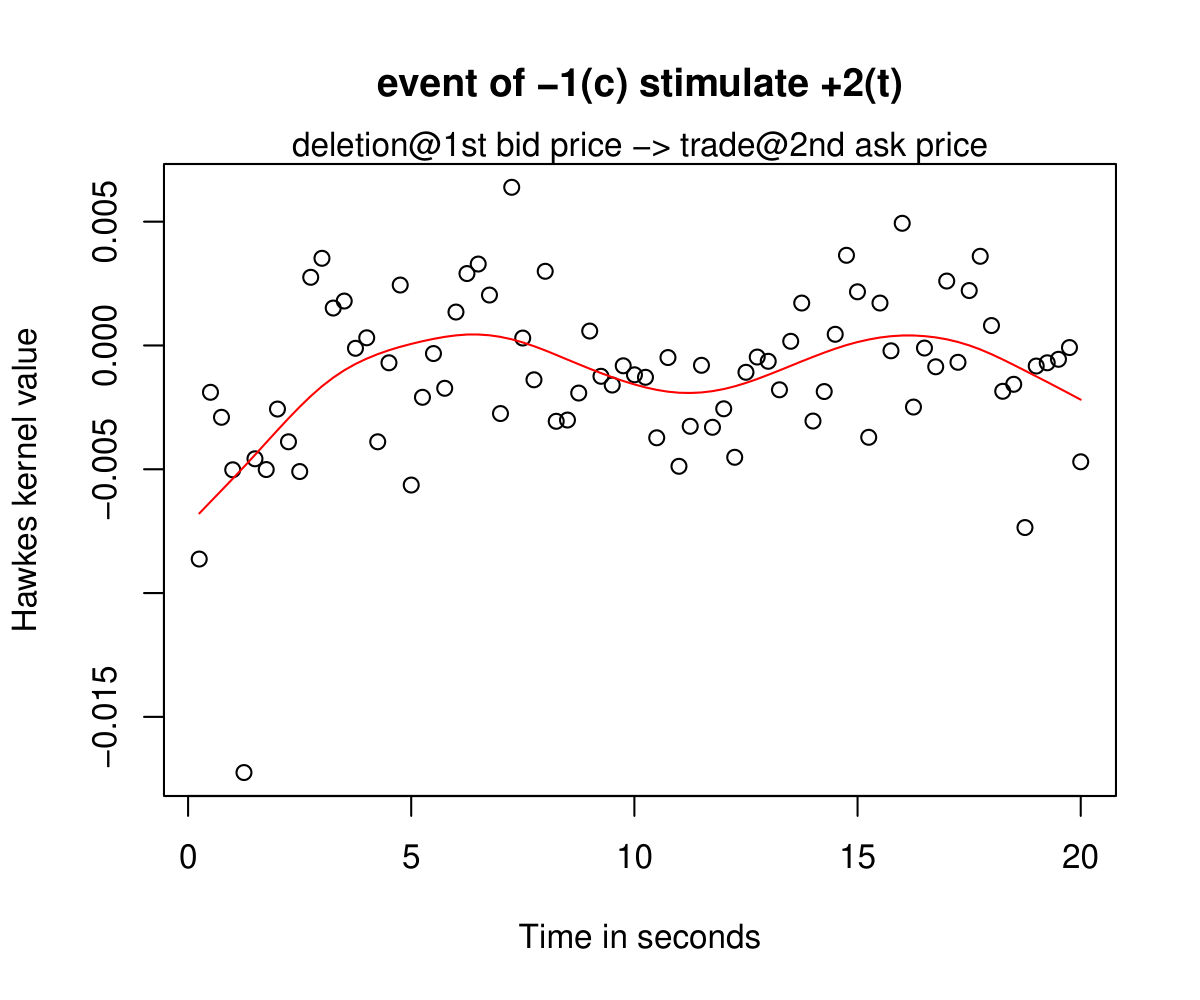} }}%
    \qquad
    \subfloat[\texttt{+2(t)} stimulate \texttt{-2(t)}]{{\includegraphics[scale =0.16]{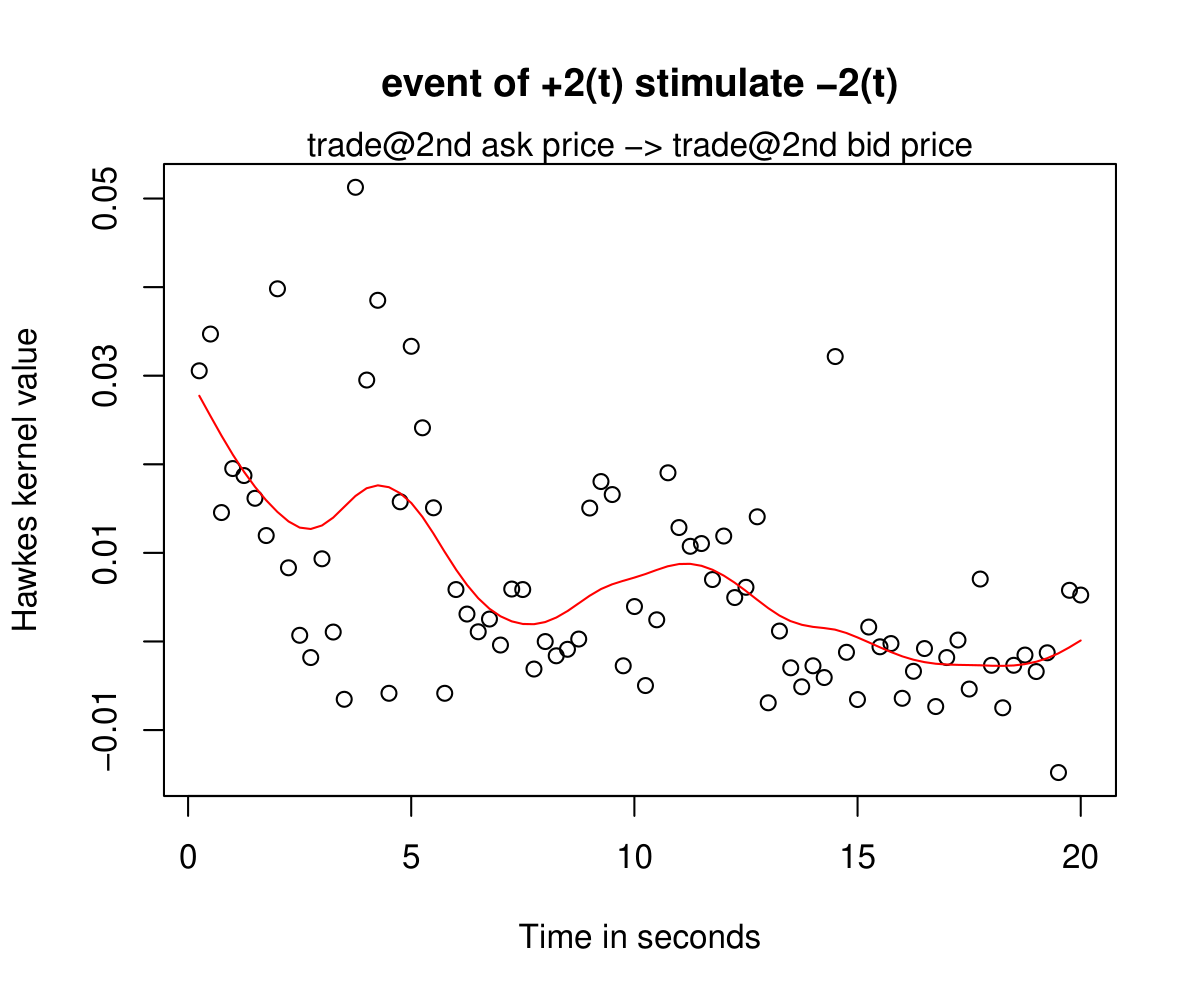} }}%
    
    \caption{Aggregated Hawkes excitement function estimation under $(s=\\20 \text{ seconds}, \Delta=0.25 \text{ seconds})$ when LASSO regularization is removed. The points illustrate the discrete function estimators. The red line illustrates the cubic smoothing spline for the points.}
    \label{fig: non-exponential kernels noLASSO}
\end{figure}
The results for the model without LASSO regularization are consistent with the results discussed in Section \ref{sec: exponential and Non-exponential shape of excitement function}: while most of the estimated Hawkes functions exhibit exponential-decaying shapes, some estimated results do exhibit non-exponential shapes.

\subsection{Liquidity state}

\begin{figure}[H]
    \centering
    \subfloat[Liquidity state for \texttt{+3(i)}]{{\includegraphics[scale = 0.43]{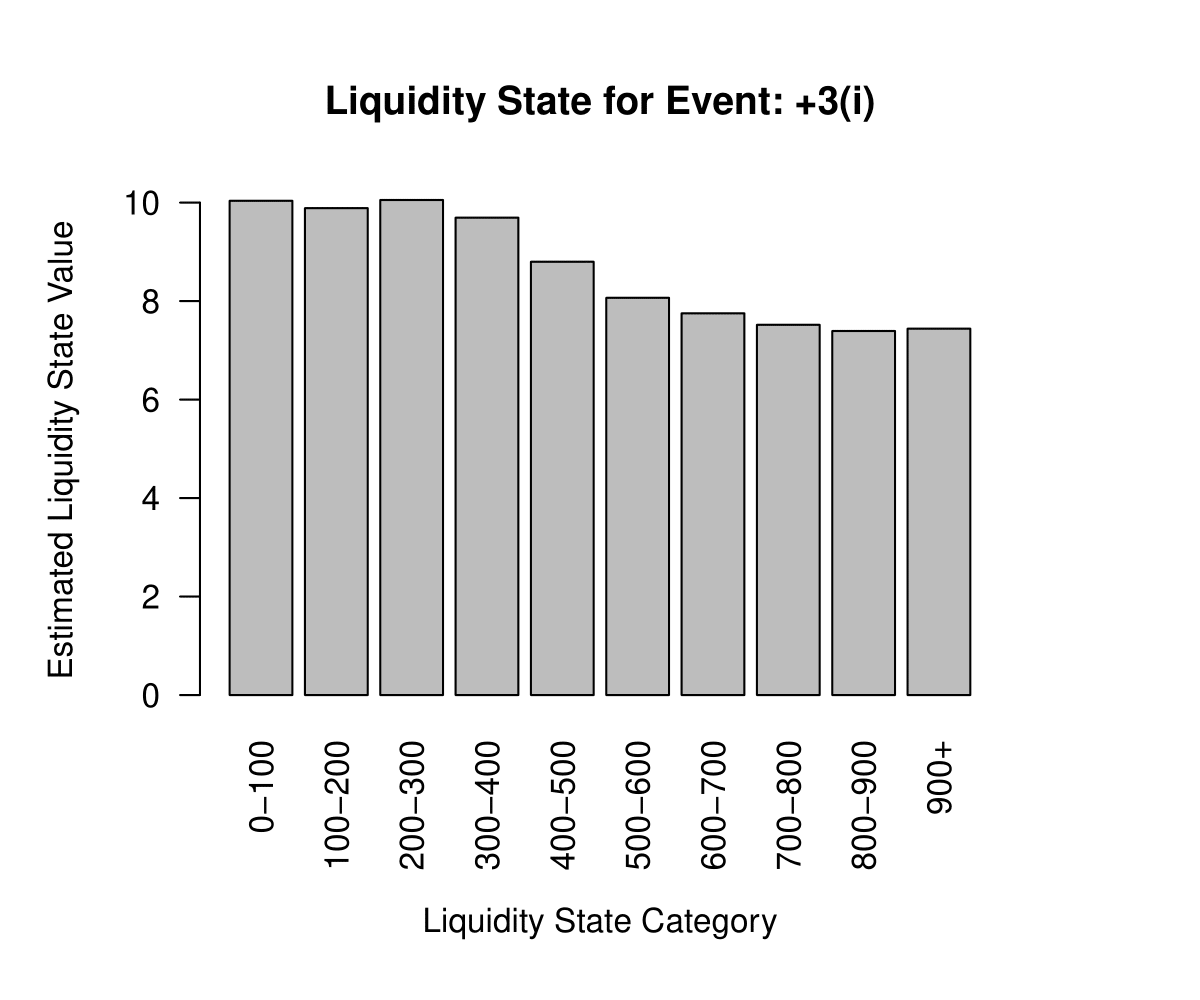} }}%
    \qquad
    \subfloat[Liquidity state for \texttt{-3(i)} ]{{\includegraphics[scale = 0.43]{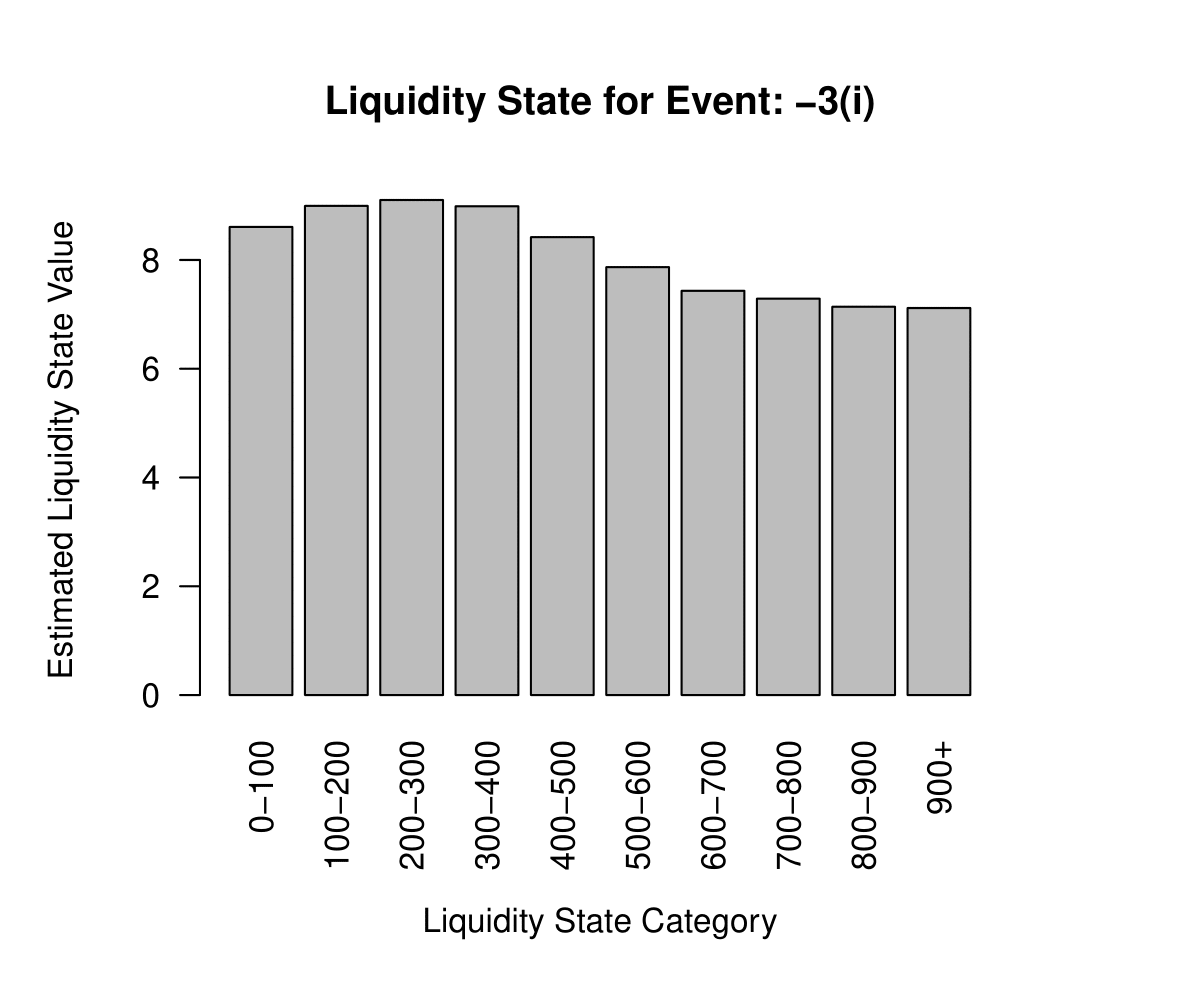} }}%
    \qquad
    \subfloat[Liquidity state for \texttt{+2(i)} ]{{\includegraphics[scale = 0.43]{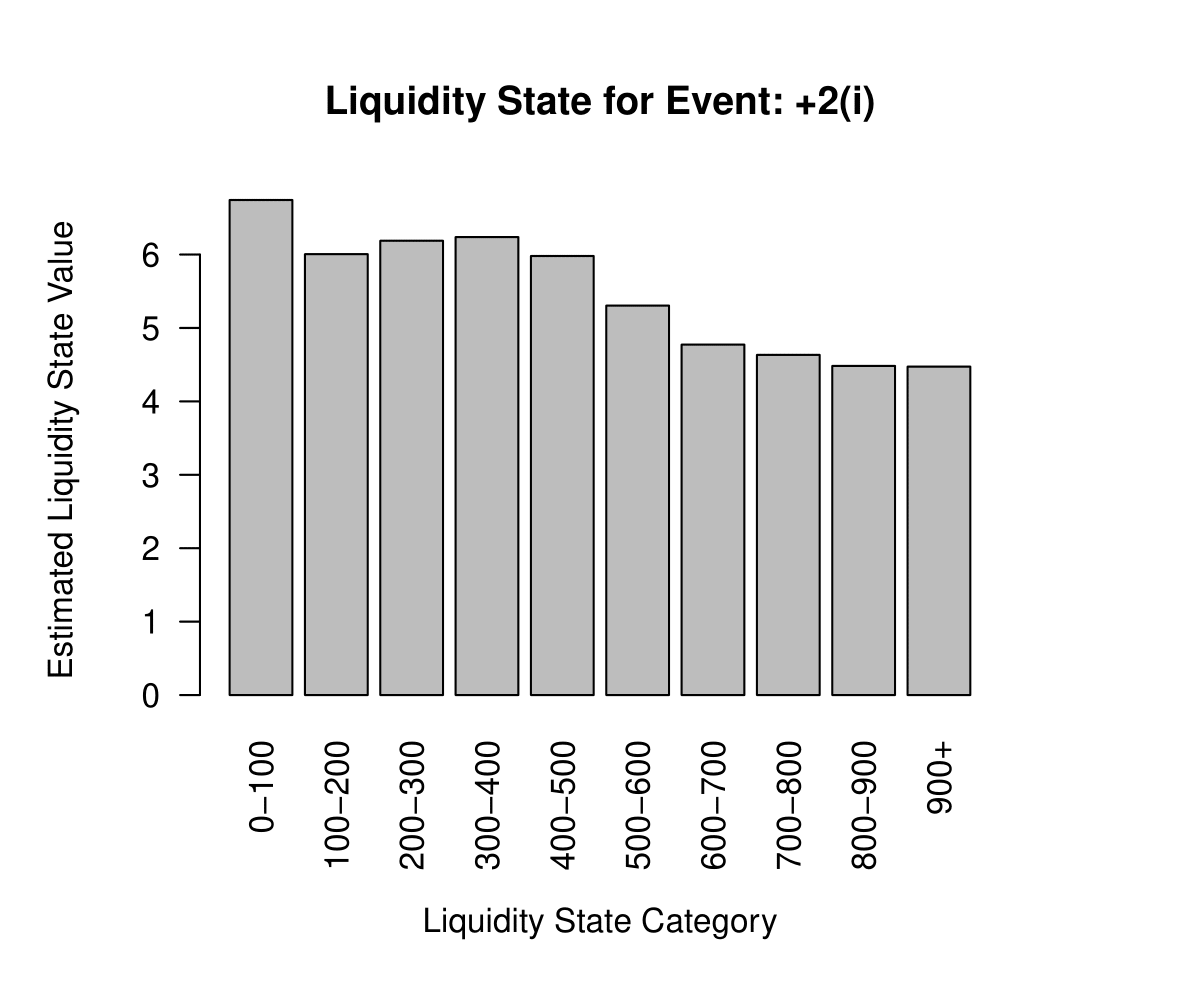} }}%
    \qquad
    \subfloat[Liquidity state for \texttt{-2(i)} ]{{\includegraphics[scale = 0.43]{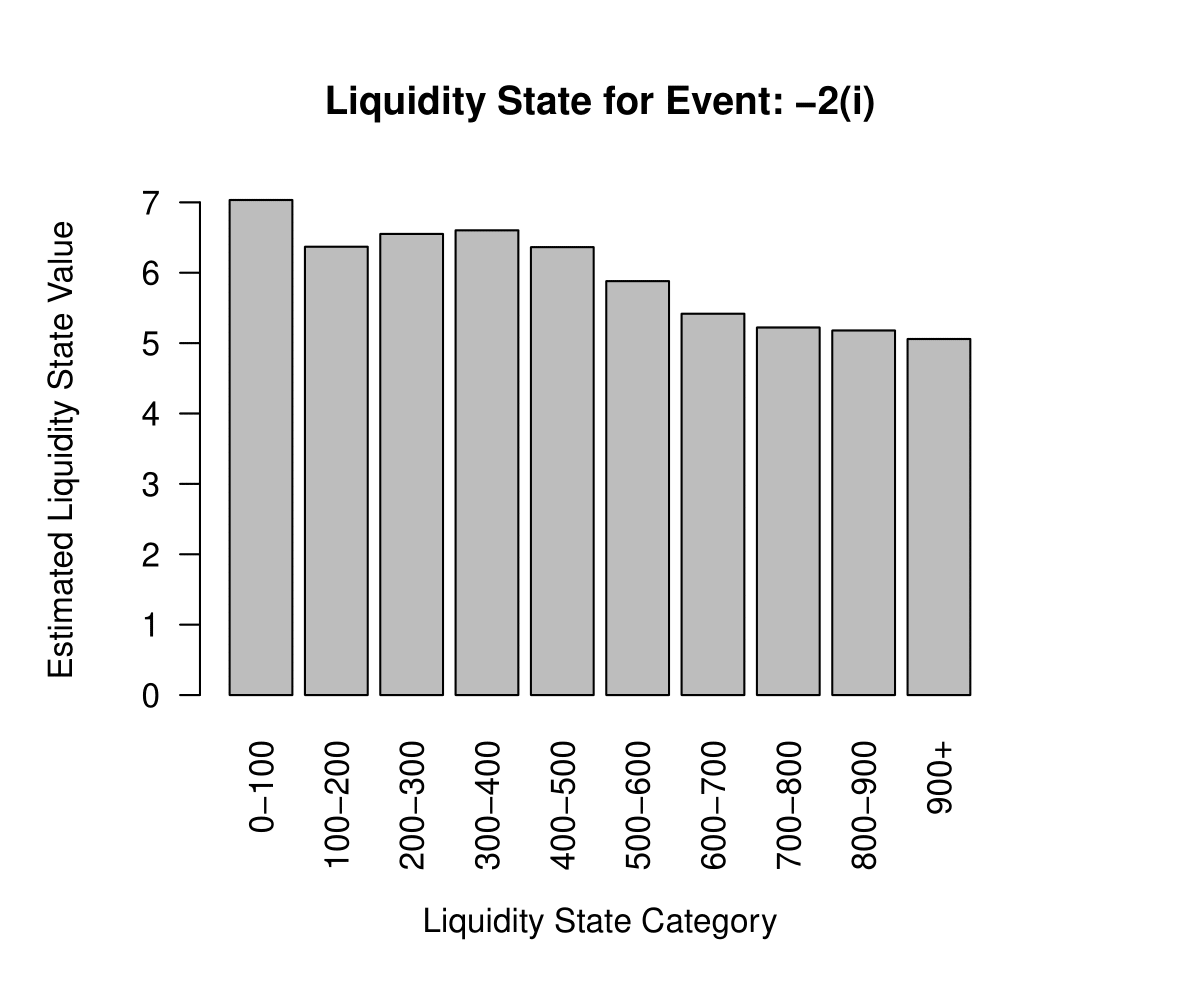} }}%
    \caption{Aggregated estimation result for liquidity state for event \texttt{+3(i)}, \-\texttt{-3(i)}, \-\texttt{+2(i)}, \-\texttt{-2(i)} under $(s=20 \text{ seconds}, \Delta=0.25 \text{ seconds})$ when LASSO regularization is removed. For these events the event arrival intensity decreases as liquidity state increases.}
    \label{fig: state_decrease noLASSO}
\end{figure}

Based on Figure \ref{fig: state_increase} and Figure \ref{fig: state_decrease} in Section \ref{sec: liquidity state}, the following Figure \ref{fig: state_increase noLASSO} and Figure \ref{fig: state_decrease noLASSO} demonstrate the liquidity state estimations of the model when the LASSO regularization is removed.

\begin{figure}[H]
    \centering
    \subfloat[Liquidity state for \texttt{+1(i)}]{{\includegraphics[scale = 0.43]{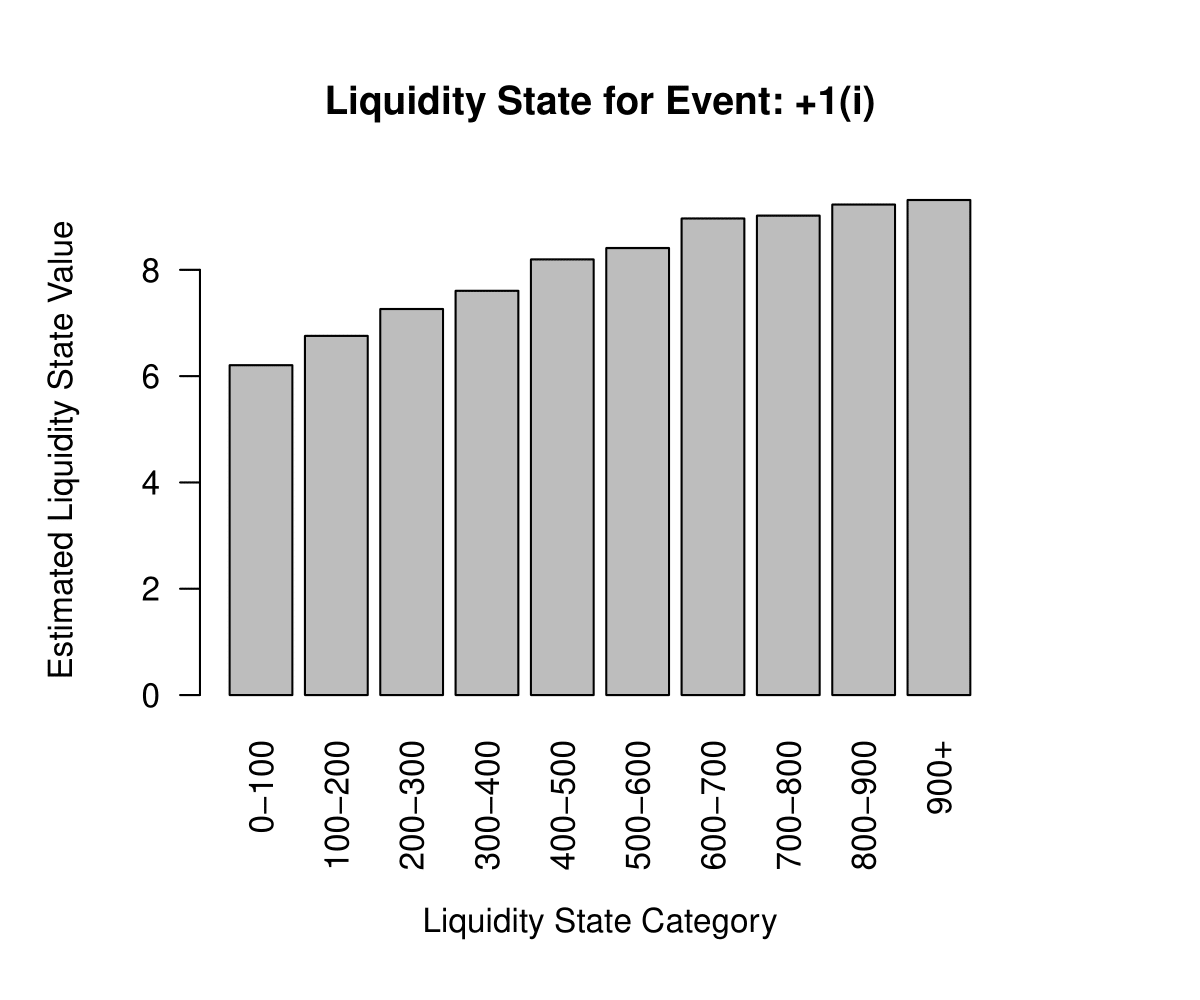} }}%
    \qquad
    \subfloat[Liquidity state for \texttt{+1(c)} ]{{\includegraphics[scale = 0.43]{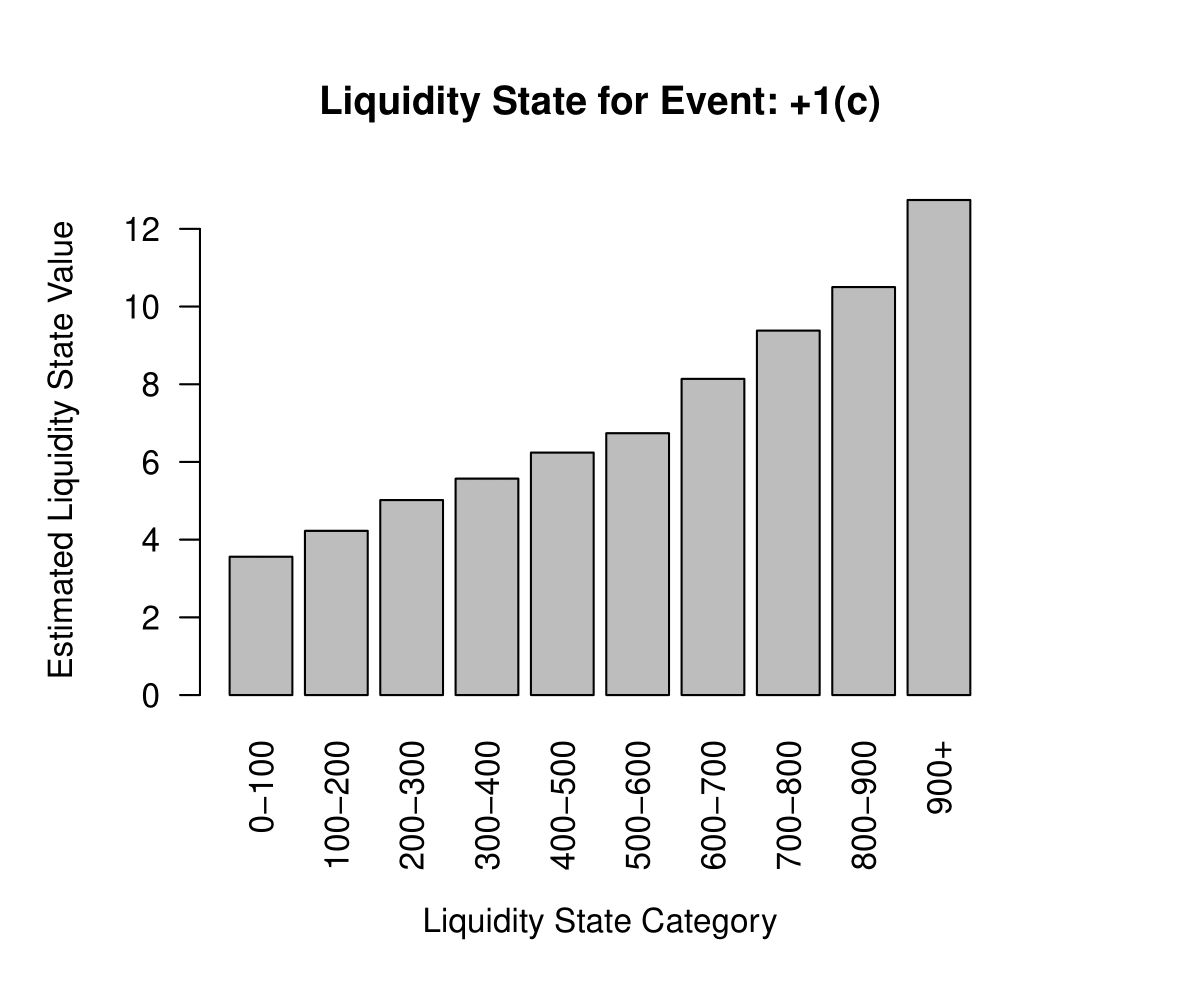} }}%
    \qquad
    \subfloat[Liquidity state for \texttt{+1(t)} ]{{\includegraphics[scale = 0.43]{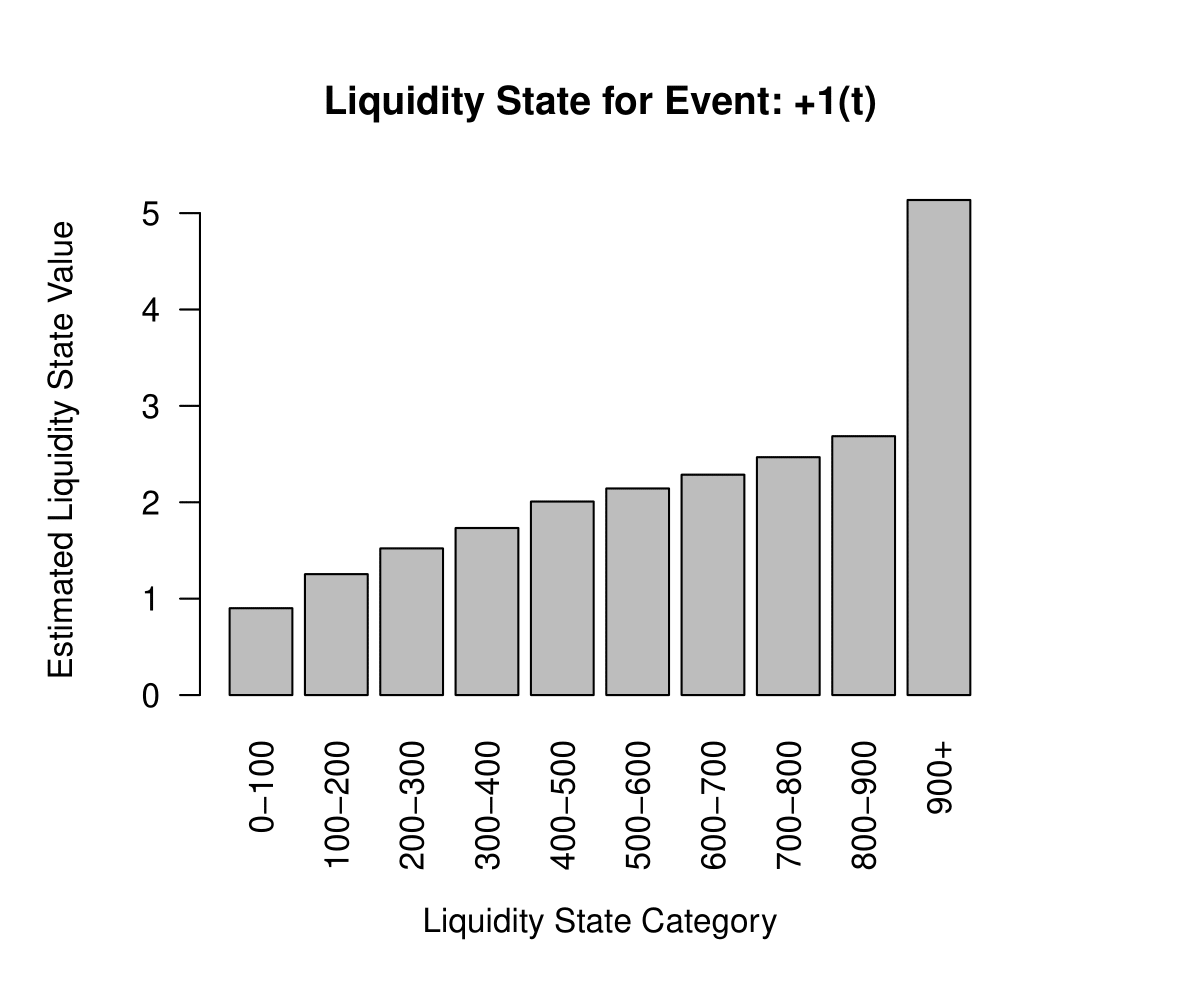} }}%
    \qquad
    \subfloat[Liquidity state for \texttt{-1(c)} ]{{\includegraphics[scale = 0.43]{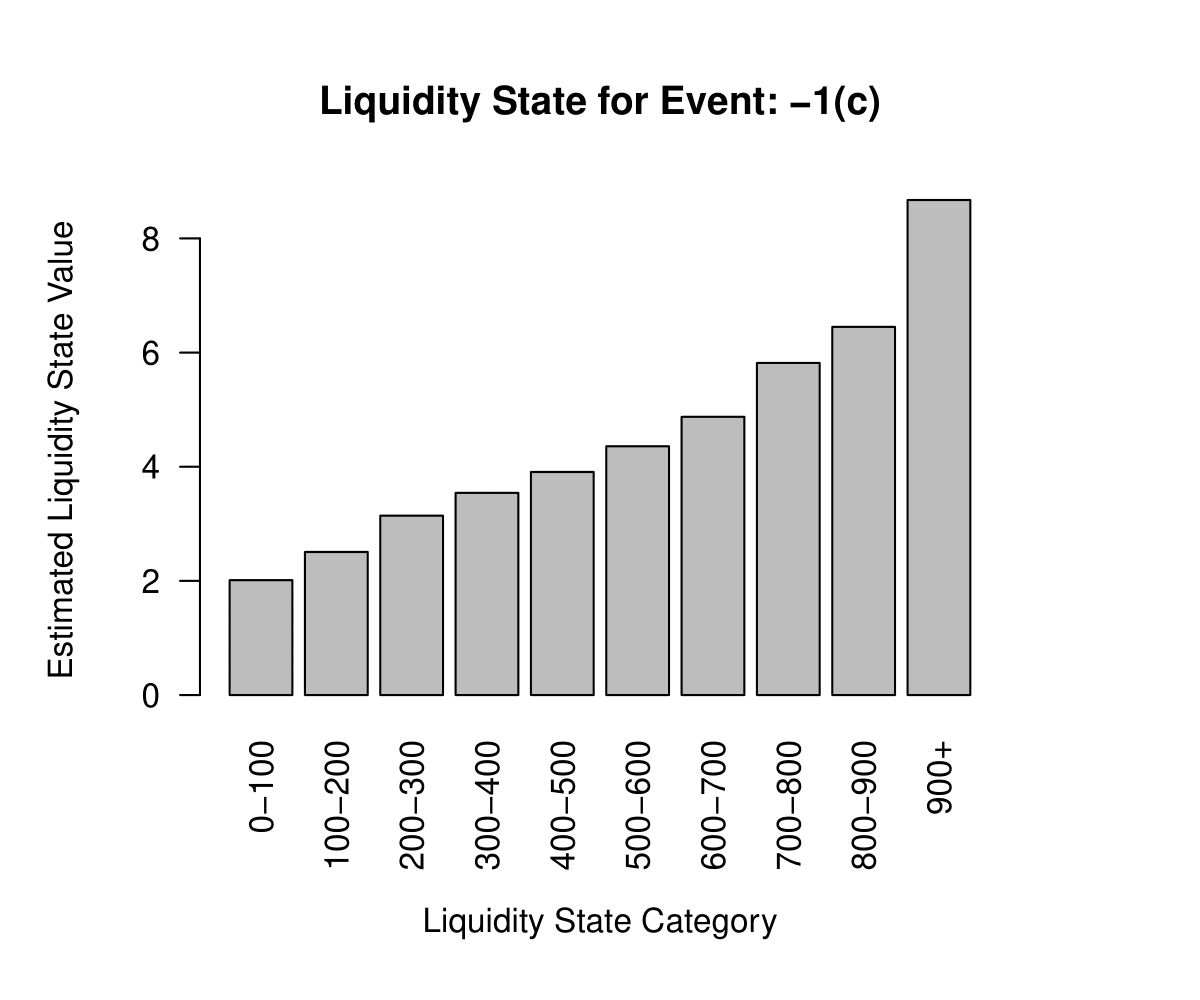} }}%
    \caption{Aggregated estimation result for liquidity state for event \texttt{+1(i)}, \-\texttt{+1(c)}, \-\texttt{+1(t)}, \-\texttt{-1(c)} under $(s=20 \text{ seconds}, \Delta=0.25 \text{ seconds})$ when LASSO regularization is removed. For these events the event arrival intensity increases as liquidity state increases.}
    \label{fig: state_increase noLASSO}
\end{figure}

The demonstrated liquidity state estimation results of the model without LASSO regularization is consistent with the results discussed in Section \ref{sec: liquidity state}.

\subsection{Time factor}

Based on Figure \ref{fig: time factor} in Section \ref{sec: time factor}, the following Figure \ref{fig: time factor noLASSO} demonstrates the time factor estimations of the model when the LASSO regularization is removed.

\begin{figure}[H]
    \centering
    \subfloat[Time factor for event \texttt{+1(i)}]{{\includegraphics[scale = 0.43]{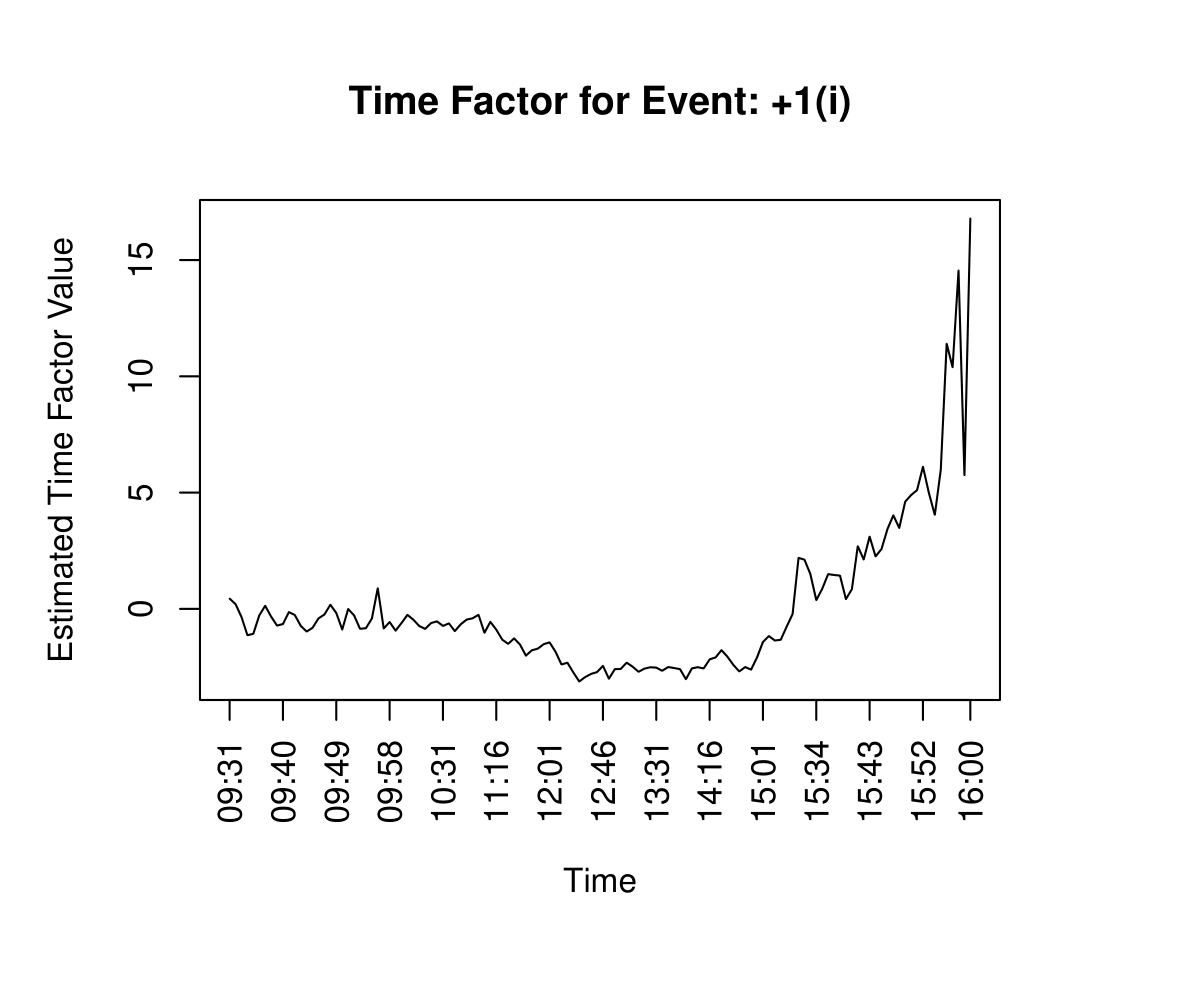} }}%
    \qquad
    \subfloat[Time factor for event \texttt{+3(t)}]{{\includegraphics[scale = 0.43]{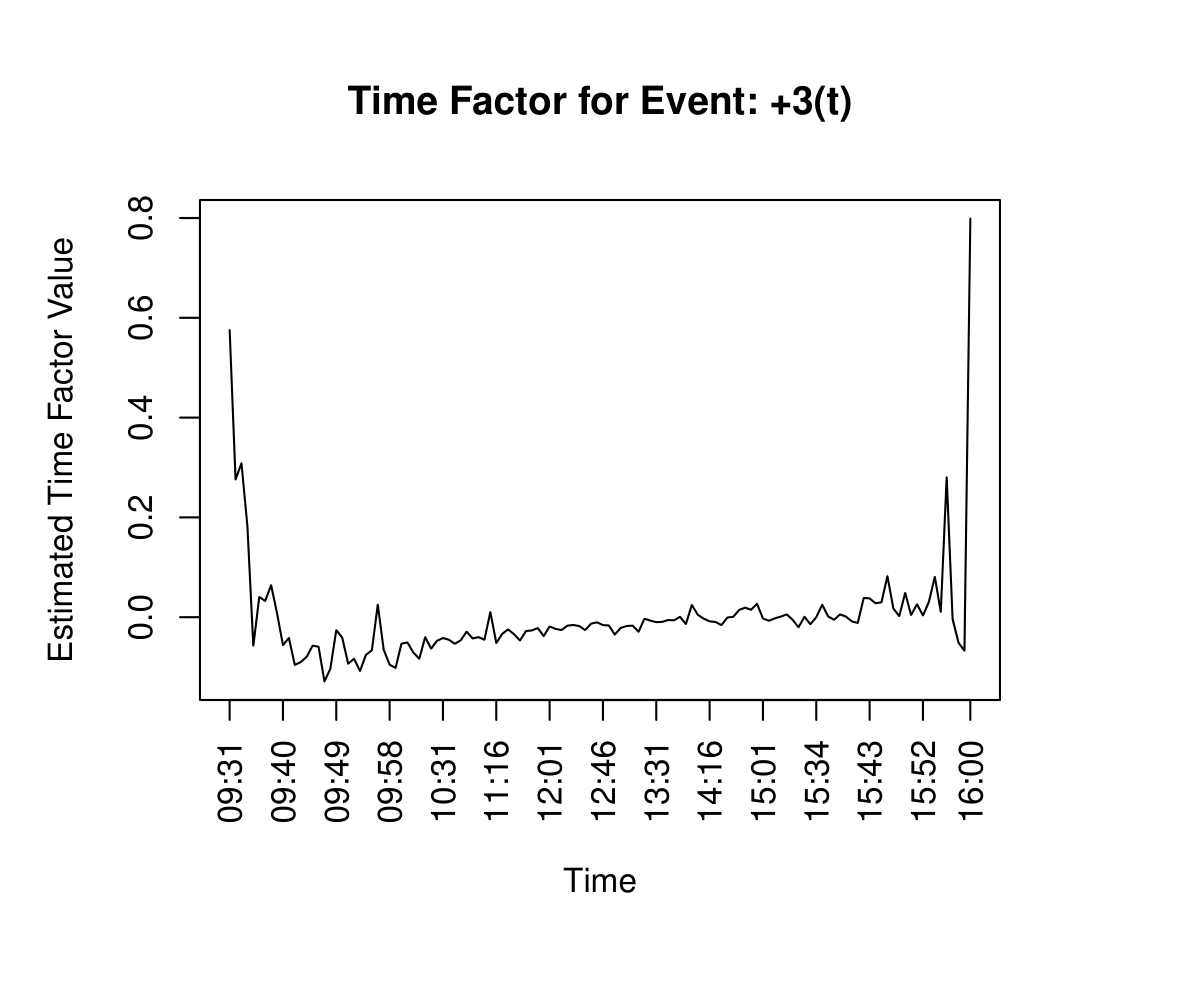} }}%
    \caption{Aggregated estimation result for time factor between 9:30 am and 4:00 pm under $(s=20 \text{ seconds}, \Delta=0.25 \text{ seconds})$ when LASSO regularization is removed. All orders are considered to have size 1.}
    \label{fig: time factor noLASSO}
\end{figure}

The demonstrated time factor estimation results of the model without LASSO regularization is consistent with the results discussed in Section \ref{sec: time factor}.

\section{Empirical results with enlarged bin-size}\label{sec_support: result large bin size}
This supporting section demonstrates the empirical estimation results when the bin-size $\Delta$ is enlarged from 0.25 seconds to 0.5 seconds, as mentioned in Section \ref{sec: sensitivity analysis}. The demonstrations will be presented in the same format as the demonstrations from Section \ref{sec:estimated excitement functions: ask} to Section \ref{sec: time factor}.  As a whole, the results on excitement function, liquidity state, and time factor still hold qualitatively. This meets our expectation that enlarging the bin-size won't change the estimated result significantly as estimations are obtained from the same dataset and the $\Delta = 0.5s$ estimation result is just a coarse version of the $\Delta = 0.25s$ result.

\subsection{Estimated excitement functions}
\begin{figure}[H]
    \centering
    \subfloat[\texttt{-1(i)} stimulate \texttt{-1(i)}]{{\includegraphics[scale = 0.16]{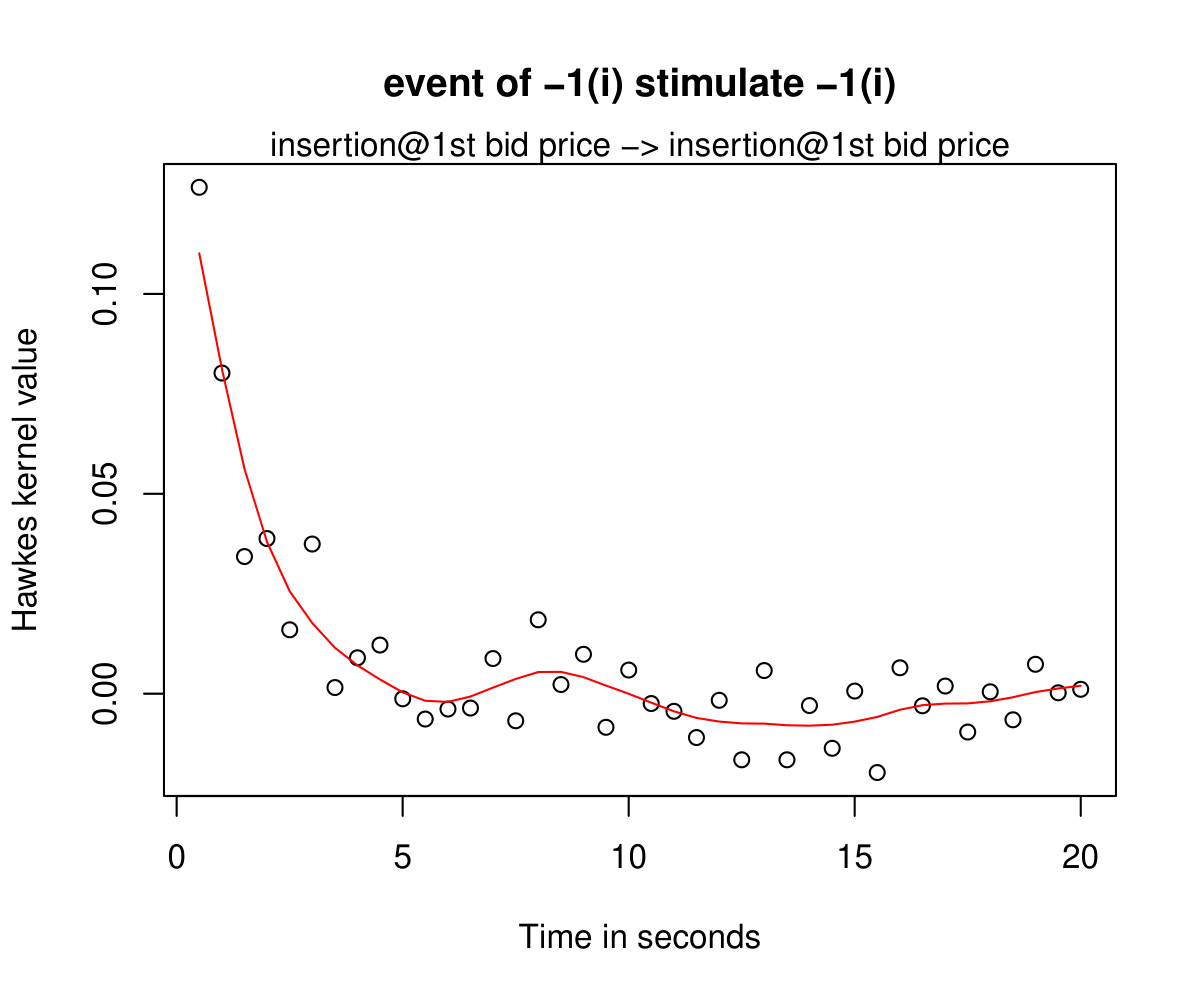} }}%
    \qquad
    \subfloat[\texttt{-1(i)} stimulate \texttt{-1(c)}]{{\includegraphics[scale =0.16]{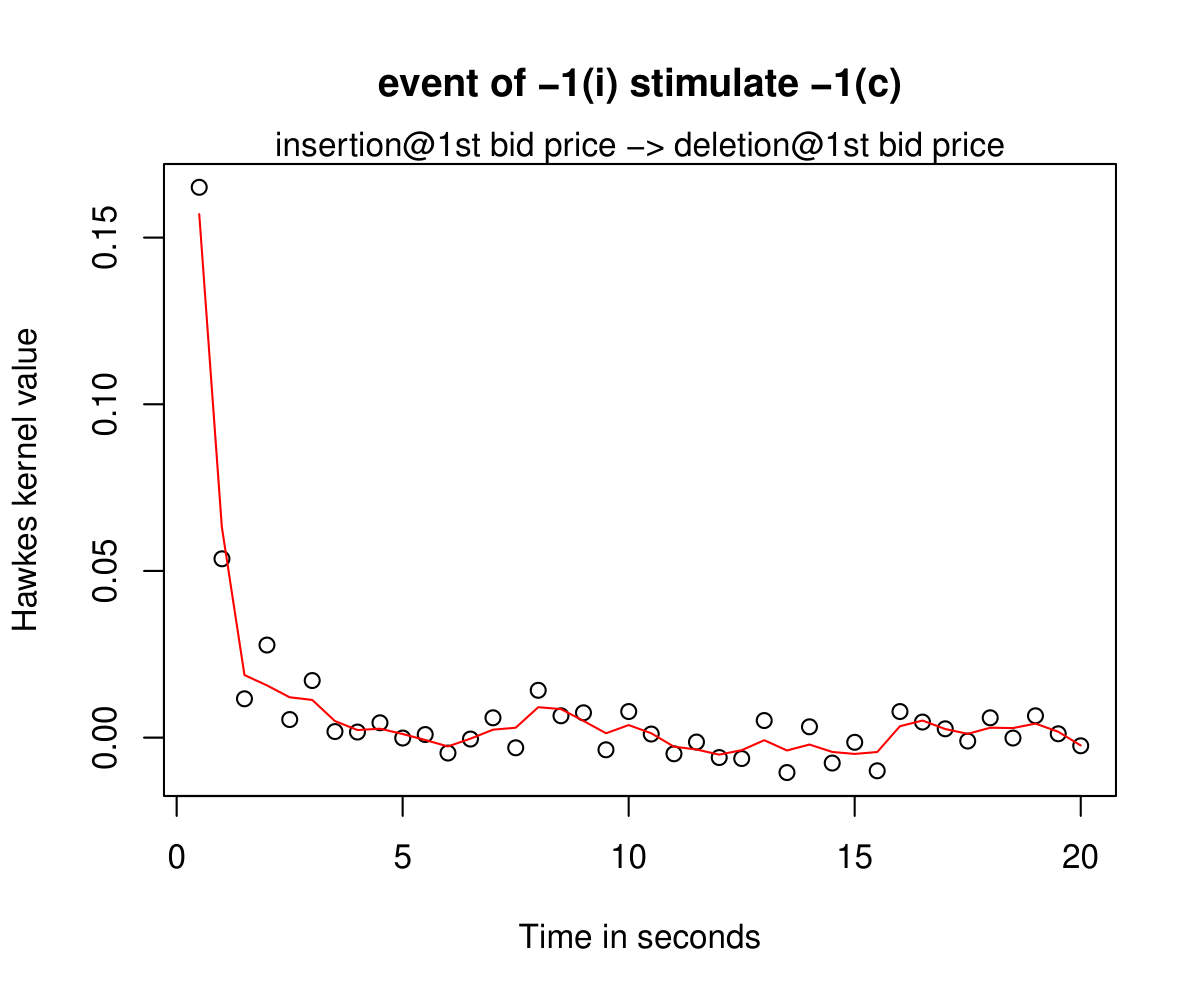} }}%
    \caption{Aggregated Hawkes excitement function estimation under $(s=\\20 \text{ seconds}, \Delta=0.5 \text{ seconds})$ with LASSO regularization. The points illustrate the discrete function estimators. The red line illustrates the cubic smoothing spline for the points.}
    \label{fig: first-level similar bid largeDelta}
\end{figure}

Based on Figure \ref{fig: first-level similar} and Figure \ref{fig: first-level similar bid}, the following Figure \ref{fig: first-level similar largeDelta} and Figure \ref{fig: first-level similar bid largeDelta} demonstrate the estimated Hawkes excitement functions when the bin-size $\Delta$ is enlarged from 0.25 seconds to 0.5 seconds.

\begin{figure}[H]
    \centering
    \subfloat[\texttt{+1(i)} stimulate \texttt{+1(i)}]{{\includegraphics[scale = 0.16]{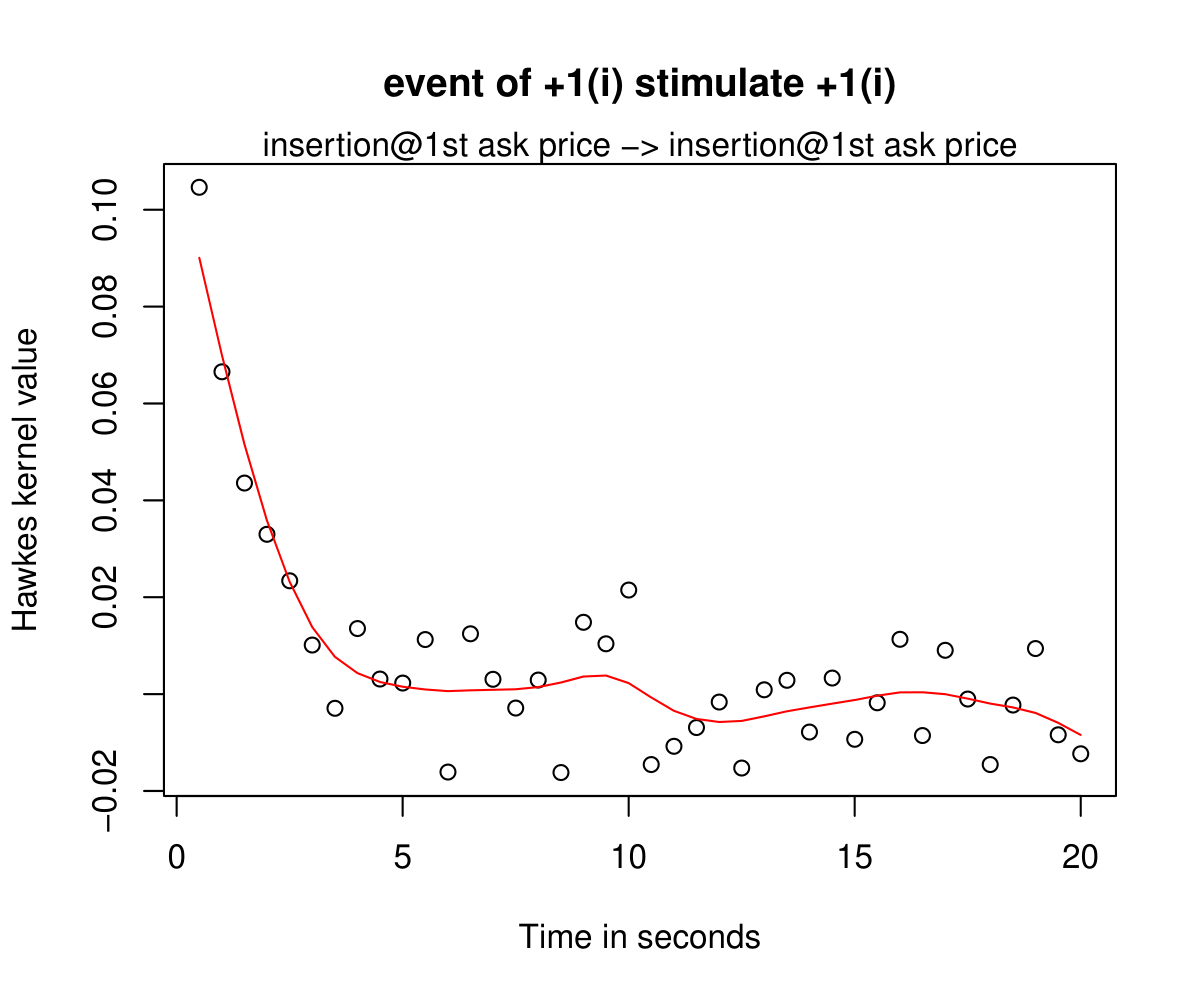} }}%
    \qquad
    \subfloat[\texttt{+1(i)} stimulate \texttt{+1(c)}]{{\includegraphics[scale =0.16]{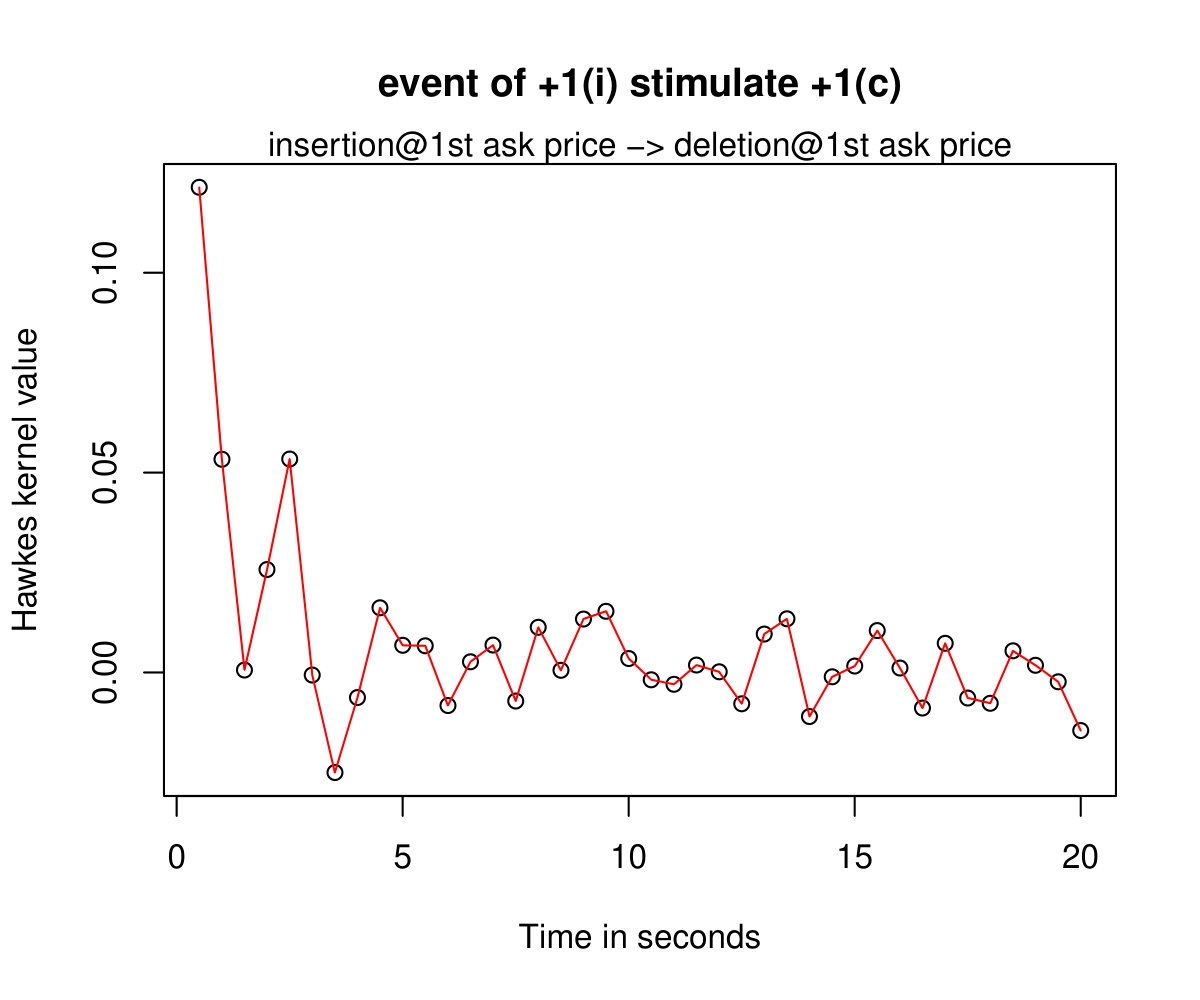} }}%
    \caption{Aggregated Hawkes excitement function estimation under $(s=\\20 \text{ seconds}, \Delta=0.5 \text{ seconds})$ with LASSO regularization. The points illustrate the discrete function estimators. The red line illustrates the cubic smoothing spline for the points.}
    \label{fig: first-level similar largeDelta}
\end{figure}

As we can observe, the above estimated functions are generally consistent with the 1st-ask and 1st-bid similarity patterns discussed in Section \ref{sec:estimated excitement functions: ask} and Section \ref{sec_support: estimated excitement functions bid}, when the bin-size is enlarged.

\subsection{Exponential and non-exponential shape of excitement functions}

Based on Figure \ref{fig: non-exponential kernels} in Section \ref{sec: exponential and Non-exponential shape of excitement function}, the following Figure \ref{fig: non-exponential kernels largeDelta} demonstrates the estimated Hawkes excitement functions with non-exponential shapes when the bin-size $\Delta$ is enlarged from 0.25 seconds to 0.5 seconds.

\begin{figure}[H]
    \centering
    \subfloat[\texttt{-1(t)} stimulate \texttt{+3(t)}]{{\includegraphics[scale = 0.16]{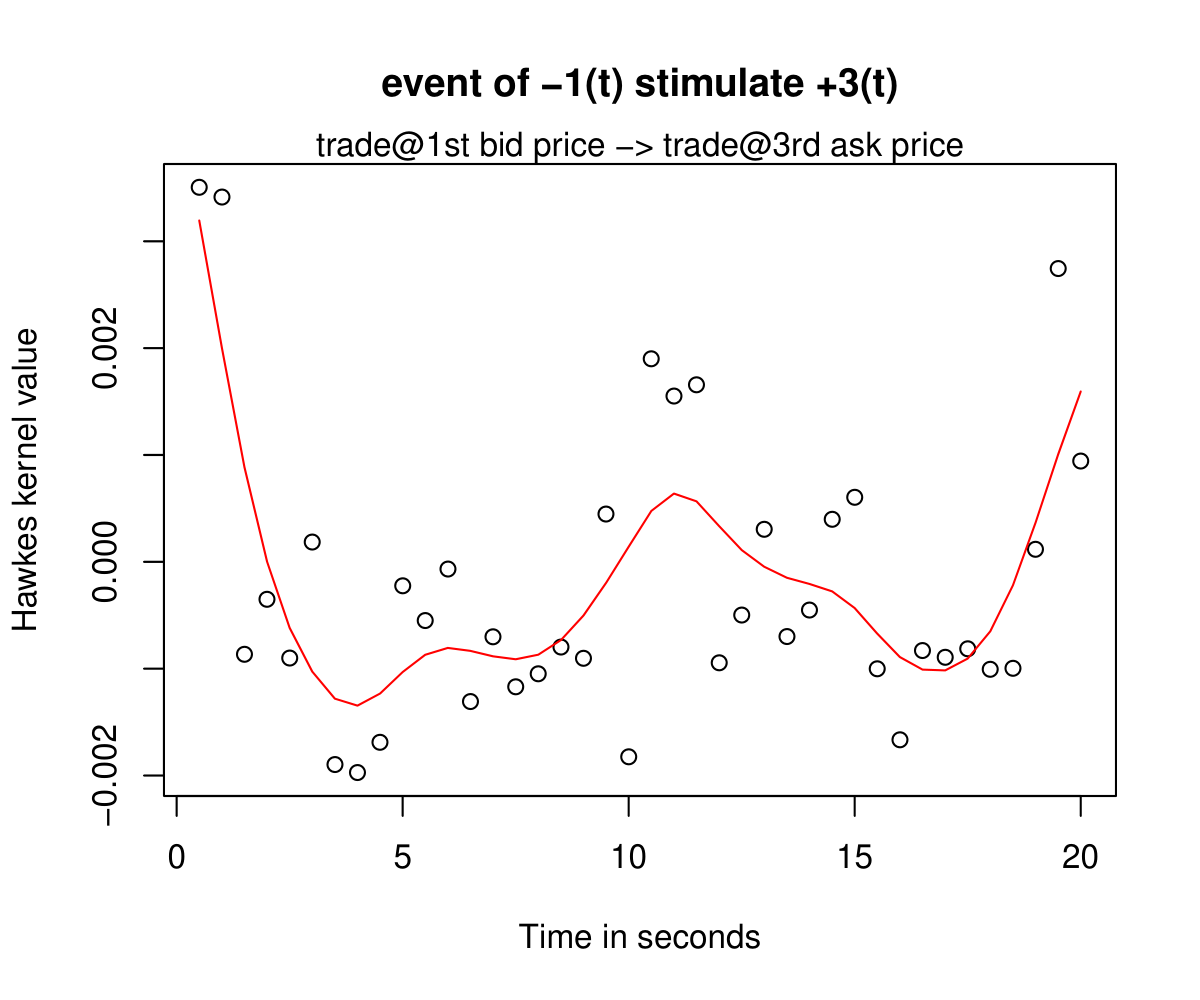} }}%
    \qquad
    \subfloat[\texttt{p+(t)} stimulate \texttt{-3(t)}]{{\includegraphics[scale =0.16]{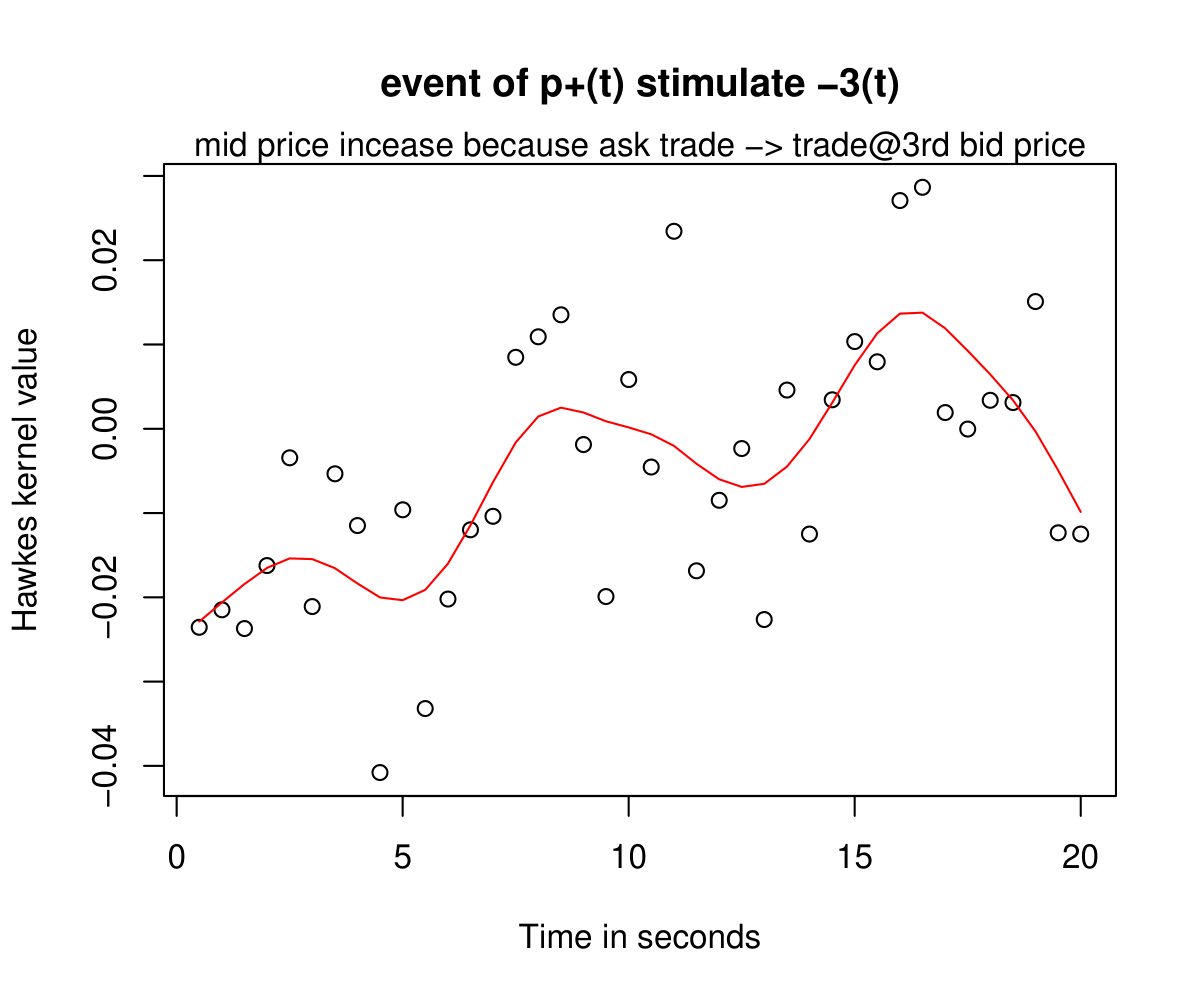} }}%
    \qquad
    \subfloat[\texttt{-1(c)} stimulate \texttt{+2(t)}]{{\includegraphics[scale =0.16]{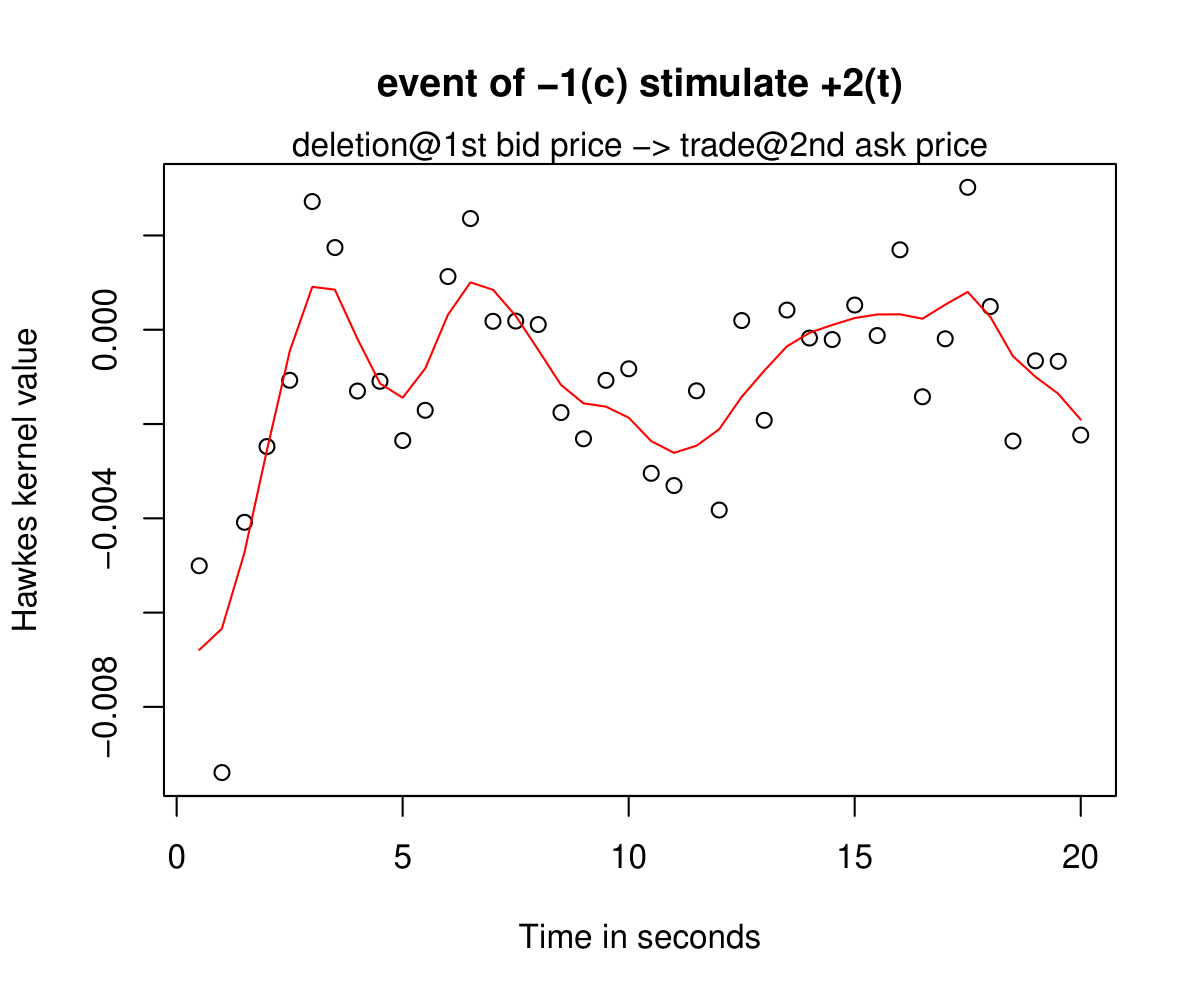} }}%
    \qquad
    \subfloat[\texttt{+2(t)} stimulate \texttt{-2(t)}]{{\includegraphics[scale =0.16]{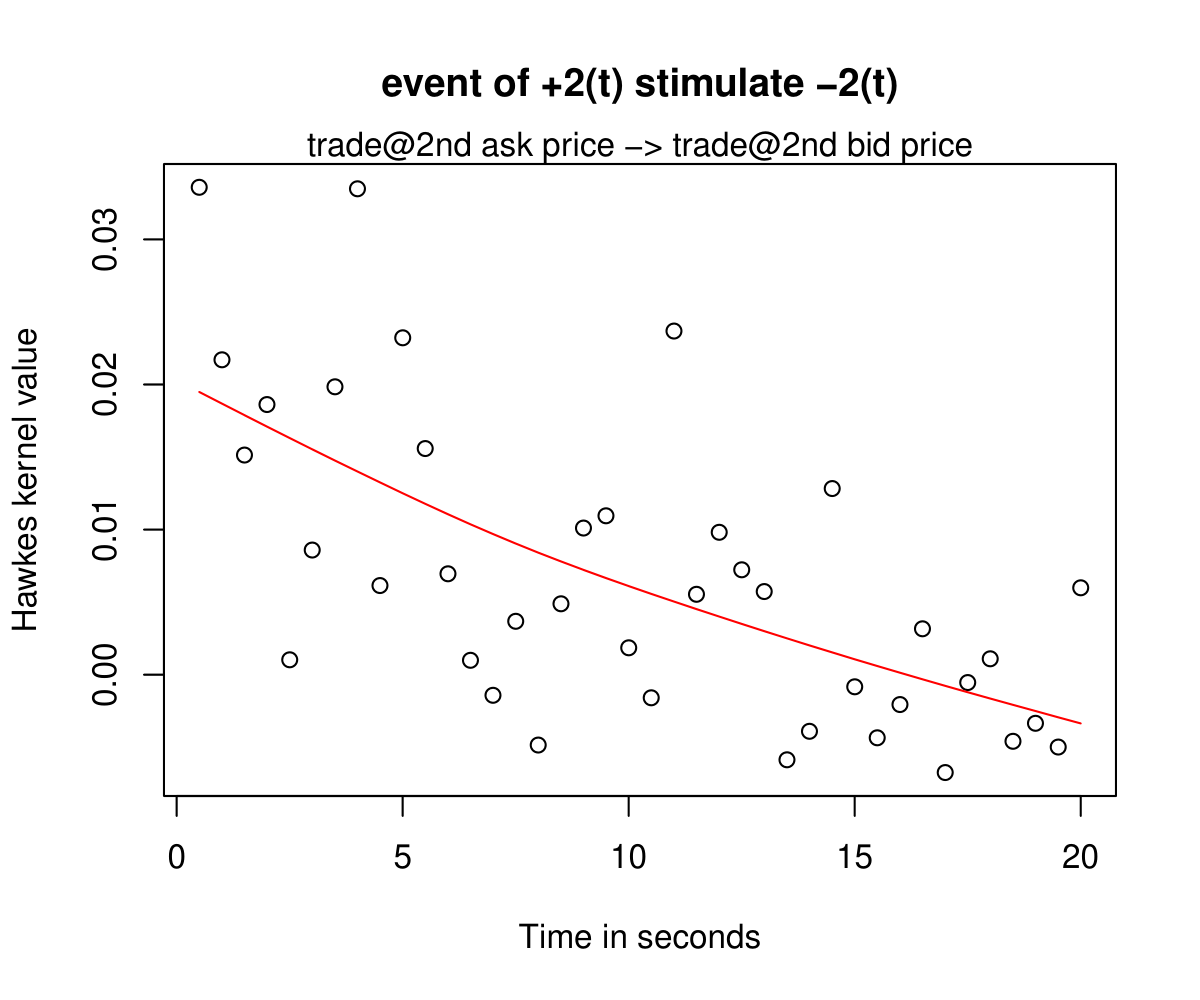} }}%
    
    \caption{Aggregated Hawkes excitement function estimation under $(s=20 \\\text{ seconds}, \Delta=0.5 \text{ seconds})$ with LASSO regularization. The points illustrate the discrete function estimators. The red line illustrates the cubic smoothing spline for the points.}
    \label{fig: non-exponential kernels largeDelta}
\end{figure}
The demonstrated results for the model without LASSO regularization are consistent with the results discussed in Section \ref{sec: exponential and Non-exponential shape of excitement function}: while most of the estimated Hawkes functions exhibit exponential time-decaying shapes, some estimated results do exhibit non-exponential shapes.

\subsection{Liquidity state}

\begin{figure}[H]
    \centering
    \subfloat[Liquidity state for \texttt{+3(i)}]{{\includegraphics[scale = 0.43]{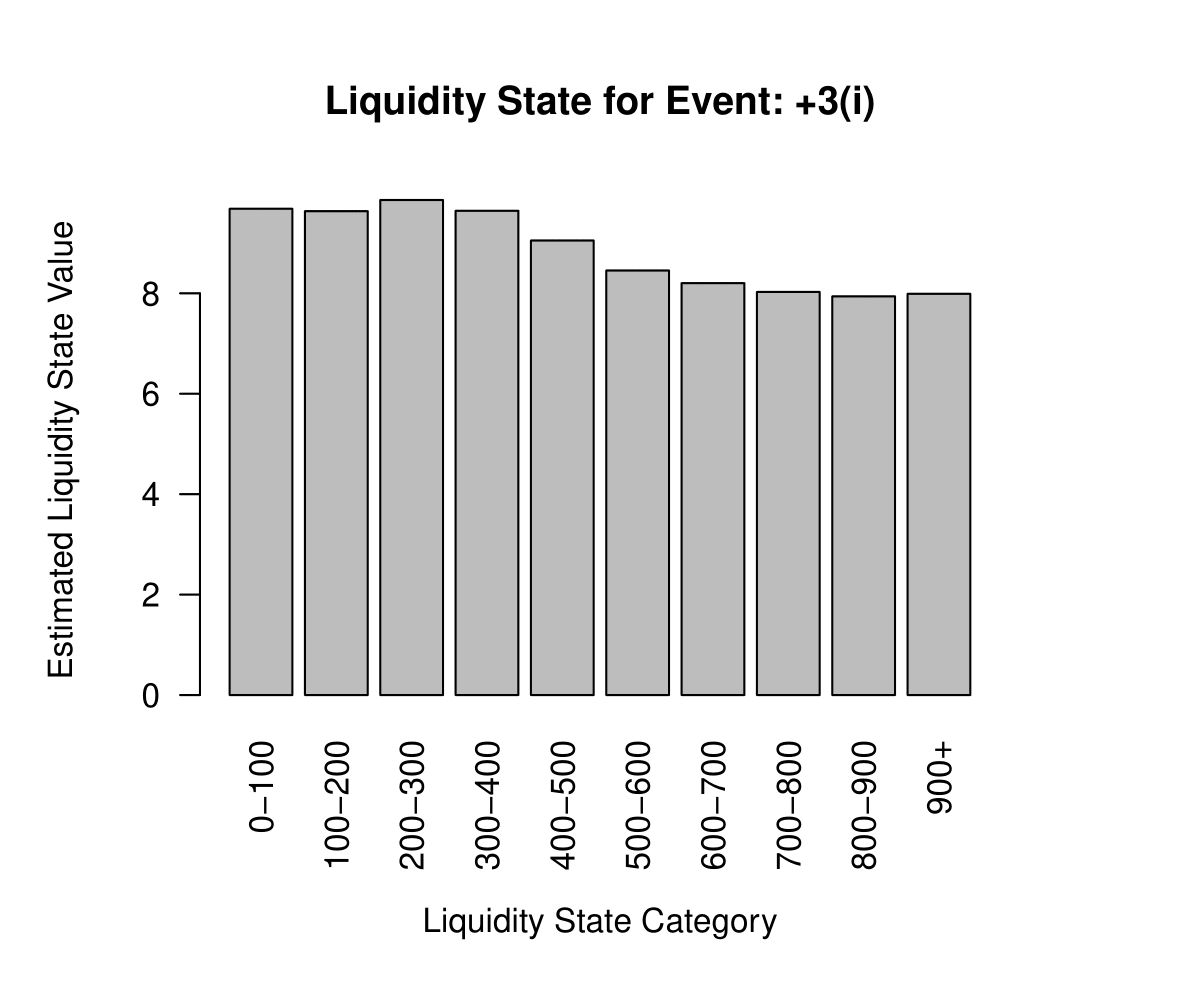} }}%
    \qquad
    \subfloat[Liquidity state for \texttt{-3(i)} ]{{\includegraphics[scale = 0.43]{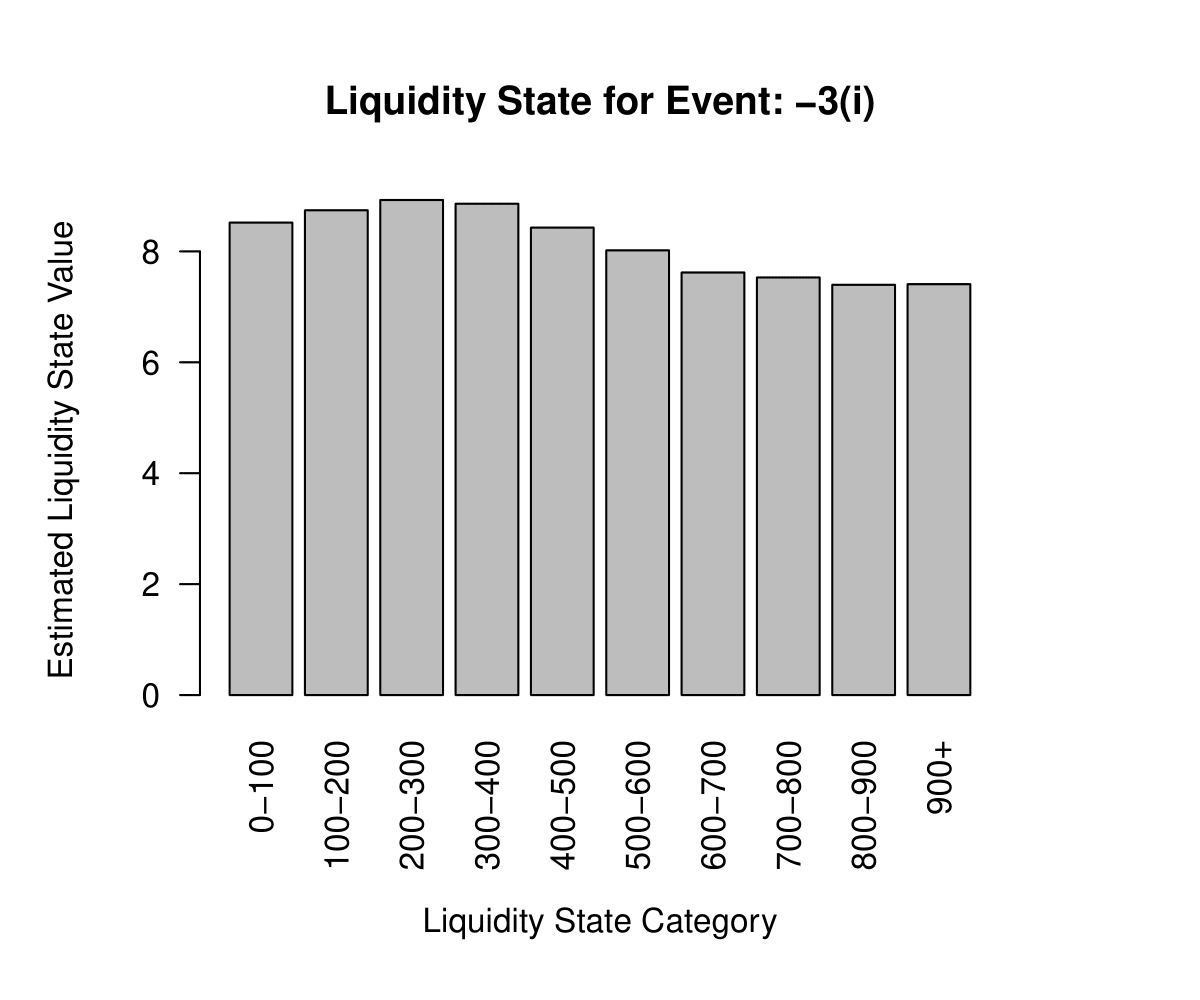} }}%
    \qquad
    \subfloat[Liquidity state for \texttt{+2(i)} ]{{\includegraphics[scale = 0.43]{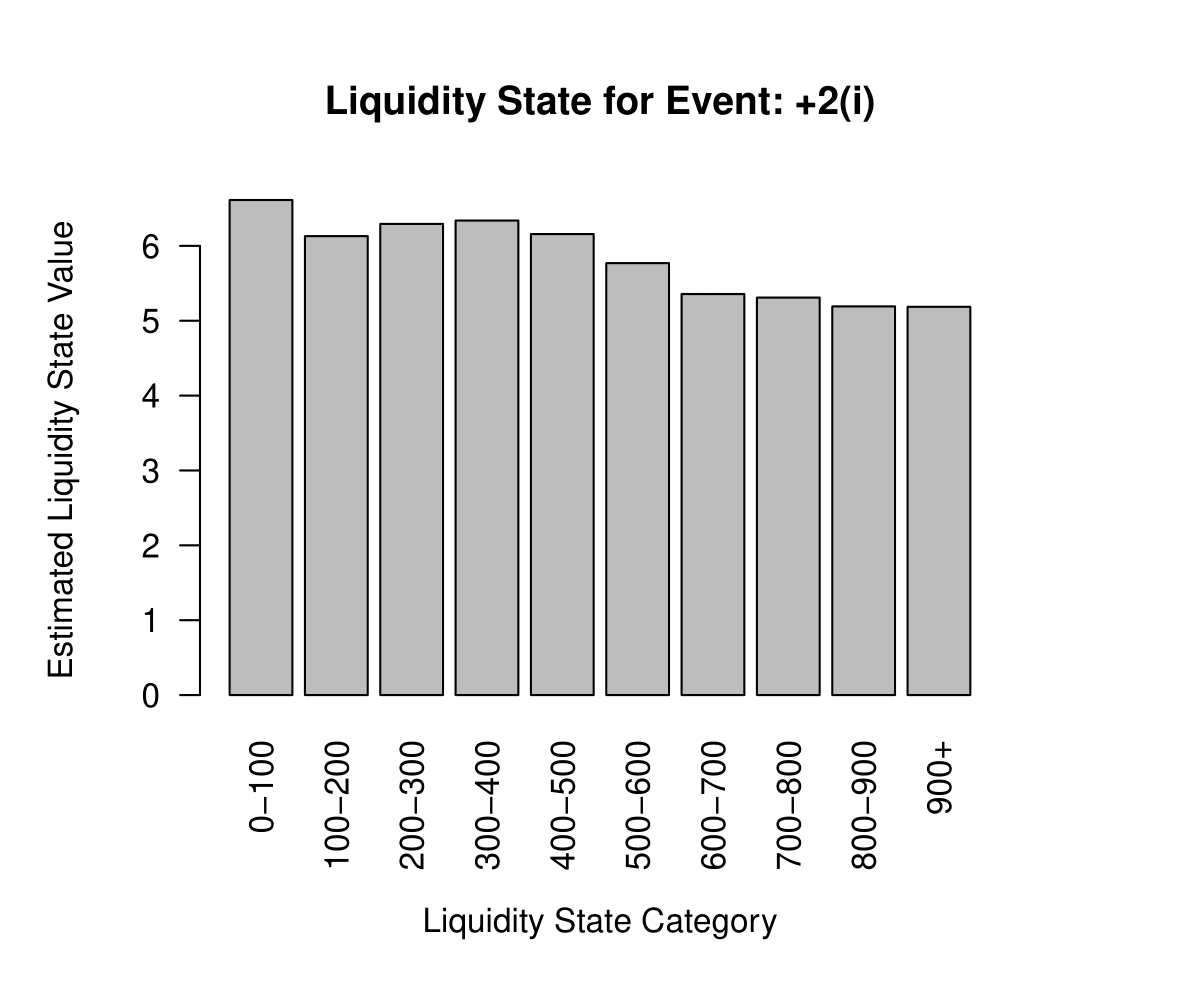} }}%
    \qquad
    \subfloat[Liquidity state for \texttt{-2(i)} ]{{\includegraphics[scale = 0.43]{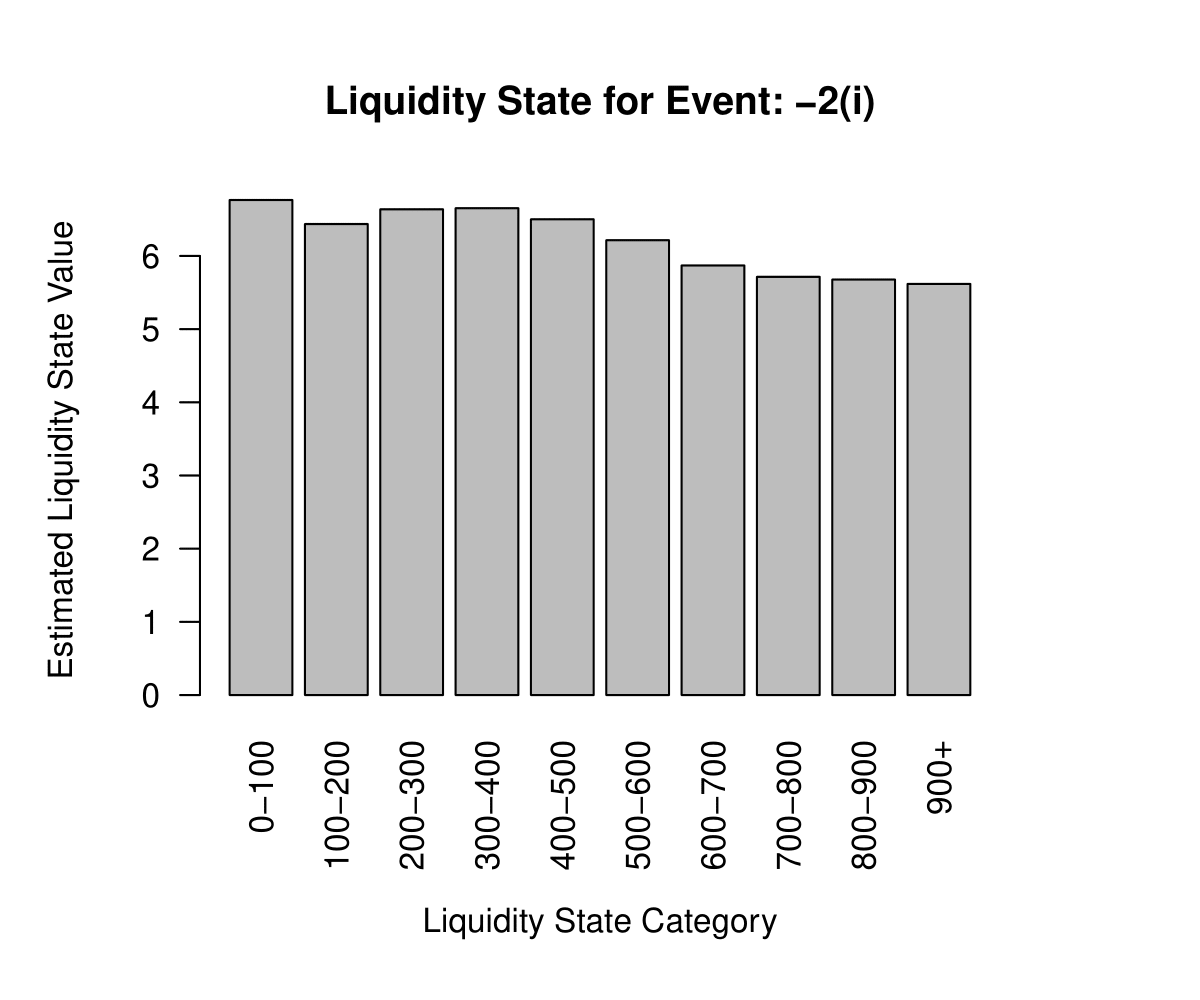} }}%
    \caption{Aggregated estimation result for liquidity state for event \texttt{+3(i)}, \-\texttt{-3(i)}, \-\texttt{+2(i)}, \-\texttt{-2(i)} under $(s=20 \text{ seconds}, \Delta=0.5 \text{ seconds})$. For these events the event arrival intensity decreases as liquidity state increases.}
    \label{fig: state_decrease largeDelta}
\end{figure}

Based on Figure \ref{fig: state_increase} and Figure \ref{fig: state_decrease} in Section \ref{sec: liquidity state}, the following Figure \ref{fig: state_increase largeDelta} and Figure \ref{fig: state_decrease largeDelta} demonstrate the liquidity state estimations of the model when the bin-size $\Delta$ is enlarged from 0.25 seconds to 0.5 seconds.

\begin{figure}[H]
    \centering
    \subfloat[Liquidity state for \texttt{+1(i)}]{{\includegraphics[scale = 0.43]{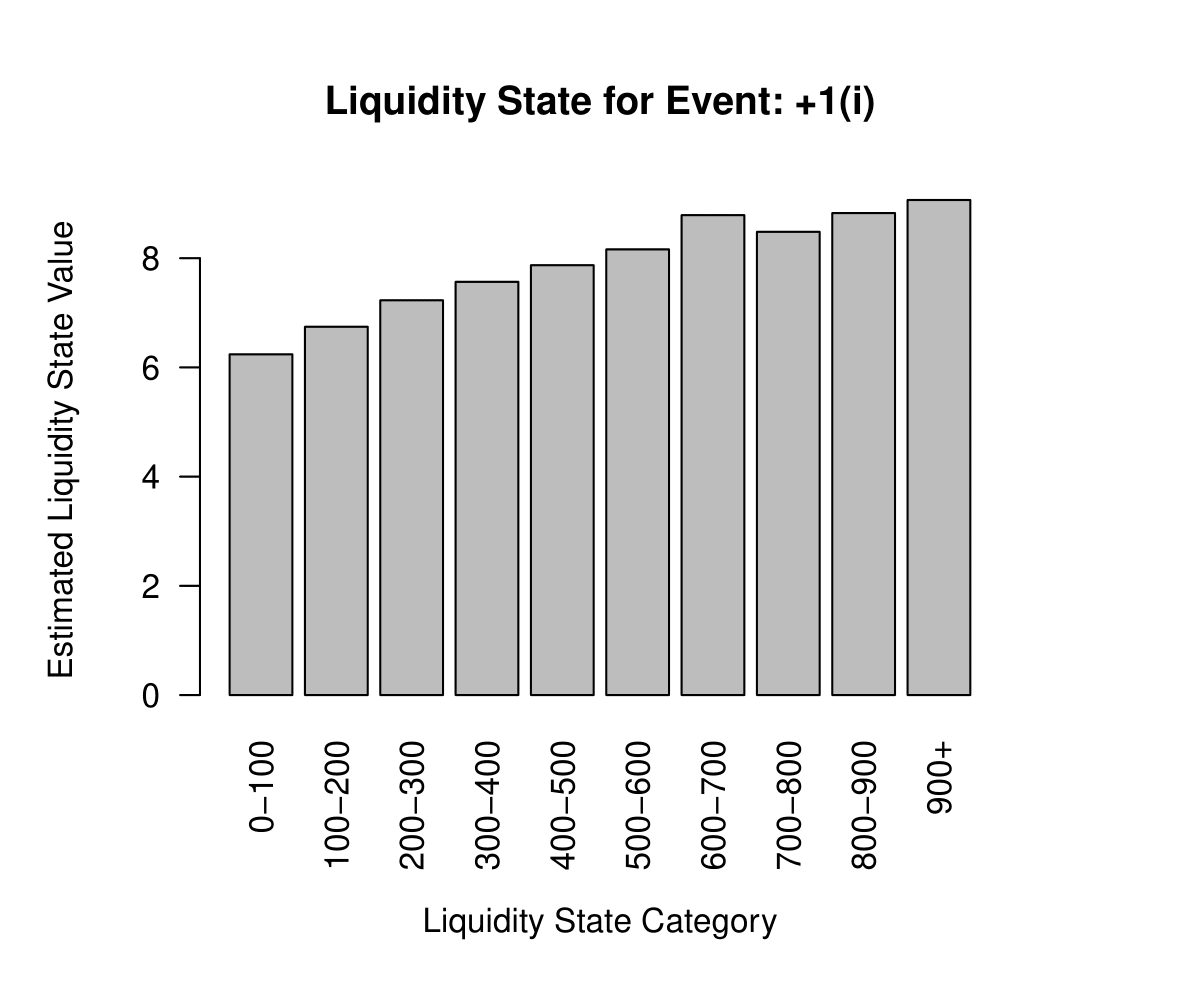} }}%
    \qquad
    \subfloat[Liquidity state for \texttt{+1(c)} ]{{\includegraphics[scale = 0.43]{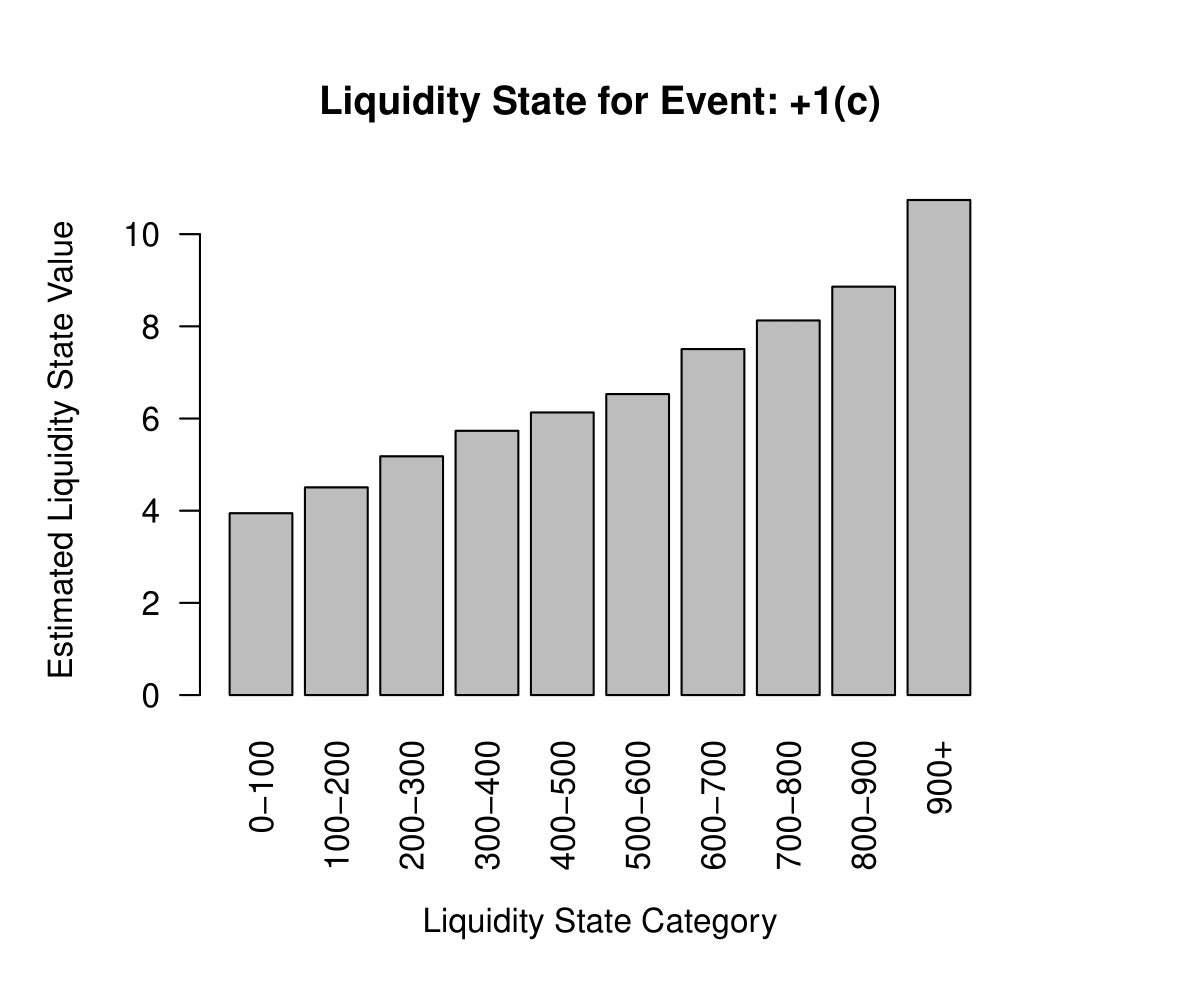} }}%
    \qquad
    \subfloat[Liquidity state for \texttt{+1(t)} ]{{\includegraphics[scale = 0.43]{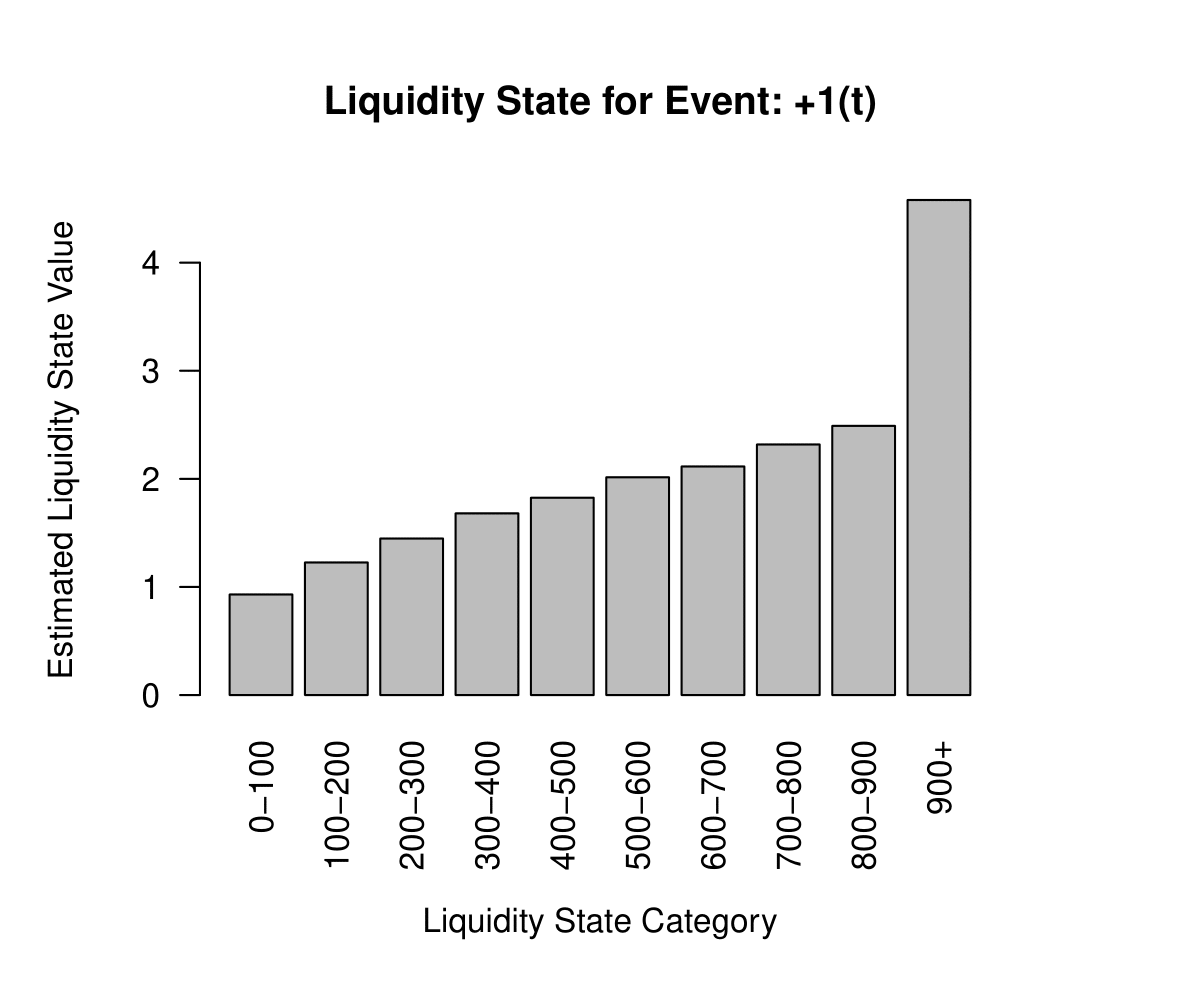} }}%
    \qquad
    \subfloat[Liquidity state for \texttt{-1(c)} ]{{\includegraphics[scale = 0.43]{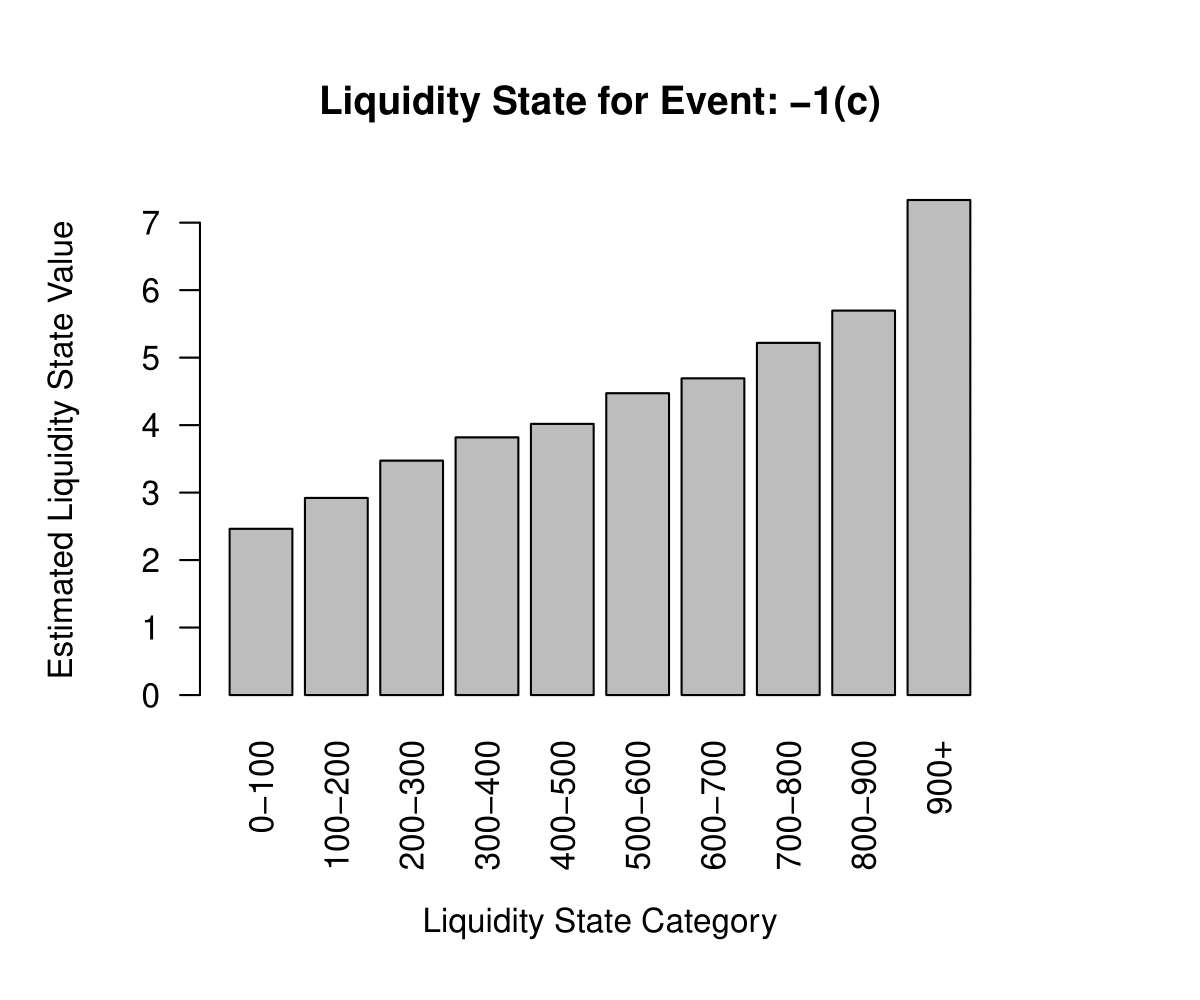} }}%
    \caption{Aggregated estimation result for liquidity state for event \texttt{+1(i)}, \-\texttt{+1(c)}, \-\texttt{+1(t)}, \-\texttt{-1(c)} under $(s=20 \text{ seconds}, \Delta=0.5 \text{ seconds})$. For these events the event arrival intensity increases as liquidity state increases.}
    \label{fig: state_increase largeDelta}
\end{figure}

The demonstrated liquidity state estimation results of the model with LASSO regularization is consistent with the results discussed in Section \ref{sec: liquidity state} when the bin-size is enlarged.

\subsection{Time factor}

Based on Figure \ref{fig: time factor} in Section \ref{sec: time factor}, the following Figure \ref{fig: time factor largeDelta} demonstrates the time factor estimations of the model with LASSO regularization.

\begin{figure}[H]
    \centering
    \subfloat[Time factor for event \texttt{+1(i)}]{{\includegraphics[scale = 0.43]{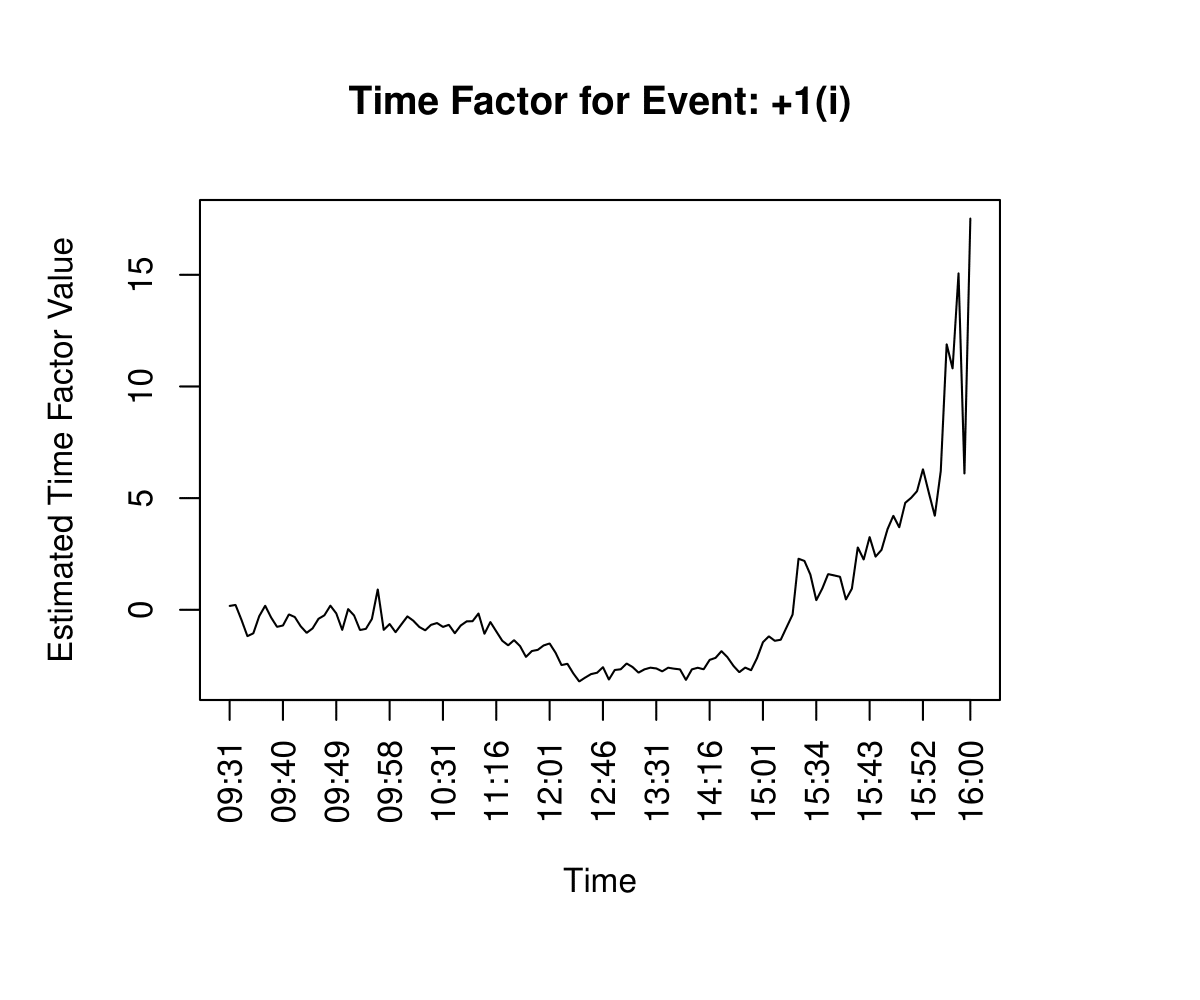} }}%
    \qquad
    \subfloat[Time factor for event \texttt{+3(t)}]{{\includegraphics[scale = 0.43]{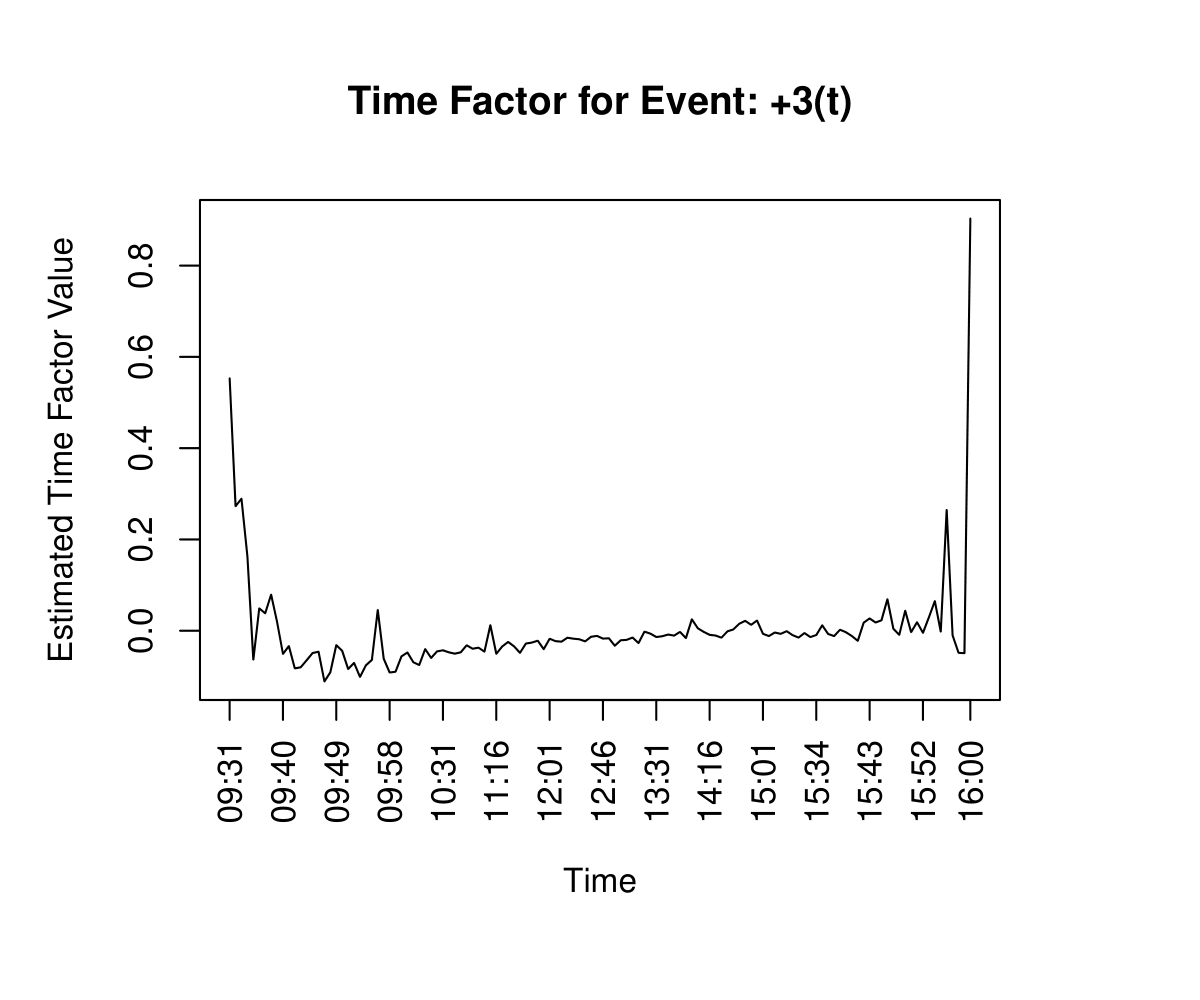} }}%
    \caption{Aggregated estimation result for time factor between 9:30 am and 4:00 pm under $(s=20 \text{ seconds}, \Delta=0.5 \text{ seconds})$. }
    \label{fig: time factor largeDelta}
\end{figure}

The demonstrated time factor estimation results of the model with LASSO regularization is consistent with the results discussed in Section \ref{sec: time factor} when the bin-size is enlarged.

\section{Additional model selection results}\label{sec_support: additional model selection}
This section presents additional model selection results based on AIC for the models when the size of order is ignored and when the bin-size is enlarged from 0.25 seconds to 0.5 seconds. The selection results are presented in the same format at Figure \ref{fig:aic_delta025_withsize} and Table \ref{table: aic diff}.

The following Figure \ref{fig:aic_delta025_nosize} and Table \ref{table: aic diff nosize} demonstrate the model selection result when the size of order is ignored. Model \cirnum{1}-\cirnum{7} are explained in Section \ref{sec: model selection result}.

\begin{figure}[H]
\centering
\includegraphics[scale = 0.20]{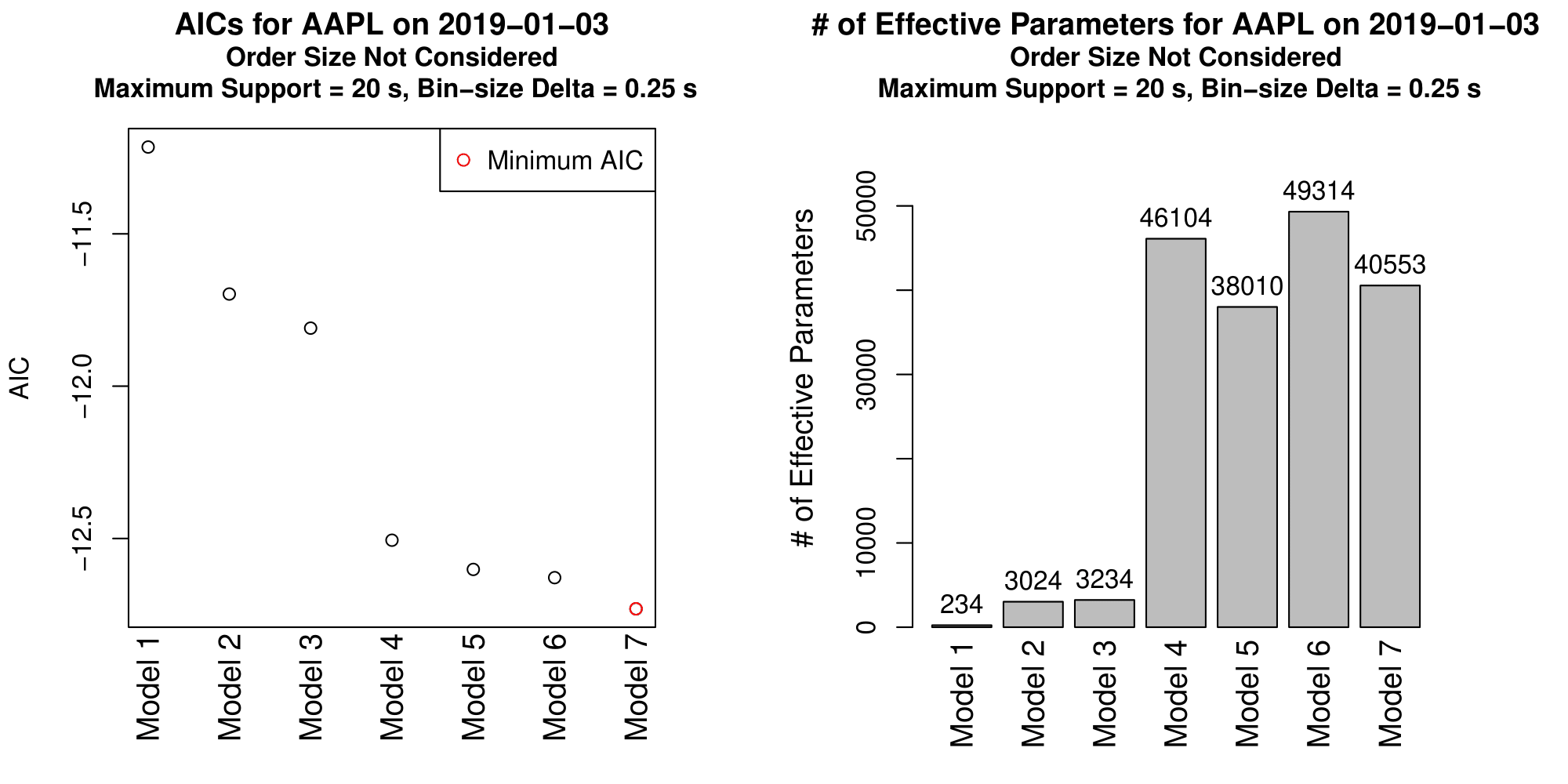}
\caption{AICs and number of effective parameters for seven model types for Apple.Inc on 2019-01-03 with maximum support $s = 20$ seconds and bin-size $\Delta=0.25$ seconds. All orders are considered to have size 1.}
\label{fig:aic_delta025_nosize}
\end{figure}

\begin{table}[H]
\centering
    \begin{threeparttable}
            \begin{tabular}{c|c|c|c|c|c|c|c} 
            % \hline\hline
            % \multicolumn{8}{c}{\begin{tabular}[c]{@{}c@{}}Summary Statistics of AIC Difference of Apple .Inc 2019-01-02 - 2019-01-31\\~Maximum Support s = 20s, bin-size $\Delta$ = 0.25s\\All orders are considered to have size 1\end{tabular}}                                                                    \\ 
            % \hline\hline
            \hline
            \begin{tabular}[c]{@{}c@{}}\textbf{AIC} \\\textbf{Difference}\end{tabular}    & \textbf{Min}   & \begin{tabular}[c]{@{}c@{}}\textbf{1st} \\\textbf{Quantile}\end{tabular} & \textbf{Median} & \textbf{Mean}  & \begin{tabular}[c]{@{}c@{}}\textbf{3rd} \\\textbf{Quantile}\end{tabular} & \textbf{Max}          & \begin{tabular}[c]{@{}c@{}} \textbf{\# of} \\\textbf{days with} \\ \textbf{decreased }\\\textbf{AIC}\end{tabular}  \\ 
            \hline
            \begin{tabular}[c]{@{}c@{}}\textbf{ \cirnum{4}} - \textbf{\cirnum{3}}\tnote{1}\end{tabular} & -1.43 & -0.30 & -0.15  & -0.23 & -0.005  & 0.13 & 15 out of 20  \\ 
            \hline
            \begin{tabular}[c]{@{}c@{}}\textbf{\cirnum{6}} - \textbf{\cirnum{4}}\tnote{2}\end{tabular}         & -0.23 & -0.19 & -0.18  & -0.18 & -0.16  & -0.12 & 20 out of 20  \\ 
            \hline
            \begin{tabular}[c]{@{}c@{}}\textbf{\cirnum{5}} - \textbf{\cirnum{4}}\tnote{3}\end{tabular}              & -0.15 & -0.12 & -0.11  & -0.10 & -0.08  & -0.02 & 20 out of 20  \\ 
            \hline
            \begin{tabular}[c]{@{}c@{}}\textbf{\cirnum{5}} - \textbf{\cirnum{3}}\tnote{4}\end{tabular}    & -1.46 & -0.38 & -0.26  & -0.33 & -0.12  & -0.016 & 20 out of 20  \\ 
            \hline
            \begin{tabular}[c]{@{}c@{}}\textbf{\cirnum{7}} - \textbf{\cirnum{6}}\tnote{5}\end{tabular}            & -0.16 & -0.12 & -0.11  & -0.10 & -0.09  & -0.02 & 20 out of 20                                                              \\
            \hline
            \end{tabular}
    
    \begin{tablenotes}
    {\setlength\itemindent{-10pt} \item Interpretations:}
        \item[1] The Hawkes part has stronger explanation power than the liquidity state and time factor part.
        \item[2] Adding the liquidity state and time factor to the Hawkes part further improves explanation power.
        \item[3] Adding LASSO (LASSO parameter 0.0005) to the Hawkes part further improves explanation power.
        \item[4] The Hawkes part with LASSO (LASSO parameter 0.0005) generates stronger explanation power than the liquidity state and time factor part.
        \item[5] Adding LASSO (LASSO parameter 0.0005) further improves the explanation power of the model with the liquidity state, time factor, and Hawkes.
    \end{tablenotes}
    \end{threeparttable}
        \caption{AIC difference summary statistics of Apple. Inc from 2019-01-02 to 2019-01-31 with maximum support $s = 20$ seconds and bin-size $\Delta=0.25$ seconds. All orders are considered to have size 1. AIC has been adjusted for sample size so that it reflects the AIC per single sample.}
    \label{table: aic diff nosize}
\end{table}

The following Figure \ref{fig:aic_delta05_withsize} and Table \ref{table: aic diff largeDelta} demonstrate the model selection result when the bin size is enlarged from 0.25 seconds to 0.5 seconds. Model \cirnum{1}-\cirnum{7} are explained in Section \ref{sec: model selection result}.

\begin{figure}[H]
\centering
\includegraphics[scale = 0.20]{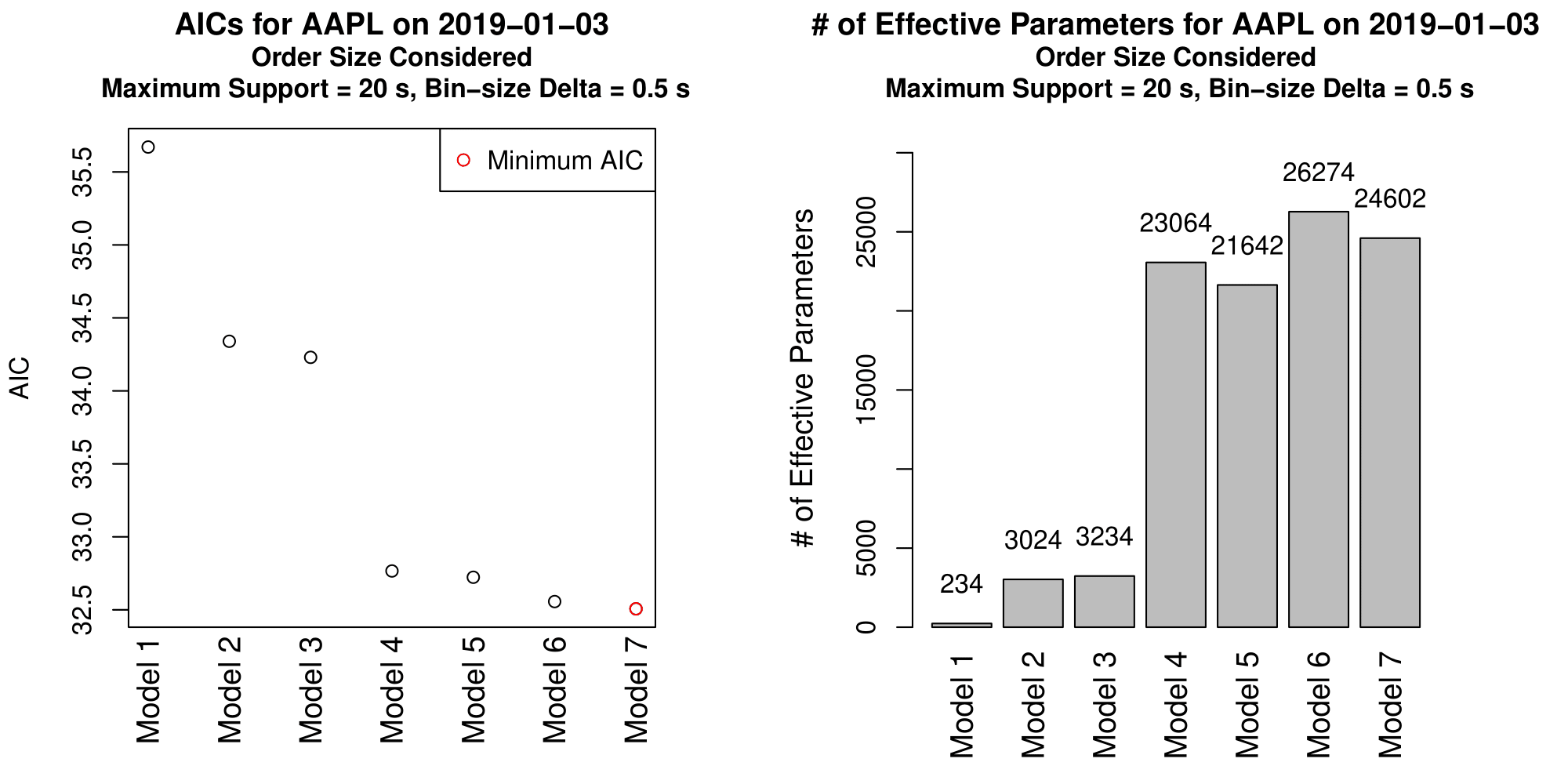}
\caption{AICs and number of effective parameters for seven model types for Apple.Inc on 2019-01-03 with maximum support $s = 20$ seconds and bin-size $\Delta=0.5$ seconds. Order sizes are considered in bin count sequence construction.}
\label{fig:aic_delta05_withsize}
\end{figure}

\begin{table}[H]
\centering
    \begin{threeparttable}
            \begin{tabular}{c|c|c|c|c|c|c|c} 
            % \hline\hline
            % \multicolumn{8}{c}{\begin{tabular}[c]{@{}c@{}}Summary Statistics of AIC Difference of Apple .Inc 2019-01-02 - 2019-01-31\\~Maximum Support s = 20s, bin-size $\Delta$ = 0.5s\\Order sizes are considered in bin count sequence construction\end{tabular}}                                                                    \\ 
            % \hline\hline
            \hline
            \begin{tabular}[c]{@{}c@{}}\textbf{AIC} \\\textbf{Difference}\end{tabular}   & \textbf{Min}   & \begin{tabular}[c]{@{}c@{}}\textbf{1st} \\\textbf{Quantile}\end{tabular} & \textbf{Median} & \textbf{Mean}  & \begin{tabular}[c]{@{}c@{}}\textbf{3rd} \\\textbf{Quantile}\end{tabular} & \textbf{Max}          & \begin{tabular}[c]{@{}c@{}} \textbf{\# of} \\\textbf{days with} \\ \textbf{decreased }\\\textbf{AIC}\end{tabular}  \\ 
            \hline
            \begin{tabular}[c]{@{}c@{}}\textbf{\cirnum{4}} - \textbf{\cirnum{3}}\tnote{1}\end{tabular} & -4.25 & -1.19 & -0.61  & -0.88 & -0.18  & 0.06 & 18 out of 20  \\ 
            \hline
            \begin{tabular}[c]{@{}c@{}}\textbf{\cirnum{6}} - \textbf{\cirnum{4}}\tnote{2}\end{tabular}         & -0.4 & -0.31 & -0.29  & -0.28 & -0.24  & -0.18 & 20 out of 20  \\ 
            \hline
            \begin{tabular}[c]{@{}c@{}}\textbf{\cirnum{5}} - \textbf{\cirnum{4}}\tnote{3}\end{tabular}              & -0.09 & -0.06 & -0.06  & -0.05 & -0.04  & -0.03 & 20 out of 20  \\ 
            \hline
            \begin{tabular}[c]{@{}c@{}}\textbf{\cirnum{5}} - \textbf{\cirnum{3}}\tnote{4}\end{tabular}    & -4.29 & -1.26 & -0.65  & -0.94 & -0.23  & 0.001 & 19 out of 20  \\ 
            \hline
            \begin{tabular}[c]{@{}c@{}}\textbf{\cirnum{7}} - \textbf{\cirnum{6}}\tnote{5}\end{tabular}            & -0.09 & -0.07 & -0.06  & -0.06 & -0.05  & -0.03 & 20 out of 20                                                              \\
            \hline
            \end{tabular}
    
    \begin{tablenotes}
    {\setlength\itemindent{-10pt} \item Interpretations:}
        \item[1] The Hawkes part has stronger explanation power than the liquidity state and time factor part.
        \item[2] Adding the liquidity state and time factor to the Hawkes part further improves explanation power.
        \item[3] Adding LASSO (LASSO parameter 0.0005) to the Hawkes part further improves explanation power.
        \item[4] The Hawkes part with LASSO (LASSO parameter 0.0005) generates stronger explanation power than the liquidity state and time factor part.
        \item[5] Adding LASSO (LASSO parameter 0.0005) further improves the explanation power of the model with the liquidity state, time factor, and Hawkes.
    \end{tablenotes}
    \end{threeparttable}
        \caption{AIC difference summary statistics of Apple. Inc from 2019-01-02 to 2019-01-31 with maximum support $s = 20$ seconds and bin-size $\Delta=0.5$ seconds. Order sizes are considered in bin count sequence construction.. AIC has been adjusted for sample size so that it reflects the AIC per single sample. }
    \label{table: aic diff largeDelta}
\end{table}

\end{appendices}

\end{document}